\documentclass[a4paper,UKenglish]{lmcs}
\usepackage{microtype}
\usepackage[utf8]{inputenc}

\usepackage{amsthm,amsmath}
\usepackage{amssymb}
\usepackage{latexsym}
\usepackage{bussproofs}
\usepackage{cmll}
\usepackage{stmaryrd}

\usepackage{comment}
\usepackage{todonotes}
\usepackage{array}
\usepackage{alltt}
\usepackage{array}
\usepackage{tikz}
\usetikzlibrary{matrix,shapes}
\usepackage{pgfplots}

\usepackage{proof}

\usepackage{lineno}
\newcommand\CMLLPAR{
\usepackage{cmll}
\newcommand\IPar{\mathord{\parr}}
}

\CMLLPAR

\theoremstyle{plain}

\theoremstyle{definition}

\theoremstyle{remark}

\renewcommand\paragraph{\subsubsection*}

\newcommand{\Endproof}{
  \ifmmode % if math mode, assume display: omit penalty etc.
  \else \leavevmode\unskip\penalty9999 \hbox{}\nobreak\hfill
  \fi
  \quad\hbox{$\Box$}
  \par\medskip}

\newcommand\Eqref[1]{(\ref{#1})}

\newcommand\Eg{\textsl{e.g.}~}

\renewcommand{\phi}{\varphi}
\renewcommand\epsilon{\varepsilon}

\newcommand{\Implies}{\Rightarrow}

\newcommand\Equiv{\Leftrightarrow}
\newcommand{\St}{\mid}

\newcommand{\Fbot}{{\mathord{\perp}}}
\newcommand{\Top}{\top}

\newcommand\cA{\mathcal{A}}
\newcommand\cB{\mathcal{B}}
\newcommand\cC{\mathcal{C}}

\newcommand\cE{\mathcal{E}}
\newcommand\cF{\mathcal{F}}

\newcommand\cI{\mathcal{I}}

\newcommand\cK{\mathcal{K}}
\newcommand\cL{\mathcal{L}}

\newcommand\cP{\mathcal{P}}

\newcommand\cT{\mathcal{T}}
\newcommand\cU{\mathcal{U}}
\newcommand\cV{\mathcal{V}}

\newcommand\cY{\mathcal{Y}}
\newcommand\cZ{\mathcal{Z}}

\newcommand\Fini{{\mathrm{fin}}}

\newcommand\Part[1]{{\mathcal P}({#1})}

\newcommand{\Linarrow}{\multimap}

\newcommand\Myleft{}
\newcommand\Myright{}

\newcommand\Web[1]{\Myleft|{#1}\Myright|}

\newcommand\Cl[1]{\mbox{\textsf{Cl}}({#1})}

\newcommand\Par[2]{{#1}\mathrel{\IPar}{#2}}
\newcommand\Parp[2]{\left({#1}\mathrel{\IPar}{#2}\right)}
\newcommand\ITens{\otimes}
\newcommand\Tens[2]{{#1}\ITens{#2}}
\newcommand\Tensp[2]{\left({#1}\ITens{#2}\right)}
\newcommand\IWith{\mathrel{\&}}
\newcommand\With[2]{{#1}\IWith{#2}}
\newcommand\IPlus{\oplus}
\newcommand\Plus[2]{{#1}\IPlus{#2}}
\newcommand\Orth[2][]{#2^{\Fbot_{#1}}}
\newcommand\Orthp[2][]{(#2)^{\Fbot_{#1}}}

\newcommand\Pair[2]{\langle{#1},{#2}\rangle}

\newcommand\Inj[1]{\overline\pi_{#1}}

\newcommand\Biorth[1]{#1^{\Fbot\Fbot}}

\newcommand\Triorth[1]{{#1}^{\Fbot\Fbot\Fbot}}

\newcommand\One{1}
\newcommand\Onelem{*}

\newcommand\Seq[1]{\vdash{#1}}

\newcommand\Partfin[1]{{\cP_\Fini}({#1})}

\newcommand\IExcl{{\mathord{!}}}

\newcommand\Locun[1]{1^J}

\newcommand\Isom\simeq

\newcommand\Comp{\mathrel\circ}

\newcommand\Funinv[1]{#1^{-1}}

\newcommand\Limpl[2]{{#1}\Linarrow{#2}}

\newcommand\Nat{{\mathbb{N}}}

\newcommand\Snat{\mathsf N}

\newcommand\Zero{0}
\newcommand\App[2]{\left({#1}\right){#2}}

\newcommand\Abst[3]{\lambda\Var{#1}^{#2}\,{#3}}

\newcommand\List[3]{#1_{#2},\dots,#1_{#3}}

\newcommand\Subst[3]{{#1}\left[{#2}/{#3}\right]}

\newcommand\Substbis[2]{{#1}\left[{#2}\right]}

\newcommand\Evlin{\operatorname{\mathsf{ev}}}
\newcommand\REL{\operatorname{\mathbf{Rel}}}

\newcommand\Rel[1]{\mathrel{#1}}
\newcommand\Reltr[1]{\mathrel{#1^*}}

\newcommand\Redst[1]{\mathop{\mathsf{Red}}}

\newcommand\Tuple[1]{\langle{#1}\rangle}

\newcommand\Msetofsubst[1]{\bar F}

\newcommand\Matapp[2]{{#1}\Compl{#2}}

\newcommand\Leftu{\lambda}
\newcommand\Rightu{\rho}
\newcommand\Assoc{\alpha}
\newcommand\Sym{\gamma}

\newcommand\Retri\zeta
\newcommand\Retrp\rho

\newcommand\Impl[2]{{#1}\Rightarrow{#2}}

\newcommand\Tsem[1]{\llbracket{#1}\rrbracket}

\newcommand\Psem[1]{\llbracket{#1}\rrbracket}

\newcommand\Tnat\iota

\newcommand\Succ[1]{\operatorname{\mathsf{succ}}(#1)}
\newcommand\Num[1]{\underline{#1}}
\newcommand\Loop\Omega

\newcommand\Tseq[3]{{#1}\vdash{#2}:{#3}}

\newcommand\Timpl\Impl
\newcommand\Simpl\Impl

\newcommand\Weak[1]{\operatorname{\mathsf{w}}_{#1}}
\newcommand\Weakmu[1]{\operatorname{\mathsf{w}}'_{#1}}
\newcommand\Contr[1]{\operatorname{\mathsf{contr}}_{#1}}
\newcommand\Contrmu[1]{\operatorname{\mathsf{contr}}'_{#1}}

\newcommand\Der[1]{\operatorname{\mathsf{der}}_{#1}}
\newcommand\Dermu[1]{\operatorname{\mathsf{der}}'_{#1}}

\newcommand\Digg[1]{\operatorname{\mathsf{dig}}_{#1}}
\newcommand\Diggmu[1]{\operatorname{\mathsf{dig}}'_{#1}}

\newcommand\Lfun{\operatorname{\mathsf{fun}}}

\newcommand\Ltrace{\mathsf{tr}}

\newcommand\Id{\operatorname{\mathsf{Id}}}

\newcommand\Proj[1]{\pi_{#1}}

\newcommand\Excl[1]{\oc{#1}}
\newcommand\Exclmu[1]{\oc'{#1}}
\newcommand\Exclp[1]{\oc({#1})}

\newcommand\Int[1]{\wn{#1}}
\newcommand\Intmu[1]{\wn'{#1}}

\newcommand\Prom[1]{#1^!}
\newcommand\Prommu[1]{#1^{!'}}

\newcommand\Relincl\eta
\newcommand\Relrestr\rho

\newcommand\Seely{\mathsf m}
\newcommand\Seelyz{\mathsf m^0}
\newcommand\Seelyt{\mathsf m^2}

\newcommand\Seelyinvz{\mathsf n^0}
\newcommand\Seelyinvt{\mathsf n^2}

\newcommand\Monoidalz{\mu^0}
\newcommand\Monoidalt{\mu^2}

\newcommand\Compl{\,}
\newcommand\Curlin{\mathsf{cur}}

\newcommand\Op[1]{{#1}^{\mathsf{op}}}

\newcommand\Em[1]{{#1}^\oc}

\newcommand\Let[3]{\mathsf{let}(#1,#2,#3)}

\newcommand\Snum[1]{\overline{#1}}

\newcommand\Vect[1]{\vec{#1}}

\newcommand\Bnfeq{\mathrel{\mathord:\mathord=}}
\newcommand\Bnfor{\,\,\mathord|\,\,}

\newcommand\Subcoh{\subseteq}
\newcommand\Coh[3]{#2\coh_{#1}#3}
\newcommand\Incoh[3]{#2\incoh_{#1}#3}
\newcommand\Scoh[3]{#2\scoh_{#1}#3}
\newcommand\Sincoh[3]{#2\sincoh_{#1}#3}

\newcommand\COH{\mathbf{Coh}}
\newcommand\COHT{\mathbf{CohT}}
\newcommand\COHPO{\COH_\subseteq}
\newcommand\Cohlub{\cup}

\newcommand\Tcohca[1]{\underline{#1}}
\newcommand\Tcoht[1]{\mathsf T{#1}}
\newcommand\Tcohtp[1]{\cT(#1)}

\newcommand\Tot[1]{\mathop{\mathsf{Tot}}(#1)}

\newcommand\Vcsnot[1]{\mathbb{#1}}
\newcommand\Vcstnot[1]{\mathbb{#1}}

\newcommand\Strfun[1]{\overline{#1}}
\newcommand\Strnat[1]{\widehat{#1}}

\newcommand\COHLIN{\mathbf{Coh}}
\newcommand\Cohemb[2]{\eta_{#1,#2}^+}
\newcommand\Cohret[2]{\eta_{#1,#2}^-}
\newcommand\Indcoh[1]{{#1}^+}
\newcommand\Procoh[1]{{#1}^-}

\newcommand\Ddvcst[1]{\Orthp{#1}}
\newcommand\Ddvcstnp[1]{\Orth{#1}}

\newcommand\Cohempty{\emptyset}
\newcommand\Vcsfp[1]{\sigma\,#1}
\newcommand\ALGFUN[2]{\mathbf{Alg}_{#1}(#2)}
\newcommand\COALGFUN[2]{\mathbf{Coalg}_{#1}(#2)}

\newcommand\COHLINT{\mathbf{CohT}}

\newcommand\Vcstca[1]{\underline{#1}}
\newcommand\Vcstot[1]{\mathsf T{#1}}
\newcommand\Vcstotp[1]{\mathsf T{(#1)}}

\newcommand\Vcstmuu[1]{\mu\,#1}
\newcommand\Vcstmu[1]{\mu\,#1}
\newcommand\Vcstnu[1]{\nu\,#1}

\newcommand\Lfpll[2]{\mu#1\,#2}
\newcommand\Gfpll[2]{\nu#1\,#2}
\newcommand\LLvars{\mathcal V}

\newcommand\Oneelem{*}

\newcommand\Naxiom{(\mathsf{ax})}
\newcommand\Ncut{(\mathsf{cut})}
\newcommand\None{(\One)}
\newcommand\Ntens{(\ITens)}
\newcommand\Nbot{(\Fbot)}
\newcommand\Npar{(\IPar)}
\newcommand\Ntop{(\Top)}
\newcommand\Nplusl{(\IPlus_1)}
\newcommand\Nplusr{(\IPlus_2)}
\newcommand\Nwith{(\IWith)}
\newcommand\Nweak{(\mathsf{w})}
\newcommand\Ncontr{(\mathsf{c})}
\newcommand\Nder{(\mathsf{d})}
\newcommand\Nprom{(\mathsf{p})}
\newcommand\Nlfp{(\mu-\mathsf{fold})}
\newcommand\Ngfp{(\nu-\mathsf{rec})}
\newcommand\Ngfpfold{(\nu-\mathsf{fold})}
\newcommand\Ngfpbis{(\nu-\mathsf{rec}')}

\newcommand\Nzero{(0)}
\newcommand\Nsucc{(\mathsf{succ})}
\newcommand\Nintit{(\mathsf{it}_\Tnat)}
\newcommand\Ngprom{(\mathsf{gp})}
\newcommand\Ngweak{(\mathsf{gw})}
\newcommand\Ngcontr{(\mathsf{gc})}

\newcommand\MULL{\mu\mathsf{LL}}
\newcommand\LL{\mathsf{LL}}

\newcommand\Sone{1}
\newcommand\Sbot{\mathord\bot}

\newcommand\Coalgca[1]{\underline{#1}}
\newcommand\Coalgstr[1]{h_{#1}}

\newcommand\LCAT{\cL}

\newcommand\LCATPO{\LCAT_{\mathord\subseteq}}

\newcommand\Obj[1]{#1}
\newcommand\Promp[1]{\Prom{(#1)}}

\newcommand\Termm[1]{\tau_{#1}}

\newcommand\Fcomod[1]{\mathsf{Comod}_{#1}}

\newcommand\Klp[2]{{#1}_{#2}}

\newcommand\Mtens[3]{#2\otimes_{#1}#3}

\newcommand\Mevlin[1]{\Evlin_{#1}}

\newcommand\Kcomod[2]{#1[#2]}

\newcommand\Morth[2]{{#2}^{\Fbot[#1]}}
\newcommand\Mbiorth[2]{{#2}^{\Fbot[#1]\Fbot[#1]}}

\newcommand\Mexcl[2]{\oc_{#1}{#2}}
\newcommand\Mder[1]{\Der{}[#1]}
\newcommand\Mdigg[1]{\Digg{}[#1]}

\newcommand\Mfun[2]{{#1}[#2]}

\newcommand\Strid{\cI}
\newcommand\Strcst[1]{\cK^{#1}}

\newcommand\Fungfp[1]{\nu#1}
\newcommand\Funlfp[1]{\mu#1}

\newcommand\Totop[1]{\Theta(#1)}

\newcommand\Lnat{\iota_l}

\newcommand\Gtnat{\mathsf{nat}}
\newcommand\Rec[3]{\mathsf{rec}(#1,#2,#3)}
\newcommand\Gredwh{\beta_{wh}}
\newcommand\Appsep{\,}
\newcommand\GODELT{\mathsf T}
\newcommand\Tgtrad[1]{#1^*}
\newcommand\Tgtradc[1]{#1^*}

\newcommand\Pprom[3]{\chi(#1,\Excl{#2}/#3)}
\newcommand\Ppromcl[1]{\chi(#1)}

\newcommand\Greduc[1]{\mathsf{Red}_{#1}}
\newcommand\Var[1]{\mathsf{#1}}

\newcommand\Exclmuext[1]{\mathsf{ext^{\oc'}}(#1)}

\newcommand\Seemu{{\mathsf m^2}'}

\newcommand\Seeinvmu{{\mathsf n^2}'}

\newcommand\Wweak{\mathsf w}
\newcommand\Wder[1]{\mathsf d(#1)}
\newcommand\Wcontr[2]{\mathsf c(#1,#2)}

\newcommand\Winl[1]{1\cdot#1}
\newcommand\Winr[1]{2\cdot#1}

\newcommand\Wprl[1]{\mathsf p_1(#1)}
\newcommand\Wprr[1]{\mathsf p_2(#1)}

\newcommand\Subtree{\mathsf{st}}
\newcommand\Emptypath{\langle\rangle}
\newcommand\Conspath[2]{#1\,#2}
\newcommand\Pathes[1]{\mathsf{paths}(#1)}

\newcommand\Tcohforg{\cU}
\newcommand\Tcohempty{\cZ}

\newcommand\ST{\mathsf{T}}
\newcommand\SF{\mathsf{F}}

\begin{document}

% \special{papersize=8.5in,11in}
% \setlength{\pdfpageheight}{\paperheight}
% \setlength{\pdfpagewidth}{\paperwidth}

\title[Semantics of $\LL$ with fixed points]{On the denotational
  semantics of Linear Logic with least and greatest fixed points of
  formulas} % 'preprint' option specified.

\author[]{Thomas Ehrhard}
\address{CNRS, IRIF,
    Université de Paris, France}
\email{ehrhard@irif.fr}
\author[]{Farzad Jafar-Rahmani}
\address{Université de Paris, IRIF, 
    CNRS, France}
\email{farzadhtg@gmail.com}
%
% \authorrunning{Lecture Notes in Computer Science: Authors' Instructions}
% (feature abused for this document to repeat the title also on left hand pages)
% the affiliations are given next; don't give your e-mail address
% unless you accept that it will be published

% DBR: pas d'auteur dans la version soumise
% \authorinfo{Thomas Ehrhard}
%            {PPS, CNRS and University Paris Diderot}
%            {thomas.ehrhard@pps.univ-paris-diderot.fr}

% \authorinfo{}{}{}

\maketitle

\begin{abstract}
  We develop a denotational semantics of Linear Logic with least and
  greatest fixed points in coherence spaces (where both fixed points are
  interpreted in the same way) and in coherence spaces with totality
  (where they have different interpretations). These constructions can
  be carried out in many different denotational models of $\LL$
  (hypercoherences, Scott semantics, finiteness spaces etc). We also
  present a natural embedding of Gödel System $\ST$ in $\LL$ with
  fixed points thus enforcing the expressive power of this system as a
  programming language featuring both normalization and a huge
  expressive power in terms of data types.
\end{abstract}

\section{Introduction}
Propositional Linear Logic is a well-established logical system
introduced by Girard in~\cite{Girard87}. It provides a fine-grain
analysis of proofs in intuitionistic and classical logic, and more
specifically of their cut-elimination. $\LL$ features indeed a
well-behaved logical account of the structural rules (weakening,
contraction) which are handled implicitly in intuitionistic and
classical logic. For this reason, $\LL$ has many useful outcomes in the
Curry-Howard based approach to the theory of programming: logical
understanding of evaluation strategies (CBN and CBV correspond to two
different translations from the lambda-calculus into $\LL$, and such
translations extend naturally to abstract machines), new presentation
of proofs/programs (proof-nets), connections with other branches of
mathematics (linear algebra, differential calculus etc), new
operational semantics (geometry of interaction)\dots

However, just as the simply typed lambda-calculus, propositional $\LL$
is not a reasonable programming language, by lack of data-types and
iteration or recursion principles. This is usually remedied by
extending propositional $\LL$ to the $2^{\mathrm{nd}}$ order, thus
defining a logical system in which Girard's System
$\SF$~\cite{Girard89} can be embedded. As explained in~\cite{Girard89}
for System $\SF$, this extension allows to represent many data-types,
for instance the type of natural numbers can be written as
$\Tnat=\forall\zeta\,\Limpl{\Excl{(\Limpl{\Excl\zeta}{\zeta}})}{\Limpl{\Excl\zeta}{\zeta}}$,
and then integers are represented as \emph{Church numerals}. The
resulting logical system is extremely expressive in terms of
representable computable functions: all functions $\Nat\to\Nat$ whose
totality can be proven in second order arithmetics are
representable. In contrast, its \emph{algorithmic} expressiveness
seems quite poor: as is well-known it is not possible to write a term
$\Limpl{\Excl\Tnat}{\Tnat}$ which computes the predecessor function in
one (or a uniformly bounded) number of reduction steps.

Another option to turn propositional $\LL$ into a programming language
is to extend it with least and greatest fixed points of formulas.  The
kind of extension has early been suggested by Girard in an unpublished
note~\cite{Girard92}, though the first comprehensive proof-theoretic
investigation of such an extension of $\LL$ is rather
recent~\cite{Baelde12}: Baelde considers an extension of
Multiplicative Additive $\LL$ sequent calculus with least and greatest
fixed points, axiomatized by means of two deduction rules. His main
motivation for introducing this extension of $\LL$ comes from the need
of powerful logical systems for formal verification of programs and
the investigations on $\MULL$ were carried out mainly from a
proof-search perspectives (cut-elimination, focusing etc). However
from a Curry-Howard viewpoint, fixed points allow to define inductive
data-types (integers, lists, trees) and coinductive types (streams and
other infinite structures). So $\MULL$ can also be considered as a
programming language and it is this approach that we develop in this
paper. Admittedly the rules associated with fixed points are complex
(this is especially true of the $\nu$-introduction rule, the Park's
rule) and lead to subtle cut-elimination proof rewrite rules for which
Baelde could prove a restricted form of cut-elimination, sufficient
for establishing for instance that a proof of the type of integers
$\Lfpll\zeta{(\Plus\One\zeta)}$ necessarily reduces to an
integer. There are also alternative proof-systems for the same logic,
involving infinite or cyclic proofs,
see~\cite{BaeldeDoumaneSaurin16}, whose connections with the
aforementioned finitary proof-system are not completely clear.

Since the proof-theory (and hence the ``operational semantics'') of
$\MULL$ is sophisticated and still under development, it is
particularly important to investigate its denotational semantics,
whose definition does not rely on the rewrite system $\MULL$ is
equipped with.
% What follows is wrong
%
% In~\cite{Clairambault09}, a game semantics of an
% extension of intuitionistic logic with least and greatest fixed points
% is studied. As far as we understand that paper, there is no
% restriction in the logical system on the polarity of the occurrences
% of the variables, offering the possibility to write non-terminating
% programs.
We develop here a semantics of $\MULL$ using coherence
spaces~\cite{Girard87} equipped with a notion of totality: our model
accounts for the termination of computations in $\MULL$ and interprets
least and greatest fixed points in different ways. It presents some
similarities with the game-theoretic model of
Clairambault~\cite{Clairambault09,Clairambault13} of intuitionistic
logic with least and greatest fixed points (see below).

Girard introduced coherence spaces in~\cite{Girard86} in his
denotational study of System $\SF$. In this model, types (with free
variables) are interpreted as continuous (actually, stable) functors
on a category of coherence spaces and embedding-retraction pairs. Our
basic observation is that such functors admit ``least fixed points''
in a categorical sense. In the case of $\MULL$ we have to depart from
Girard's setting because formulas must act on proofs and not only on
embedding-retraction pairs: given a formula $F$ and a proof $\pi$ of
$\Limpl AB$, $\MULL$ uses crucially a proof $\Subst F\pi\zeta$ of
$\Limpl{\Subst FA\zeta}{\Subst FB\zeta}$ defined by induction on $F$
(syntactic functoriality). This means that the functors interpreting
formulas must act on all stable morphisms, and not only on
embedding-retraction pairs. This is made possible by the fact that, in
$\MULL$, type variables have only positive occurrences in formulas.

$\MULL$ has a construction $\Lfpll\zeta F$ for least fixed points and
a dual construction $\Gfpll\zeta F$ for greatest fixed points. The
logical rule for introducing $\Gfpll\zeta F$ allows to deduce
$\Seq{\Orth A,\Gfpll\zeta F}$ from $\Seq{\Orth A,\Subst FA\zeta}$
(Park's Rule). We prefer to consider a slightly generalized version of
this rule: deduce $\Seq{\Int\Gamma,\Orth A,\Gfpll\zeta F}$ from
$\Seq{\Int\Gamma,\Orth A,\Subst FA\zeta}$ (which does not increase
expressiveness but makes the embedding of functional formalisms such
as Gödel's System $\ST$ in $\MULL$ much more natural). We need
therefore a version of syntactic functoriality which accommodates
these additional contexts $\Int\Gamma$. The denotational counterpart
is that our functors interpreting types are equipped with a
\emph{strength} (as in~\cite{Clairambault09,Clairambault13}, and
essentially for the same reason). All together, these requirements
lead to the definition of a \emph{variable coherence space}
(VCS). Such a functor has a fixed point which is at the same time an
initial algebra and a final coalgebra, meaning that $\Lfpll\zeta F$
and $\Gfpll\zeta F$ have the same interpretation.

This first model, based on coherence spaces, admits morphisms which
are not \emph{total} (for instance the interpretation of
$\Limpl\Tnat\Tnat$, where $\Tnat=\Lfpll\zeta{(\Plus\One\zeta)}$ is the
type of integers, contains non total functions
$\Nat\to\Nat$). Following~\cite{Girard86} we equip our coherence spaces
with a semantic notion of \emph{totality}. We introduce accordingly
\emph{variable coherence spaces with totality} which are VCS's equipped
with a functorial notion of totality. Using Knaster-Tarski Theorem, we
show that these functors have least and greatest fixed points, which now
are distinct in general. The benefit of such models is that they give an
abstract and compositional account of normalization: the
interpretation of any proof $\pi$ of $\Seq\Tnat$ is an integer, which
is the normal form of $\pi$: the model ``computes'' this integer
without reducing the proof.

\paragraph{Contents} In Section~\ref{sec:LL-models} we recall basic
categorical notions essential in the paper: Seely categories,
Eilenberg-Moore and Kleisli categories of the exponential, strong
functors and their extension to categories of comodules and, last,
initial algebras (least fixed points) and final coalgebras (greatest
fixed points) of (strong) functors. Section~\ref{sec:coherence-spaces}
is devoted to a general presentation of coherence spaces and to a well
behaved notion of strong functors on coherence spaces admitting fixed
points (Variable Coherence Spaces, VCSs). In this setting, there is no
distinction between least fixed points and greatest fixed points, due
to the presence of partial computations in the
model. Section~\ref{sec:coh-tot} endows coherence spaces with a notion
of totality rejecting such partial morphisms and introduces an adapted
notion of variables types (VCSTs), giving rise to a clear distinction
between least and greatest fixed points in this model $\COHT$. In
Section~\ref{sec:MULL} we present $\MULL$ and various associated
concepts: functorial extension of formulas, cut elimination,
interpretation of formulas and proofs in the model $\COHT$. The idea
that a semantic notion of totality enforces a distinction between
least and greatest fixed points is of course not new, it is essential
for instance in~\cite{Clairambault09,Clairambault13}.  In
Section~\ref{sec:examples} we present various examples, stressing in
particular that Gödel's System $\ST$ can be embedded in $\MULL$
(following Clairambault~\cite{Clairambault13}) and hence showing that
this system has a significant expressive power from the viewpoint of
the Curry-Howard correspondence. We also analyze in our model the
encoding of exponentials using least and greatest fixed points
proposed in~\cite{Baelde12}; we show in particular that these
exponentials do not give rise to a Seely category, thus justifying our
choice of considering the whole system $\MULL$, with exponentials. As
an outcome of the paper, Section~\ref{sec:generalizations} introduces
a quite simple and general notion of categorical model of classical
$\MULL$, of which the model $\COHT$ is an instance.

\paragraph{Related work} Most importantly, we want to mention again
the work of Pierre Clairambault (see~\cite{Clairambault09} and the
long version~\cite{Clairambault13}) who investigates the denotational
semantics of an extension of intuitionistic logic with least and
greatest fixed points. Our categorical notion of model of classical
$\MULL$ of Definition~\ref{def:categorical-muLL-models} can probably
be seen as a ``linearized version'' of his notion of $\mu$-closed
category (an extension of the notion of cartesian closed categories
with least and greatest fixed points) ---~perhaps through some kind of
Kleisli construction~--- though our setting seems simpler as we do not
need contravariant strong functors, only covariant ones.  The concrete
model considered by Clairambault is based on games and features a
notion of totality guaranteeing that all computations terminate and
enforcing the distinction between least and greatest fixed points, just
as in the present work. Due to technical limitations intrinsically
related to game semantics, Clairambault's model is restricted to the
\emph{pseudo-polynomial} fragment where free variables never appear at
the right hand side of an implication. No such limitations are
necessary in our coherence space interpretation.

\begin{rem}
  We use the following notational conventions: $\Vect a$ stands for a
  list $(\List a1n)$. When we write natural transformations, we very
  often omit the objects where they are taken and prefer to keep these
  objects implicit for the sake of readability, because they can
  easily be retrieved from the context. If $\cF:\cA\times\cB\to\cC$ is
  a functor and $A$ is an object of $\cA$ (notation $A\in\cA$) then
  $\cF_A:\cB\to\cC$ is the functor defined by $\cF_A(B)=\cF(A,B)$ and
  $\cF_A(f)=\cF(\Id_A,f)$.
\end{rem}

\section{Categorical models of $\LL$}\label{sec:LL-models}

A model of $\LL$ consists of the following data (our main reference is
the notion of a \emph{Seely category} as presented
in~\cite{Mellies09}. We refer to that survey for all the technical
material that we do not record here).

A symmetric monoidal closed category
$(\LCAT,\ITens,\One,\Leftu,\Rightu,\Assoc,\Sym)$ where
$\Leftu_X\in\cL(\Tens\One X,X)$, $\Rightu_X\in\cL(\Tens X\One,X)$,
$\Assoc_{X,Y,Z}\in\cL(\Tens{(\Tens XY)}{Z},\Tens{X}{(\Tens{Y}{Z})})$
and $\Sym_{X,Y}\in\cL(\Tens XY,\Tens YX)$ are natural isomorphisms
satisfying coherence diagrams that we do not record here. We use
$\Limpl XY$ for the object of linear morphisms from $X$ to $Y$,
$\Evlin\in\LCAT(\Tens{(\Limpl XY)}{X},Y)$ for the evaluation morphism
and $\Curlin$ for the linear curryfication map
$\LCAT(\Tens ZX,Y)\to\LCAT(Z,\Limpl XY)$. We assume this SMCC to be
$*$-autonomous with dualizing object $\Sbot$. We use $\Orth X$ for the
object $\Limpl X\Sbot$ of $\LCAT$ (the dual, or linear negation, of
$X$).

$\LCAT$ is cartesian with final object $\Top$, product
$(\With {X_1}{X_2},(\Proj i:\With{X_1}{X_2}\to X_i)_{i=1,2})$. By
$*$-autonomy $\LCAT$ is cocartesian with initial object $\Zero$,
coproduct $\IPlus$ and injections $\Inj i$.

We are given a comonad $\Excl\_:\LCAT\to\LCAT$ with counit
$\Der X\in\LCAT(\Excl X,X)$ (\emph{dereliction}) and comultiplication
$\Digg X\in\LCAT(\Excl X,\Excl{\Excl X})$ (\emph{digging}) together
with a strong symmetric monoidal structure (Seely natural isos
$\Seelyz:\Sone\to\Excl\Top$ and $\Seelyt$ with
$\Seelyt_{X_1,X_2}:\Tens{\Excl{X_1}}{\Excl{X_2}}\to\Exclp{\With{X_1}{X_2}}$,
we use $\Seelyinvz$ and $\Seelyinvt_{X_1,X_2}$ for the inverses of
these isos) for the functor $\Excl\_$, from the symmetric monoidal
category $(\LCAT,\IWith)$ to the symmetric monoidal category
$(\LCAT,\ITens)$ satisfying an additional coherence condition
wrt.~$\Digg{}$. It is also important to remember that this strong
monoidal structure allows to define a weak monoidal structure
$(\Monoidalz,\Monoidalt)$ of ``$\oc$'' from $(\LCAT,\ITens)$ to
itself. More precisely $\Monoidalz\in\LCAT(\Sone,\Excl\Sone)$ and
$\Monoidalt_{X_1,X_2}\in
\LCAT(\Tens{\Excl{X_1}}{\Excl{X_2}},\Exclp{\Tens{X_1}{X_2}})$ are
defined using $\Seelyz$ and $\Seelyt$. Also, for each object $X$ of
$\cL$, there is a canonical structure of commutative $\ITens$-comonoid
on $\Excl X$ given by $\Weak X\in\cL(\Excl X,\One)$ and
$\Contr X\in\cL(\Excl X,\Tens{\Excl X}{\Excl X})$. The definition of
these morphisms involves all the structure of ``$\oc$'' explained
above, and in particular the Seely isos. In Section~\ref{sec:exclmu},
we will use the fact that the following equation holds
\begin{align}
  \Seelyinvt_{X_1,X_2}
  =\Tensp{\Excl{\Proj 1}}{\Excl{\Proj 2}}\Compl\Contr{\With{X_1}{X_2}}
\end{align}
and also, as a consequence:
\begin{equation}
  \begin{split}
    \Weak{\With{X_1}{X_2}}\Compl\Seelyt_{X_1,X_2}
    &=\Tens{\Weak{X_1}}{\Weak{X_2}}\\
    \Der{\With{X_1}{X_2}}\Compl\Seelyt_{X_1,X_2}
    &=\Tuple{\Tens{\Der X}{\Weak Y},\Tens{\Weak X}{\Der Y}}\\
    \Contr{\With{X_1}{X_2}}\Compl\Seelyt_{X_1,X_2}
    &=\Tensp{\Seelyt_{X_1,X_2}}{\Seelyt_{X_1,X_2}}\Compl\Sym_{2,3}
    \Compl\Tensp{\Contr X}{\Contr Y}
  \end{split}
\end{equation}

We use $\Int\_$ for the ``De Morgan dual'' of $\Excl\_$:
$\Int X=\Orthp{\Exclp{\Orth X}}$ and similarly for morphisms. It is a
monad on $\LCAT$.

\subsection{Eilenberg-Moore category and free comodules}
\label{sec:EM-Kl-category}
It is then standard to define the category $\Em\LCAT$ of
$\IExcl$-coalgebras. An object of this category is a pair
$P=(\Coalgca P,\Coalgstr P)$ where $\Coalgca P\in\Obj\LCAT$ and
$\Coalgstr P\in\LCAT(\Coalgca P,\Excl{\Coalgca P})$ is such that
$\Der{\Coalgca P}\Compl\Coalgstr P=\Id$ and
$\Digg{\Coalgca P}\Compl\Coalgstr P=\Excl{\Coalgstr P}\Compl\Coalgstr P$.
Then $f\in\Em\LCAT(P,Q)$ iff $f\in\LCAT(\Coalgca P,\Coalgca Q)$ such
that $\Coalgstr Q\Compl f=\Excl f\Compl\Coalgstr P$.  The functor
$\Excl\_$ can be seen as a functor from $\LCAT$ to $\Em\LCAT$ mapping
$X$ to $(\Excl X,\Digg X)$ and $f\in\LCAT(X,Y)$ to $\Excl f$. It is
right adjoint to the forgetful functor
$\Em\LCAT\to\LCAT$. Given $f\in\LCAT(\Coalgca P,X)$, we use
$\Prom f\in\Em\LCAT(P,\Excl X)$ for the morphism associated with $f$
by this adjunction, one has $\Prom f=\Excl f\Compl\Coalgstr P$. If
$g\in\Em\LCAT(Q,P)$, we have $\Prom f\Compl g=\Promp{f\Compl g}$.
% \begin{align}\label{eq:prom-nat}
%   \Prom f\Compl g=\Promp{f\Compl g}
% \end{align}

Then $\Em\LCAT$ is cartesian (with product of shape
$\Tens PQ=(\Tens{\Coalgca P}{\Coalgca Q},\Coalgstr{\Tens PQ})$ and
final object $(\One,\Coalgstr\One)$, still denoted as $\One$). Here is
the definition of $\Coalgstr{\Tens PQ}$:
\begin{center}
\begin{tikzpicture}[->, >=stealth]
  \node (1) {$\Tens{\Coalgca P}{\Coalgca Q}$};
  \node (2) [ right of=1, node distance=2.7cm ]
  {$\Tens{\Excl{\Coalgca P}}{\Excl{\Coalgca Q}}$};
  \node (3) [ right of=2, node distance=2.8cm ]
  {$\Exclp{\Tens{\Coalgca P}{\Coalgca Q}}\,.$};
  % \node (4) [ right of=3, node distance=2.8cm ]
  % {$\Tens{\Excl{\Coalgca P}}{\Excl{\Coalgca P}}$};
  % \node (5) [ right of=4, node distance=2.8cm ]
  % {$\Tens{\Coalgca P}{\Coalgca P}$};
  \tikzstyle{every node}=[midway,auto,font=\scriptsize]
  \draw (1) -- node {$\Tens{\Coalgstr P}{\Coalgstr Q}$} (2);
  \draw (2) -- node {$\Monoidalt_{\Coalgca P,\Coalgca Q}$} (3);
%  \draw (3) -- node {$\Funinv{(\Seelyt)}$} (4);
%  \draw (4) -- node {$\Tens{\Der{\Coalgca P}}{\Der{\Coalgca P}}$} (5);
\end{tikzpicture}    
\end{center}

This category is also cocartesian with coproduct of shape
$\Plus PQ=(\Plus{\Coalgca P}{\Coalgca Q},\Coalgstr{\Plus PQ})$ and
initial object $(\Zero,\Coalgstr\Zero)$ still denoted as $\Zero$.
Here is the definition of $\Coalgstr{\Plus PQ}$. One first defines
$h_1:\Coalgca P\to\Exclp{\Plus{\Coalgca P}{\Coalgca Q}}$ as
\begin{center}
\begin{tikzpicture}[->, >=stealth]
  \node (1) {$\Coalgca P$};
  \node (2) [ right of=1, node distance=1.2cm ]
  {$\Excl{\Coalgca P}$};
  \node (3) [ right of=2, node distance=1.8cm ]
  {$\Exclp{\Plus{\Coalgca P}{\Coalgca Q}}$};
  % \node (4) [ right of=3, node distance=2.8cm ]
  % {$\Tens{\Excl{\Coalgca P}}{\Excl{\Coalgca P}}$};
  % \node (5) [ right of=4, node distance=2.8cm ]
  % {$\Tens{\Coalgca P}{\Coalgca P}$};
  \tikzstyle{every node}=[midway,auto,font=\scriptsize]
  \draw (1) -- node {$\Coalgstr P$} (2);
  \draw (2) -- node {$\Excl{\Inj 1}$} (3);
%  \draw (3) -- node {$\Funinv{(\Seelyt)}$} (4);
%  \draw (4) -- node {$\Tens{\Der{\Coalgca P}}{\Der{\Coalgca P}}$} (5);
\end{tikzpicture}    
\end{center}
and similarly one defines
$h_2:\Coalgca Q\to\Exclp{\Plus{\Coalgca P}{\Coalgca Q}}$ and then
$\Coalgstr{\Plus PQ}$ is defined as the unique morphism
$\Plus{\Coalgca P}{\Coalgca Q}\to\Exclp{\Plus{\Coalgca P}{\Coalgca
    Q}}$ such that $\Coalgstr{\Plus PQ}\Compl\Inj i=h_i$ for $i=1,2$.

More details can be found in~\cite{Ehrhard16a}. We use
$\Contr P\in\Em\LCAT(P,\Tens PP)$ (\emph{contraction}) for the
diagonal and $\Weak P\in\Em\LCAT(P,\One)$ (\emph{weakening}) for the
unique morphism to the final object. These morphisms turn $\Coalgca P$
into a commutative $\ITens$-comonoid, and are defined as
%\vspace{-3.5mm}
\begin{center}
\begin{tikzpicture}[->, >=stealth]
  \node (1) {$\Coalgca P$};
  \node (2) [ right of=1, node distance=1cm ]
  {$\Excl{\Coalgca P}$};
  \node (3) [ right of=2, node distance=2cm ]
  {$\Exclp{\With{\Coalgca P}{\Coalgca P}}$};
  \node (4) [ right of=3, node distance=2.3cm ]
  {$\Tens{\Excl{\Coalgca P}}{\Excl{\Coalgca P}}$};
  \node (5) [ right of=4, node distance=2.8cm ]
  {$\Tens{\Coalgca P}{\Coalgca P}$};
  \tikzstyle{every node}=[midway,auto,font=\scriptsize]
  \draw (1) -- node {$\Coalgstr P$} (2);
  \draw (2) -- node {$\Excl{\Pair{\Id}{\Id}}$} (3);
  \draw (3) -- node {$\Funinv{(\Seelyt)}$} (4);
  \draw (4) -- node {$\Tens{\Der{\Coalgca P}}{\Der{\Coalgca P}}$} (5);
\end{tikzpicture}    
\end{center}
%\vspace{-3.5mm}
and 
\raisebox{-6pt}{\begin{tikzpicture}[->, >=stealth]
  \node (1) {$\Coalgca P$};
  \node (2) [ right of=1, node distance=1cm ]
  {$\Excl{\Coalgca P}$};
  \node (3) [ right of=2, node distance=1.3cm ]
  {$\Exclp{\Top}$};
  \node (4) [ right of=3, node distance=1.5cm ]
  {$\Sone$};
  \tikzstyle{every node}=[midway,auto,font=\scriptsize]
  \draw (1) -- node {$\Coalgstr P$} (2);
  \draw (2) -- node {$\Excl{\Termm{\Coalgca P}}$} (3);
  \draw (3) -- node {$\Funinv{(\Seelyz)}$} (4);
\end{tikzpicture}}.

\subsubsection{The model of free comodules on a given
  coalgebra.}\label{sec:free-comodules-model}
Given an object $P$ of $\Em\LCAT$, we can define a functor
$\Fcomod P:\LCAT\to\LCAT$ which maps an object $X$ to
$\Tens{\Coalgca P}{X}$ and a morphism $f$ to $\Tens{\Coalgca
  P}{f}$. This functor is clearly a comonad (with structure maps
defined using $\Weak P$, $\Contr P$ and the monoidal structure of
$\LCAT$). A coalgebra for this comonad is a
\emph{$P$-comodule}\footnote{This is just the dual notion of the
  standard algebraic notion of an \emph{$M$-module} which can be
  defined as soon as a commutative $\ITens$-monoid $M$ is given.}. It
was observed by Girard in~\cite{Girard98c} that the Kleisli category
$\Kcomod\LCAT P=\Klp\LCAT{\Fcomod P}$ of this comonad (that is, the
category of free $P$-comodules) is in turn a model of $\LL$ with
operations on objects defined in the same way as in $\LCAT$, and using
the coalgebra structure of $P$ on morphisms. Let us summarize this
construction which will be used in the sequel. First let
$f_i\in\Kcomod\LCAT P(X_i,Y_i)$ for $i=1,2$. Then we define
$\Mtens P{f_1}{f_2}\in\Kcomod\LCAT P(\Tens{X_1}{X_2},\Tens{Y_1}{Y_2})$
as
\begin{center}
\begin{tikzpicture}[->, >=stealth]
  \node (1) {$\Coalgca P\ITens X_1\ITens X_2$};
  \node (2) [ right of=1, node distance=4.5cm ]
  {$\Coalgca P\ITens \Coalgca P\ITens X_1\ITens X_2$};
  \node (3) [ right of=2, node distance=3.7cm ]
  {$\Coalgca P\ITens X_1\ITens \Coalgca P\ITens X_2$};
  \node (4) [ right of=3, node distance=3.4cm ]
  {$\Tens{Y_1}{Y_2}\,.$};
  \tikzstyle{every node}=[midway,auto,font=\scriptsize]
  \draw (1) -- node {$\Contr{\Coalgca P}\ITens\Id$} (2);
  \draw (2) -- node {$\sim$} (3);
  \draw (3) -- node {$\Tens{f_1}{f_2}$} (4);
\end{tikzpicture}    
\end{center}
The object of linear morphisms from $X$ to $Y$ in $\Kcomod\LCAT P$ is
$\Limpl XY$, and the evaluation morphism
$\Mevlin P\in\Kcomod\LCAT P(\Tens{(\Limpl XY)}{X},Y)$ is simply
(keeping implicit the $\ITens$-monoidality isos)
\raisebox{-6pt}{\begin{tikzpicture}[->, >=stealth] \node (1)
    {$\Coalgca P\ITens(\Limpl XY)\ITens X$}; \node (2) [ right of=1,
    node distance=3.8cm ] {$(\Limpl XY)\ITens X$}; \node (3) [ right
    of=2, node distance=2cm ] {$Y$}; \tikzstyle{every
      node}=[midway,auto,font=\scriptsize] \draw (1) -- node
    {$\Weak P\ITens Id$} (2); \draw (2) -- node {$\Evlin$} (3);
  \end{tikzpicture}}.  Then it is easy to check that if
$f\in\Kcomod\LCAT P(\Tens ZX,Y)$, that is
$f\in\LCAT(\Coalgca P\ITens Z\ITens X,Y)$, the morphism
$\Curlin f\in\Kcomod\LCAT P(Z,\Limpl XY)$ satisfies the required
monoidal closeness equations. With these definitions, the category
$\Kcomod\LCAT P$ is *-autonomous, with $\Sbot$ as dualizing
object. Specifically, given $f\in\Kcomod\LCAT P(X,Y)$, then $\Morth Pf$
is the following composition of morphisms:
\raisebox{-6pt}{
  \begin{tikzpicture}[->, >=stealth]
    \node (1) {$\Tens{\Coalgca P}{\Orth Y}$};
    \node (2) [ right of=1, node distance=3.5cm ]
    {$\Coalgca P\ITens(\Limpl{\Coalgca P}{\Orth X})$};]
    \node (3) [ right of=2, node distance=2.4cm ]  {$\Orth X$};
    \tikzstyle{every node}=[midway,auto,font=\scriptsize]
    \draw (1) -- node {$\Tens{\Coalgca P}{\Orth f}$} (2);
    \draw (2) -- node {$\Evlin$} (3);
  \end{tikzpicture}
}, using implicitly the iso between $\Orthp{\Tens{Z}{X}}$ and
$\Limpl Z{\Orth X}$, and the *-autonomy of $\LCAT$ allows to prove
that indeed $\Mbiorth Pf=f$.

The category $\Kcomod\LCAT P$ is easily seen to be cartesian with
$\Top$ as final object, $\With{X_1}{X_2}$ as cartesian product (and
projections defined in the obvious way, using the projections of
$\LCAT$ and the counit $\Weak P$). Last we define a functor
$\Mexcl P\_:\Kcomod\LCAT P\to\Kcomod\LCAT P$ by $\Mexcl PX=\Excl X$
and, given $f\in\Kcomod\LCAT P(X,Y)$, we define
$\Mexcl Pf\in\Kcomod\LCAT P(\Excl X,\Excl Y)$ as 
\raisebox{-6pt}{\begin{tikzpicture}[->, >=stealth]
  \node (1) {$\Tens{\Coalgca P}{\Excl X}$};
  \node (2) [ right of=1, node distance=2.5cm ]
  {$\Tens{\Excl{\Coalgca P}}{\Excl X}$};
  \node (3) [ right of=2, node distance=2cm ]
  {$\Exclp{\Tens{{\Coalgca P}}{X}}$};
  \node (4) [ right of=3, node distance=1.5cm ]
  {$\Excl Y$};
  \tikzstyle{every node}=[midway,auto,font=\scriptsize]
  \draw (1) -- node {$\Tens{\Coalgstr P}{\Excl X}$} (2);
  \draw (2) -- node {$\Monoidalt$} (3);
  \draw (3) -- node {$\Excl f$} (4);
\end{tikzpicture}} and this functor is equipped with a comonad
structure $(\Mder P,\Mdigg P)$ easily defined using $\Der{}$,
$\Digg{}$ and $\Weak P$ (for this crucial construction, we need $P$ to
be a $\oc$-coalgebra and not simply a commutative $\ITens$-comonoid;
notice however that if $\oc$ is the free exponential, as
in~\cite{Girard98c}, the latter condition implies the former).

\subsection{Strong functors on $\LCAT$}\label{sec:gen-strong-functors}

Given $n\in\Nat$, an \emph{$n$-ary strong functor} on $\LCAT$ is a
pair $\Vcsnot F=(\Strfun{\Vcsnot F},\Strnat{\Vcsnot F}))$ where
$\Strfun{\Vcsnot F}:\LCAT^n\to\LCAT$ is a functor and
$\Strnat{\Vcsnot F}_{X,\Vect Y}\in\LCAT(\Tens{\Excl X}{\Strfun{\Vcsnot
    F}(\Vect Y)},\Strfun{\Vcsnot F}(\Tens{\Excl X}{\Vect Y})$ is a
natural transformation, called the \emph{strength} of $\Vcsnot F$. We
use the notation
$\Tens Z{(\List Y1n)}=(\Tens{Z}{Y_1},\dots,\Tens{Z}{Y_n})$. It is
assumed moreover that the diagrams of
Figure~\ref{fig:strength-monoidality} commute, expressing a
monoidality of this strength.

\begin{figure}
  {\footnotesize
\begin{tikzpicture}[->, >=stealth]
  \node (1) {$(\Excl{X_1}\ITens\Excl{X_2})
    \ITens{\Strfun{\Vcsnot F}}(\Vect{Y})$};
  \node (2) [ right of=1, node distance=4.3cm ]
  {$\Exclp{\With{X_1}{X_2}}\ITens \Strfun{\Vcsnot F}(\Vect Y)$};
  \node (3) [ below of=1, node distance=1.2cm ]
  {$\Excl{X_1}\ITens \Strfun{\Vcsnot F}(\Excl{X_2}\ITens\Vect Y)$};
  \node (4) [ below of=3, node distance=1.2cm ]
  {$\Strfun{\Vcsnot F}(\Excl{X_1}\ITens\Excl{X_2}
    \ITens \Strfun{\Vcsnot F}(\Vect Y))$};
  \node (5) [ right of=4, node distance=4.3cm ]
  {$\Strfun{\Vcsnot F}(\Exclp{\With{X_1}{X_2}}\ITens\Vect Y)$};
  \tikzstyle{every node}=[midway,auto,font=\scriptsize]
  \draw (1) -- node {$\Tens{\Seelyt}{\Strfun{\Vcsnot F}(\Vect Y)}$} (2);
  \draw (4) -- node {$\Strfun{\Vcsnot F}(\Tens{\Seelyt}{\Vect Y})$} (5);
  \draw (1) -- node {$\Tens{\Excl{X_1}}
    {\Strnat{\Vcsnot F}_{X_2,\Vect Y}}$} (3);
  \draw (3) -- node [swap] {$\Strnat{\Vcsnot F}_{X_1,
      \Tens{\Excl{X_2}}{\Vect Y}}$} (4);
  \draw (2) -- node {$\Strnat{\Vcsnot F}_{\With{X_1}{X_2},\Vect Y}$} (5);
\end{tikzpicture}
\raisebox{0.6cm}{
  \begin{tikzpicture}[->, >=stealth]
    \node (1) {$\Tens{\Sone}{\Strfun{\Vcsnot F}(\Vect Y)}$};
    \node (2) [ right of=1, node distance=2.9cm ]
    {$\Tens{\Excl\Top}{\Strfun{\Vcsnot F}(\Vect Y)}$};
    \node (3) [ below of=1, node distance=1.2cm ]
    {$\Strfun{\Vcsnot F}(\Tens\Sone{\Vect Y})$};
    \node (4) [ right of=3, node distance=2.9cm ]
    {$\Strfun{\Vcsnot F}(\Excl{\Top}\ITens\Vect Y)$};
    \tikzstyle{every node}=[midway,auto,font=\scriptsize]
    \draw (1) -- node {$\Tens{\Seelyz}{\Strfun{\Vcsnot F}(\Vect Y)}$} (2);
    \draw (3) -- node {$\Strfun{\Vcsnot F}(\Tens{\Seelyz}{\Vect Y})$} (4);
    \draw (1) -- node [above, rotate=90] {$\mathord\sim$} (3);
    \draw (2) -- node {$\Strnat{\Vcsnot F}_{\Top,\Vect Y}$} (4);
  \end{tikzpicture}}}
\caption{Monoidality diagrams for strong functors}
\label{fig:strength-monoidality}
\end{figure}

The main purpose of this definition is that one can then define a
functor $\Mfun{\Vcsnot F}P:\Kcomod\LCAT P^n\to\Kcomod\LCAT P$ as
follows. First one sets
$\Mfun{\Vcsnot F}P(\Vect X)=\Strfun{\Vcsnot F}(\Vect X)$. Then, given
$\Vect f\in\Kcomod\LCAT P^n(\Vect X,\Vect Y)$ we define
$\Mfun{\Vcsnot F}{P}(\Vect f)\in\Kcomod\LCAT P({\Vcsnot
  F}(\Vect X),{\Vcsnot F}(\Vect Y))$ as 
\begin{center}
  \begin{tikzpicture}[->, >=stealth]
    \node (1) {$\Tens{\Coalgca P}{\Vcsnot F(\Vect X)}$};
    \node (2) [ right of=1, node distance=3.2cm ]
    {$\Tens{\Excl{\Coalgca P}}{\Vcsnot F(\Vect X)}$};
    \node (3) [ right of=2, node distance=2.5cm ]
    {$\Vcsnot F(\Tens{\Excl{\Coalgca P}}{\Vect X})$};
    \node (4) [ right of=3, node distance=3.7cm ]
    {$\Vcsnot F(\Tens{{\Coalgca P}}{\Vect X})$};
    \node (5) [ right of=4, node distance=2.3cm ]  {$\Vcsnot F(\Vect Y)\,.$};
    \tikzstyle{every node}=[midway,auto,font=\scriptsize]
    \draw (1) -- node {$\Tens{\Coalgstr P}{\Id}$} (2);
    \draw (2) -- node {$\Strnat{\Vcsnot F}$} (3);
    \draw (3) --
    node {$\Strfun{\Vcsnot F}(\Der{\Coalgca P}\ITens\Vect X)$} (4);
    \draw (4) -- node {$\Vcsnot F(\Vect f)$} (5);
  \end{tikzpicture}  
\end{center}
The fact that we have defined a functor results from the two
monoidality diagrams of Figure~\ref{fig:strength-monoidality} and from
the definition of $\Weak P$ and $\Contr P$ based on the Seely
isomorphisms.

\paragraph{Operations on strong functors.} There is an obvious unary
identity strong functor $\Strid$ and for each object $Y$ of $\LCAT$
there is an $n$-ary $Y$-valued constant strong functor $\Strcst X$; in
the first case the strength natural transformation is the identity
morphism and in the second case, it is defined using $\Weak{\Excl
  X}$. Let $\Vcsnot F$ be an $n$-ary strong functor and
$\List{\Vcsnot G}1n$ be $k$-ary strong functors. Then one defines a
$k$-ary strong functor $\Vcsnot H=\Vcsnot F\Comp(\List{\Vcsnot G}1n)$:
the functorial component $\Strfun{\Vcsnot H}$ is defined in the
obvious compositional way. The strength is defined as follows
%\vspace{-5mm}
\begin{center}
  \begin{tikzpicture}[->, >=stealth]
    \node (1) {$\Tens{\Excl X}{\Strfun{\Vcsnot H}(\Vect Y)}$};
    \node (2) [ right of=1, node distance=3.2cm ]
    {$\Strfun{\Vcsnot F}((\Tens{\Excl X}
      {\Strfun{\Vcsnot G_i}(\Vect Y)})_{i=1}^n)$};
    \node (3) [ right of=2, node distance=6.4cm ]
    {$\Strfun{\Vcsnot F}((
      {\Strfun{\Vcsnot G_i}(\Tens{\Excl X}{\Vect Y})})_{i=1}^n)
      =\Strfun{\Vcsnot H}(\Tens{\Excl X}{\Vect Y})$};
    \tikzstyle{every node}=[midway,auto,font=\scriptsize]
    \draw (1) -- node {$\Strnat{\Vcsnot F}$} (2);
    \draw (2) -- node {$\Strfun{\Vcsnot F}
      ((\Strnat{\Vcsnot G_i})_{i=1}^k)$} (3);
  \end{tikzpicture}  
\end{center}
and is easily seen to satisfy the required monoidality commutations.

Given an $n$-ary strong functor, we can define its \emph{De Morgan
  dual} $\Orth{\Vcsnot F}$ which is also an $n$-ary strong functor. On
objects, we set
$\Strfun{\Orth{\Vcsnot F}}(\Vect Y)=\Orth{\Strfun{\Vcsnot
    F}(\Orth{\Vect Y})}$ and similarly for morphisms. The strength of
$\Orth{\Vcsnot F}$ is defined as the Curry transpose of the following
morphism (remember that
$\Limpl{\Excl X}{\Orth{\Vect Y}}=\Orthp{\Tens{\Excl X}{\Vect Y}}$ up
to canonical iso):
\begin{center}
  \begin{tikzpicture}[->, >=stealth]
    \node (1) {$\Excl X\ITens{\Orth{\Strfun{\Vcsnot F}(\Orth{\Vect Y})}}
      \ITens\Strfun{\Vcsnot F}(\Limpl{\Excl X}{\Orth{\Vect Y}})$};
    \node (2) [ below of=1, node distance=1cm ]
    {$\Excl X
      \ITens\Strfun{\Vcsnot F}(\Limpl{\Excl X}{\Orth{\Vect Y}})
      \ITens{\Orth{\Strfun{\Vcsnot F}(\Orth{\Vect Y})}}$};
    \node (3) [ below of=2, node distance=1cm ]
    {$\Strfun{\Vcsnot F}(\Excl X
      \ITens(\Limpl{\Excl X}{\Orth{\Vect Y}}))
      \ITens{\Orth{\Strfun{\Vcsnot F}(\Orth{\Vect Y})}}$};
    \node (4) [ below of=3, node distance=1cm ]
    {$\Strfun{\Vcsnot F}(\Orth{\Vect Y})
      \ITens{\Orth{\Strfun{\Vcsnot F}(\Orth{\Vect Y})}}$};
    \node (5) [ below of=4, node distance=1cm ]
    {$\Sbot$};
    \tikzstyle{every node}=[midway,auto,font=\scriptsize]
    \draw (1) -- node {$\sim$} (2);
    \draw (2) -- node {$\Strnat{\Vcsnot F}\ITens\Id$} (3);
    \draw (3) -- node [swap] {$\Strfun{\Vcsnot F}(\Evlin)\ITens\Id$} (4);
    \draw (4) -- node [swap] {$\Evlin\Compl\Sym$} (5);
  \end{tikzpicture}  
\end{center}
Then it is possible to prove, using the *-autonomy of $\LCAT$, that
$\Biorth{\Vcsnot F}$ and $\Vcsnot F$ are canonically isomorphic (as
strong functors)\footnote{In the concrete settings considered in this
  paper, these canonical isos are actuality identity maps.}.

\begin{lem}\label{lemma:strfun-comp-orth}
  $\Orthp{\Vcsnot F\Comp(\List{\Vcsnot G}1n)}=\Orth{\Vcsnot
    F}\Comp(\List{\Orth{\Vcsnot G}}1n)$ up to canonical iso.
\end{lem}
\proof Results straightforwardly from the definition of
$\Orth{\Vcsnot F}$ and from the canonical iso between
$\Biorth{\Vcsnot F}$ and $\Vcsnot F$.  \qed

The bifunctor $\mathord\ITens$ can be turned into a strong functor:
one defines the strength as 
\raisebox{-6pt}{
\begin{tikzpicture}[->, >=stealth]
  \node (1) {$\Excl X\ITens Y_1\ITens Y_2$};
  \node (2) [ right of=1, node distance=4.6cm ]
  {$\Excl X\ITens \Excl X\ITens Y_1\ITens Y_2$};
  \node (3) [ right of=2, node distance=3.7cm ]
  {$\Excl X\ITens Y_1\ITens \Excl X\ITens Y_2$};
  \tikzstyle{every node}=[midway,auto,font=\scriptsize]
  \draw (1) -- node {$\Contr{\Excl X}\ITens\Id$} (2);
  \draw (2) -- node {$\sim$} (3);
\end{tikzpicture}}.  By De Morgan duality, this endows $\IPar$ with a
strength as well. The bifunctor $\IPlus$ is also endowed with a
strength, simply using the distributivity of $\ITens$ over $\IPlus$
(which in turn results from the fact that $\LCAT$ is symmetric
monoidal closed). By duality again, $\IWith$ inherits a strength as
well. Last the unary functor ``$\oc$'' can be equipped with a strength
as follows \raisebox{-6pt}{\begin{tikzpicture}[->, >=stealth]
  \node (1) {$\Tens{\Excl X}{\Excl Y}$};
  \node (2) [ right of=1, node distance=3.4cm ]
  {$\Tens{\Excl{\Excl X}}{\Excl Y}$};
  \node (3) [ right of=2, node distance=2.2cm ]
  {$\Exclp{\Tens{{\Excl X}}{Y}}$};
  \tikzstyle{every node}=[midway,auto,font=\scriptsize]
  \draw (1) -- node {$\Tens{\Digg X}{\Excl X}$} (2);
  \draw (2) -- node {$\Monoidalt$} (3);
\end{tikzpicture}}.  The functors on $\Kcomod\LCAT P$ that these
strong functors induce coincide with the corresponding operations
defined in Section~\ref{sec:free-comodules-model}.

% Given an $n$-ary strong functor $\Vcsnot F$, one can define another
% $n$-ary strong functor $\Orth{\Vcsnot F}$, called the \emph{dual} of
% $\Vcsnot F$, as follows. First we set
% $\Strfun{\Orth{\Vcsnot F}}(\Vect E)=\Orth{\Strfun{\Vcsnot
%     F}(\Orth{\Vect E})}$ and similarly for morphisms, so that
% $\Orth{\Strfun{\Vcsnot F}}$ is actually a functor $\LCAT^n\to\LCAT$;

\subsection{Fixed Points of functors.}
\label{sec:fixpoints-functors}

The following definitions and properties are quite standard in the
literature on fixed points of functors.
\begin{defi}
  Let $\cA$ be a category and let $\cF:\cA\to\cA$ be a functor. A
  \emph{coalgebra} of $\cF$ is a pair $(A,f)$ where $A$ is an object of $\cA$
  and $f\in\cA(A,\cF(A))$. Given two coalgebras $(A,f)$ and $(A',f')$ of
  $\cF$, a coalgebra morphism from $(A,f)$ to $(A',f')$ is an
  $h\in\cA(A,A')$ such that the following diagram commutes
  \begin{center}
    \begin{tikzpicture}[->, >=stealth]
      \node (1) {$A$};
      \node (2) [right of=1, node distance=2cm] {$A'$};
      \node (3) [below of=1, node distance=1.2cm] {$\cF(A)$};
      \node (4) [below of=2, node distance=1.2cm] {$\cF(A')$};
      \tikzstyle{every node}=[midway,auto,font=\scriptsize]
      \draw (1) -- node {$h$} (2);
      \draw (1) -- node [swap] {$f$} (3);
      \draw (2) -- node {$f'$} (4);
      \draw (3) -- node {$\cF(h)$} (4);
    \end{tikzpicture}
  \end{center}
  The category of coalgebras of the functor $\cF$ will be denoted as
  $\COALGFUN\cA\cF$. The notion of algebra of an endofunctor is
  defined dually (reverse the directions of the arrows $f$ and $f'$)
  and the corresponding category is denoted as $\ALGFUN\cA\cF$.
\end{defi}

\begin{lem}[Lambek's Theorem]
  If $(A,f)$ is a final object in $\COALGFUN\cA\cF$ then $f$ is an iso.
\end{lem}
% \proof Since $(\cF(A),\cF(f))$ is also a coalgebra, there is a
% unique $f'\in\cA(\cF(A), A)$ such that
% $f\Compl f'=\cF(f')\Compl\cF(f)=\cF(f'\Compl f)$. Then $f'\Compl f$ is
% a coalgebra automorphism on $(A,f)$ and so $f'\Compl f=\Id$. Hence
% $f\Compl f'=\cF(f'\Compl f)=\Id$.  \qed

In the sequel, we will always assume that this iso is the identity
(because this holds in the concrete situations we consider in this
paper) so that this final object $(\Fungfp\cF,\Id)$ satisfies
$\cF(\Fungfp\cF)=\Fungfp\cF$. We focus on coalgebras rather than
algebras for reasons which will become clear when we shall deal with
fixed points of strong functors.

This universal property of $\Fungfp{\cF}$ gives us a powerful tool for
proving equalities of morphisms.
\begin{lem}\label{lemma:equations-final-coalgebra}
  Let $A$ be an object of $\cA$ and let
  $f_1,f_2\in\cA(A,\Fungfp{\cF})$. If there exists
  $l\in\cA(A,\cF(A))$ such that
  $\cF(f_i)\Compl l=f_i$ for $i=1,2$, then $f_1=f_2$.
\end{lem}
\proof The assumption means that both $f_1$ and $f_2$ are
coalgebra morphisms from $(A,l)$ to the final coalgebra, so they must
be equal.  \qed

\begin{lem}\label{lemma:functor-gfp-general}
  Let $\cF:\cB\times\cA\to\cA$ be a functor such that, for all
  $B\in\cB$, the category $\COALGFUN \cA{\cF_B}$ has a final
  object. Then there is a functor $\Fungfp\cF$ such that
  $(\Fungfp\cF(B),\Id)$ is the final object of
  $\COALGFUN \cA{\cF_B}$ (so that
  $\cF(B,\Fungfp\cF(B))=\Fungfp\cF(B)$) for each $B\in\cB$, and, for
  each $g\in\cB(B,B')$, $\Fungfp\cF(g)$ is uniquely characterized by
  $\cF(g,\Fungfp\cF(g))=\Fungfp\cF(g)$.
\end{lem}
\proof We have
$\cF(g,\Fungfp\cF(B))\in\cA(\Fungfp\cF(B),\cF(B',\Fungfp\cF(B)))$ thus
defining a $\cF_{B'}$-coalgebra structure on $\Fungfp\cF(B)$ and hence
there exists a unique morphism $\Fungfp\cF(g)$ such that
\[
  \cF(B',\Fungfp\cF(g))\Compl\cF(g,\Fungfp\cF(B))=\Fungfp\cF(g)\,,
\]
that is $\cF(g,\Fungfp\cF(g))=\Fungfp\cF(g)$.

Functoriality follows: consider also $g'\in\cB(B',B'')$, then we know
that $h=\Fungfp\cF(g'\Compl g)$ satisfies $\cF(g'\Compl g,h)=h$ by the
definition above. Now $h'=\Fungfp\cF(g')\Compl\Fungfp\cF(g)$ satisfies
the same equation by functoriality of $\cF$ and because
$\cF(g,\Fungfp\cF(g))=\Fungfp\cF(g)$ and
$\cF(g',\Fungfp\cF(g'))=\Fungfp\cF(g')$, and hence $h'=h$ by
Lemma~\ref{lemma:functor-gfp-general}, taking
$l=\cF(g'\Compl g,\Fungfp\cF(B))$. In the same way one proves that
$\Fungfp\cF(\Id)=\Id$. \qed

We consider now the same $\Fungfp\cF$ operation applied to strong
functors on a model $\LCAT$ of $\LL$. Let $\Vcsnot F$ be an
$n+1$-ary strong functor on $\LCAT$ (so that $\Strfun{\Vcsnot F}$ is a
functor $\LCAT^{n+1}\to\LCAT$). Assume that for each
$\Vect X\in\LCAT^n$ the category
$\COALGFUN{\LCAT}{\Strfun{\Vcsnot F}_{\Vect X}}$ has a final
object. We have defined a functor
$\Fungfp{\Strfun{\Vcsnot F}}:\LCAT^n\to\LCAT$ uniquely characterized by
$\Strfun{\Vcsnot F}(\Vect X,\Fungfp{\Strfun{\Vcsnot F}}(\Vect
X))=\Fungfp{\Strfun{\Vcsnot F}}(\Vect X)$ and
$\Strfun{\Vcsnot F}(\Vect f,\Fungfp{\Strfun{\Vcsnot F}}(\Vect
f))=\Fungfp{\Strfun{\Vcsnot F}}(\Vect f)$ for all
$\Vect f\in\LCAT^n(\Vect X,\Vect{X'})$
(Lemma~\ref{lemma:functor-gfp-general}). For each $Y,\Vect X\in\LCAT$,
we define
$\Strnat{\Vcsnot{\Fungfp F}}_{Y,\Vect X}\in\LCAT(\Tens{\Excl
  Y}{\Fungfp{\Strfun{\Vcsnot F}}(\Vect X)},\Fungfp{\Strfun{\Vcsnot
    F}}(\Tens{\Excl Y}{\Vect X}))$. We have 
\begin{center}
    \begin{tikzpicture}[->, >=stealth]
      \node (1) {$\Tens{\Excl Y}{\Strfun{\Fungfp{\Vcsnot F}}(\Vect X)}
        =\Excl Y\ITens
          \Strfun{\Vcsnot F}(\Vect X,\Strfun{\Fungfp{\Vcsnot F}}(\Vect X))$};
      \node (2) [right of=1, node distance=6.9cm]
      {$\Strfun{\Vcsnot F}(\Tens{\Excl Y}{\Vect X},
        \Tens{\Excl Y}{\Strfun{\Fungfp{\Vcsnot F}}(\Vect X)})$};
       \tikzstyle{every node}=[midway,auto,font=\scriptsize]
       \draw (1) -- node {$\Strnat{\Vcsnot F}_{Y,(\Vect X,
           \Strfun{\Fungfp{\Vcsnot F}}(\Vect X))}$} (2);
     \end{tikzpicture}   
\end{center}
exhibiting a $\Strfun{\Vcsnot F}_{\Tens{\Excl Y}{\Vect X}}$-coalgebra
structure on $\Tens{\Excl Y}{\Strfun{\Fungfp{\Vcsnot F}}(\Vect
  X)}$. Since $\Strfun{\Fungfp{\Vcsnot F}}(\Tens{\Excl Y}{\Vect X})$ is
the final coalgebra of the functor
$\Strfun{\Vcsnot F}_{\Tens{\Excl Y}{\Vect X}}$, we define
$\Strnat{\Fungfp{\Vcsnot F}}_{Y,\Vect X}$ as the unique morphism
$\Tens{\Excl Y}{\Strfun{\Fungfp{\Vcsnot F}}(\Vect
  X)}\to\Strfun{\Fungfp{\Vcsnot F}}(\Tens{\Excl Y}{\Vect X})$ such that
the following diagram commutes
\begin{equation}\label{diag:strnat-vcsfp-charact}
  \begin{tikzpicture}[->, >=stealth,baseline=(current bounding box.center)]
      \node (1) {$\Tens{\Excl Y}{\Strfun{\Fungfp{\Vcsnot F}}(\Vect X)}$};
      \node (2) [right of=1, node distance=5.8cm]
      {$\Strfun{\Vcsnot F}(\Tens{\Excl Y}{\Vect X},
        \Tens{\Excl Y}{\Strfun{\Fungfp{\Vcsnot F}}(\Vect X)})$};
      \node (3) [below of=2, node distance=1.2cm] 
      {$\Strfun{\Vcsnot F}(\Tens{\Excl Y}{\Vect X},
        \Strfun{\Fungfp{\Vcsnot F}}(\Tens{\Excl Y}{\Vect X}))
       =\Strfun{\Fungfp{\Vcsnot F}}(\Tens{\Excl Y}{\Vect X})$};
       \tikzstyle{every node}=[midway,auto,font=\scriptsize]
       \draw (1) -- node {$\Strnat{\Vcsnot F}_{Y,(\Vect X,
           \Strfun{\Fungfp{\Vcsnot F}}(\Vect X))}$} (2);
       \draw (2) -- node 
       {$\Strfun{\Vcsnot F}(\Tens{\Excl Y}{\Vect X},
         \Strnat{\Fungfp{\Vcsnot F}}_{Y,\Vect X})$} (3);
       \draw (1) -- node [swap] 
       {$\Strnat{\Fungfp{\Vcsnot F}}_{(Y,\Vect X})$} (3);
  \end{tikzpicture}   
\end{equation}

% \begin{lem}\label{lemma:strength-gfp-general}
%   $\Fungfp{\Vcsnot F}=(\Strfun{\Fungfp{\Vcsnot F}},\Strnat{\Fungfp{\Vcsnot
%       F}})$ is a strong functor.
% \end{lem}
\begin{lem}\label{lemma:strfun-gfp-general}
  Let $\Vcsnot F$ be an $n+1$-ary strong functor on $\LCAT$ such that
  for each $\Vect X\in\LCAT^n$, the category
  $\COALGFUN\LCAT{\Strfun{\Vcsnot F}_{\Vect X}}$ has a final object
  $\Fungfp{\Strfun{\Vcsnot F}_{\Vect X}}$. Then there is a unique
  $n$-ary strong functor $\Fungfp{\Vcsnot F}$ such that
  $\Strfun{\Fungfp{\Vcsnot F}}(\Vect X)=\Fungfp{\Strfun{\Vcsnot
      F}_{\Vect X}}$ (and hence
  $\Strfun{\Vcsnot F}(\Vect X,\Strfun{\Fungfp{\Vcsnot F}}(\Vect
  X))=\Strfun{\Fungfp{\Vcsnot F}}(\Vect X)$),
  \begin{itemize}
  \item
    $\Strfun{\Vcsnot F}(\Vect f,\Strfun{\Fungfp{\Vcsnot F}}(\Vect
    f))=\Strfun{\Fungfp{\Vcsnot F}}(\Vect f)$ for all
    $\Vect f\in\LCAT^n(\Vect X,\Vect{X'})$
  \item and
    $\Strfun{\Vcsnot F}(\Tens{\Excl Y}{\Vect
      X},\Strnat{\Fungfp{\Vcsnot F}}_{Y,\Vect X})\Compl\Strnat{\Vcsnot
      F}_{Y,(\Vect X,\Strfun{\Fungfp{\Vcsnot F}}(\Vect
      X))}=\Strnat{\Fungfp{\Vcsnot F}}_{Y,\Vect X}$.
  \end{itemize}
%  Moreover $\Orth{(\Fungfp{\Vcsnot F})}=\Fungfp{(\Orth{\Vcsnot F})}$
\end{lem}
\proof
Let us prove the naturality of $\Strnat{\Fungfp{\Vcsnot F}}$ so let
$\Vect f\in\LCAT^n(\Vect X,\Vect{X'})$ and $g\in\LCAT(Y,Y')$, we must
prove that the following diagram commutes
\begin{center}
  \begin{tikzpicture}[->, >=stealth]
      \node (1) {$\Tens{\Excl Y}{\Strfun{\Fungfp{\Vcsnot F}}(\Vect X)}$};
      \node (2) [right of=1, node distance=5.4cm]
      {$\Strfun{\Fungfp{\Vcsnot F}}(\Tens{\Excl Y}{\Vect X})$};
      \node (3) [below of=1, node distance=1.2cm] 
      {$\Tens{\Excl {Y'}}{\Strfun{\Fungfp{\Vcsnot F}}(\Vect{X'})}$};
      \node (4) [below of=2, node distance=1.2cm] 
      {$\Strfun{\Fungfp{\Vcsnot F}}(\Tens{\Excl {Y'}}{\Vect{X'}})$};
       \tikzstyle{every node}=[midway,auto,font=\scriptsize]
       \draw (1) -- node {$\Strnat{\Fungfp{\Vcsnot F}}_{Y,\Vect X}$} (2);
       \draw (1) -- node [swap]  
       {$\Tens{\Excl g}{\Strfun{\Fungfp{\Vcsnot F}}(\Vect f)}$} (3);
       \draw (2) -- node
       {$\Strfun{\Fungfp{\Vcsnot F}}(\Tens{\Excl g}{\Vect f})$} (4);
       \draw (3) -- node {$\Strnat{\Fungfp{\Vcsnot F}}_{Y',\Vect{X'}}$} (4);
  \end{tikzpicture}   
\end{center}
Let
$h_1=\Strnat{\Fungfp{\Vcsnot F}}_{Y',\Vect{X'}} \Compl(\Tens{\Excl
  g}{\Strfun{\Fungfp{\Vcsnot F}}(\Vect f)})$ and
$h_2=\Strfun{\Fungfp{\Vcsnot F}}(\Tens{\Excl g}{\Vect
  f})\Compl\Strnat{\Fungfp{\Vcsnot F}}_{Y,\Vect X}$ be the two
morphisms we must prove equal. We
use Lemma~\ref{lemma:equations-final-coalgebra}, taking the following
morphism $l$.
\begin{equation*}
    \begin{tikzpicture}[->, >=stealth]
      \node (1) {$\Tens{\Excl Y}{\Strfun{\Fungfp{\Vcsnot F}}(\Vect X)}
        =\Excl Y\ITens
          \Strfun{\Vcsnot F}(\Vect X,\Strfun{\Fungfp{\Vcsnot F}}(\Vect X))$};
      \node (2) [below of=1, node distance=1.2cm]
      {$\Strfun{\Vcsnot F}(\Tens{\Excl Y}{\Vect X},
        \Tens{\Excl Y}{\Strfun{\Fungfp{\Vcsnot F}}(\Vect X)})$};
      \node (3) [below of=2, node distance=1.2cm]
      {$\Strfun{\Vcsnot F}(\Tens{\Excl {Y'}}{\Vect{X'}},
        \Tens{\Excl Y}{\Strfun{\Fungfp{\Vcsnot F}}(\Vect X)})$};
       \tikzstyle{every node}=[midway,auto,font=\scriptsize]
       \draw (1) -- node {$\Strnat{\Vcsnot F}_{Y,(\Vect X,
           \Strfun{\Fungfp{\Vcsnot F}}(\Vect X))}$} (2);
       \draw (2) -- node {$\Strfun{\Vcsnot F}(\Tens{\Excl g}
         {\Vect f},\Id)$} (3);
     \end{tikzpicture}   
\end{equation*}
With these notations we have
\begin{align*}
  \Strfun{\Vcsnot F}(\Tens{\Excl {Y'}}{\Vect{X'}},h_1)\Compl l
  &= \Strfun{\Vcsnot F}
    (\Tens{\Excl {Y'}}{\Vect{X'}},\Strnat{\Fungfp{\Vcsnot F}}_{Y',\Vect{X'}})
    \Compl \Strfun{\Vcsnot F}(\Tens{\Excl {Y'}}{\Vect{X'}},
    \Tens{\Excl g}{\Strfun{\Fungfp{\Vcsnot F}}(\Vect f)})\\
  &\quad\quad \Compl \Strfun{\Vcsnot F}(\Tens{\Excl g}{\Vect f},
    \Tens{\Excl Y}{\Strfun{\Fungfp{\Vcsnot F}}(\Vect X)})
    \Compl \Strnat{\Vcsnot F}_{Y,(\Vect X,\Strfun{\Fungfp{\Vcsnot F}}(\Vect X))}\\
  &= \Strfun{\Vcsnot F}
    (\Tens{\Excl {Y'}}{\Vect{X'}},\Strnat{\Fungfp{\Vcsnot F}}_{Y',\Vect{X'}})
    \Compl \Strfun{\Vcsnot F}(\Tens{\Excl g}{\Vect f},
    \Tens{\Excl g}{\Strfun{\Fungfp{\Vcsnot F}}(\Vect f)})
    \Compl \Strnat{\Vcsnot F}_{Y,(\Vect X,\Strfun{\Fungfp{\Vcsnot F}}(\Vect X))}\\
  &= \Strfun{\Vcsnot F}
    (\Tens{\Excl {Y'}}{\Vect{X'}},\Strnat{\Fungfp{\Vcsnot F}}_{Y',\Vect{X'}})
    \Compl \Strnat{\Vcsnot F}_{Y',(\Vect{X'},\Strfun{\Fungfp{\Vcsnot F}}(\Vect{X'}))}
    \Compl (\Tens{\Excl g}{\Strfun{\Vcsnot F}}
    (\Vect f,\Strfun{\Fungfp{\Vcsnot F}}(\Vect f)))\\
  &\quad\quad\quad\quad\quad\quad \text{ by naturality of }\Strnat{\Vcsnot F}\\
  &= \Strnat{\Fungfp{\Vcsnot F}}_{Y',\Vect{X'}}
    \Compl(\Tens{\Excl g}{\Strfun{\Vcsnot F}}
    (\Vect f,\Strfun{\Fungfp{\Vcsnot F}}(\Vect f)))
    \text{ by~\Eqref{diag:strnat-vcsfp-charact}}\\
  &= \Strnat{\Fungfp{\Vcsnot F}}_{Y',\Vect{X'}}
    \Compl(\Tens{\Excl g}{\Strfun{\Fungfp{\Vcsnot F}}(\Vect f)})
    \text{ by~Lemma~\ref{lemma:functor-gfp-general}}
\end{align*}
so that $\Strfun{\Vcsnot F}(\Tens{\Excl {Y'}}{\Vect{X'}},h_1)\Compl l=h_1$
as required. On the other hand we have
\begin{align*}
  \Strfun{\Vcsnot F}(\Tens{\Excl {Y'}}{\Vect{X'}},h_2)\Compl l
  &= \Strfun{\Vcsnot F}(\Tens{\Excl {Y'}}{\Vect{X'}},
    \Strfun{\Fungfp{\Vcsnot F}}(\Tens{\Excl g}{\Vect f}))
  \Compl \Strfun{\Vcsnot F}(\Tens{\Excl {Y'}}{\Vect{X'}},
    \Strnat{\Fungfp{\Vcsnot F}}_{Y,\Vect X})\\
  &\quad\quad \Compl \Strfun{\Vcsnot F}(\Tens{\Excl g}{\Vect f},
    \Tens{\Excl Y}{\Strfun{\Fungfp{\Vcsnot F}}(\Vect X)})
    \Compl \Strnat{\Vcsnot F}_{Y,(\Vect X,\Strfun{\Fungfp{\Vcsnot F}}(\Vect X))}\\
  &= \Strfun{\Vcsnot F}(\Tens{\Excl {Y'}}{\Vect{X'}},
    \Strfun{\Fungfp{\Vcsnot F}}(\Tens{\Excl g}{\Vect f}))
    \Compl \Strfun{\Vcsnot F}(\Tens{\Excl g}{\Vect f},
    \Tens{\Excl Y}{\Strfun{\Fungfp{\Vcsnot F}}(\Vect X)})\\
  &\quad\quad \Compl \Strfun{\Vcsnot F}(\Tens{\Excl Y}{\Vect X},
    \Strnat{\Fungfp{\Vcsnot F}}_{Y,\Vect X})
    \Compl \Strnat{\Vcsnot F}_{Y,(\Vect X,\Strfun{\Fungfp{\Vcsnot F}}(\Vect X))}\\
  &= \Strfun{\Vcsnot F}(\Tens{\Excl g}{\Vect f},
    \Strfun{\Fungfp{\Vcsnot F}}(\Tens{\Excl g}{\Vect f}))
    \Compl\Strnat{\Fungfp{\Vcsnot F}}_{Y,\Vect X}
    \text{ by~\Eqref{diag:strnat-vcsfp-charact}}\\
  &=  \Strfun{\Fungfp{\Vcsnot F}}(\Tens{\Excl g}{\Vect f})
    \Compl\Strnat{\Fungfp{\Vcsnot F}}_{Y,\Vect X}
    \text{ by~Lemma~\ref{lemma:functor-gfp-general}}
\end{align*}
so that $\Strfun{\Vcsnot F}(\Tens{\Excl {Y'}}{\Vect{X'}},h_2)\Compl l=h_2$
which proves our contention. The monoidality condition on
$\Strnat{\Fungfp{\Vcsnot F}}$ is proved similarly.  \qed

\begin{lem}\label{lemma:strfun-lfp-general}
  Let $\Vcsnot F$ be an $n+1$-ary strong functor on $\LCAT$ such that
  for each $\Vect X\in\LCAT^n$, the category
  $\ALGFUN\LCAT{\Strfun{\Vcsnot F}_{\Vect X}}$ has an initial object
  $\Funlfp{\Strfun{\Vcsnot F}_{\Vect X}}$. Then there is a unique
  $n$-ary strong functor $\Funlfp{\Vcsnot F}$ such that
  $\Strfun{\Funlfp{\Vcsnot F}}(\Vect X)=\Funlfp{\Strfun{\Vcsnot
      F}_{\Vect X}}$ (and hence
  $\Strfun{\Vcsnot F}(\Vect X,\Strfun{\Funlfp{\Vcsnot F}}(\Vect
  X))=\Strfun{\Funlfp{\Vcsnot F}}(\Vect X)$),
  \begin{itemize}
  \item
    $\Strfun{\Vcsnot F}(\Vect f,\Strfun{\Funlfp{\Vcsnot F}}(\Vect
    f))=\Strfun{\Funlfp{\Vcsnot F}}(\Vect f)$ for all
    $\Vect f\in\LCAT^n(\Vect X,\Vect{X'})$
  \item and
    $\Strfun{\Vcsnot F}(\Tens{\Excl Y}{\Vect
      X},\Strnat{\Funlfp{\Vcsnot F}}_{Y,\Vect X})\Compl\Strnat{\Vcsnot
      F}_{Y,(\Vect X,\Strfun{\Funlfp{\Vcsnot F}}(\Vect
      X))}=\Strnat{\Funlfp{\Vcsnot F}}_{Y,\Vect X}$.
  \end{itemize}
  Moreover $\Orth{(\Funlfp{\Vcsnot F})}=\Fungfp{(\Orth{\Vcsnot F})}$
\end{lem}
\proof Apply Lemma~\ref{lemma:strfun-gfp-general} to the strong
functor $\Orth{\Vcsnot F}$.  \qed

\section{Coherence spaces}\label{sec:coherence-spaces}
We consider now the case where $\LCAT$ is the category $\COH$ of
coherence spaces and linear maps, a well-known model of $\LL$
introduced in~\cite{Girard86,Girard87}.

\renewcommand\LCAT{\COH}

A \emph{coherence space} is a structure $E=(\Web E,\Coh E{}{})$ where
$\Web E$ is a set (that we always assume to be at most countable since
this property is preserved by all the constructions presented in this
paper) called the \emph{web} of $E$ and $\Coh E{}{}$ is a binary
reflexive and symmetric relation on $\Web E$. A \emph{clique} of $E$
is a subset $u$ of $\Web E$ such that
$\forall a_1,a_2\in u\ \Coh E{a_1}{a_2}$. We use $\Cl E$ for the
set of all cliques of $E$ that we consider as a domain, the order
relation on $\Cl E$ being always inclusion. Observe indeed that
$\emptyset\in\Cl E$ (that is $\Cl E$ has a least element),
if $u\subseteq v$ and $v\in\Cl E$ then $u\in\Cl E$
and last if $D\subseteq\Cl E$ is directed then $\cup D\in\Cl E$.

\subsection{Coherence spaces as a model of $\LL$}

Given coherence spaces $E$ and $F$ we define a coherence space
$\Limpl EF$ whose web is $\Web E\times\Web F$ and coherence is:
$\Coh{\Limpl EF}{(a_1,b_1)}{(a_2,b_2)}$ if
\(
  \Coh{E}{a_1}{a_2} \Implies
  (\Coh F{b_1}{b_2}\text{ and }b_1=b_2\Implies a_1=a_2)
\).

The category $\LCAT$ has coherence spaces as objects, and homsets
$\LCAT(E,F)=\Cl{\Limpl EF}$. In this category the identities are the
diagonal relations and composition is the ordinary composition of
relations.

\begin{rem}
  It can be useful to keep in mind that these morphisms can be
  considered as \emph{linear functions}: a function $f:\Cl E\to\Cl F$
  is linear if it is stable (that is
  $\forall u_1,u_2\in\Cl E\ u_1\cup u_2\in\Cl E\Implies f(u_1\cap
  u_2)=f(u_1)\cap f(u_2)$) and commutes with arbitrary well-defined
  unions of cliques. Such a function $f$ has a \emph{trace}
  $\Ltrace f=\{(a,b)\in\Web E\times\Web F\St b\in f(\{a\})\}$ and this
  trace operation defines a bijection between $\Cl{\Limpl EF}$ and the
  set of all linear functions from $\Cl E$ to $\Cl F$. The converse of
  this operation maps $t\in\Cl{\Limpl EF}$ to the function
  $\Lfun(t):\Cl E\to\Cl F$ defined by
  $\Lfun(t)(u)=\{b\in\Web F\St\exists a\in u\ (a,b)\in t\}$. We will
  always write $t\Compl u$ instead of $\Lfun(t)(u)$. In this paper we
  stick to the relational point of view on morphisms.
\end{rem}

This category is monoidal, with tensor product $\Tens{E_1}{E_2}$
having $\Web{E_1}\times\Web{E_2}$ as web and
$\Coh{\Tens{E_1}{E_2}}{(a_1,a_2)}{(a'_1,a'_2)}$ if
$\Coh{E_i}{a_i}{a'_i}$ for $i=1,2$. Given $t_i\in\LCAT(E_i,F_i)$ for
$i=1,2$, one defines $\Tens{t_1}{t_2}$ as
$\{((a_1,a_2),(b_1,b_2))\St (a_i,b_i)\in t_i\text{ for
}i=1,2\}\in\LCAT(\Tens{E_1}{E_2},\Tens{F_1}{F_2})$ as easily
checked. So $\ITens$ is a functor $\LCAT^2\to\LCAT$, which equips $\LCAT$
with an obvious symmetric monoidal structure that we will not make
explicit here, for a unit object
$\Sone=(\{\Oneelem\},\mathord=)$. This category is monoidal closed
with $\Limpl EF$ object of morphisms from $E$ to $F$ (and evaluation
morphism $\Evlin\in\LCAT(\Tens{(\Limpl EF)}{E},F)$ defined by
$\Evlin=\{(((a,b),a),b)\St a\in\Web E\text{ and }b\in\Web
F\}$). Taking $\Sbot=\One$ as dualizing object, $\LCAT$ is easily seen
to be *-autonomous and the corresponding orthogonality is a functor
$\Orth\_:\Op\LCAT\to\LCAT$ where
$\Orth E=(\Web E,\Incoh E{}{})\Isom(\Limpl E\Sbot)$ (by a trivial
iso), the \emph{incoherence} binary relation $\Incoh{E}{}{}$ being
defined by $\Incoh E{a_1}{a_2}$ if $\Coh E{a_1}{a_2}\Implies
a_1=a_2$. The transpose $\Orth t$ of $t\in\LCAT(E,F)$ is simply
$\{(b,a)\St (a,b)\in t\}$. Under this linear negation, the De Morgan
dual (par or cotensor) of the tensor product is
$\Par{E_1}{E_2}=\Orthp{\Tens{\Orth{E_1}}{\Orth{E_2}}}$ whose web is
$\Web{E_1}\times\Web{E_2}$ and whose coherence relation is
characterized by: $\Scoh{\Par{E_1}{E_2}}{(a_1,a_2)}{(a'_1,a'_2)}$ iff
$\Scoh{E_i}{a_i}{a'_i}$ for $i=1$ or $i=2$ (where $\Scoh Eab$ means
$\Coh Eab$ and $a\not=b$ and is called \emph{strict coherence};
\emph{strict incoherence} $\Sincoh E{}{}$ is defined
similarly). Remember that, with these notations,
$\Limpl EF=\Par{\Orth E}F$.

$\LCAT$ has a final object $\Top=(\emptyset,\emptyset)$ and a
cartesian product
$\With{E_1}{E_2}=
(\{1\}\times\Web{E_1}\cup\{2\}\times\Web{E_2},\Coh{\With{E_1}{E_2}}{}{})$
where the coherence relation is defined by:
$\Coh{\With{E_1}{E_2}}{(i,a)}{(j,a')}$ if
$i=j\Implies\Coh{E_i}{a}{a'}$, the associated projections
$\Proj i\in\LCAT(\With{E_1}{E_2},E_i)$ being
$\Proj i=\{((i,a),a)\St a\in\Web{E_i}\}$. Dually the initial object is
$\Zero=\Orth\Top=\Top$ and the coproduct is
$\Plus{E_1}{E_2}=\Orthp{\With{\Orth{E_1}}{\Orth{E_2}}}$ whose web is
$\{1\}\times\Web{E_1}\cup\{2\}\times\Web{E_2}$ and whose coherence is
characterized by $\Coh{\Plus{E_1}{E_2}}{(i,a)}{(j,a')}$ if $i=j$ and
$\Coh{E_i}{a}{a'}$. There are canonical injections
$E_i\to\Plus{E_1}{E_2}$ which are the transposes of the projections
defined above.

We define $\Excl E$ as the coherence space whose web is the set of all
finite elements of $\Cl E$ and the coherence is:
$\Coh{\Excl E}{u_1}{u_2}$ if $u_1\cup u_2\in\Cl E$ (that is
$\forall a_1\in u_1\forall a_2\in u_2\ \Coh E{a_1}{a_2}$). This
operation is a functor: given $t\in\LCAT(E,F)$ one sets
$\Excl t=\{(\{\List a1n\},\{\List b1n\}\in\Web{\Excl
  E}\times\Web{\Excl F}\St \forall i\ (a_i,b_i)\in t\}$. The comonad
structure of this functor is given by the natural transformations
$\Der E=\{(\{a\},a)\St a\in\Web E\}\in\LCAT(\Excl E,E)$ (dereliction)
and
$\Digg E=\{(u_1\cup\dots\cup u_n,\{\List u1n\})\St \List
u1n\in\Web{\Excl E})\text{ with }u_1\cup\dots\cup u_n\in\Cl E\}$
(digging). Last there is an obvious isomorphism
$\Seelyz\in\LCAT(\Sone,\Excl\Top)$ and a natural isomorphism
$\Seelyt_{E_1,E_2}
\in\LCAT(\Tens{\Excl{E_1}}{\Excl{E_2}},\Excl{(\With{E_1}{E_2})})$
(these isos defining a strong monoidal structure), satisfying an
additional technical condition explained in~\cite{Mellies09} for
instance.

\subsection{Coherence spaces form a cpo}\label{sec:subcoh}
Let $E$ and $F$ be coherence spaces, we write $E\Subcoh F$ if
$\Web E\subseteq\Web F$ and
\(
  \forall a,a'\in\Web E\quad \Coh E{a}{a'}\Equiv\Coh F{a}{a'}
\).

Observe that when $E\subseteq F$, one has two linear morphisms 
$\Cohemb EF\in\LCAT(E,F)$ and $\Cohret EF\in\LCAT(F,E)$ given by
\(
  \Cohemb EF=\Cohret EF=\{(a,a)\St a\in\Web E\}
\).

They satisfy $\Matapp{\Cohret EF}{\Cohemb EF}=\Id_E$ and
$\Matapp{\Cohemb EF}{\Cohret EF}\subseteq\Id_F$, defining an
embedding-retraction pair of coherence spaces as considered for
instance in~\cite{Girard86} (all embedding-retraction pairs are of that
shape, up to isomorphism of coherence spaces). One major feature of
this order relation is that it makes linear negation
monotonic\footnote{and not antitonic as one might expect having
  \Eg~intersection types in mind}, making life quite easy when one
needs to compute fixed points of arbitrary $\LL$ formulas in this
model.

\begin{lem}
  The relation $\Subcoh$ is a partial order relation on coherence
  spaces, and we have $E\Subcoh F\Equiv\Orth E\Subcoh\Orth F$.
\end{lem}

We use $\LCATPO$ for the class of coherence spaces ordered under the
$\Subcoh$ partial order relation. This partially ordered class has a
least element denoted as $\Cohempty$ (the coherence space which has an
empty web).

Any countable directed subset $\cE$
of $\LCATPO$ has a lub $\Cohlub\cE$, which is the coherence space
defined by $\Web{\Cohlub\cE}=\cup_{E\in\cE}\Web E$ and, for all
$a,a'\in\Web{\Cohlub\cE}$, one has $\Coh{\Cohlub\cE}{a}{a'}$ iff
$\Coh E{a}{a'}$ for some $E\in\cE$.

The family $\cE$ gives rise to two diagrams in the category $\LCAT$:
\begin{itemize}
\item the \emph{inductive diagram} $\Indcoh{\cE}$ with morphisms
  $\Cohemb EF\in\LCAT(E,F)$ when $E,F\in\cE$ with $E\Subcoh F$
\item and the \emph{projective diagram} $\Procoh\cE$ with morphisms
  $\Cohret EF\in\LCAT(F,E)$ when $E,F\in\cE$ with $E\Subcoh F$.
\end{itemize}

\begin{lem}\label{lemma:coh-union-limits}
  Let $G=\Cohlub\cE$. Then, in $\LCAT$, the cocone
  $(E,\Cohemb EG)_{E\in\cE}$ is the colimit of the inductive diagram
  $\Indcoh\cE$ and the cone $(E,\Cohret EG)_{E\in\cE}$ is the limit of
  the projective diagram $\Procoh\cE$.
\end{lem}
\proof We prove the first statement, the second one following by
duality. Let $F$ be a coherence space and, for each $E\in\cE$ let
$t_E\in\LCAT(E,F)$ defining a cocone based on $\Indcoh\cE$, which
means
$\forall E,E'\in\cE\ E\Subcoh E'\Implies t_{E'}\Compl\Cohemb
E{E'}=t_E$, that is
$\forall E,E'\in\cE\ E\Subcoh E'\Implies t_E=t_{E'}\cap\Web
E\times\Web F$. Then the unique morphism $t\in\LCAT(G,F)$ such that
$\forall E\in\cE\ t\Compl\Cohemb EG=t_E$ is given by
$t=\cup_{E\in\cE}t_E$ as easily checked. \qed

\begin{defi}\label{def:coh-cont-obj}
  A functor $\cF:\LCAT^n\to\LCAT$ is \emph{continuous on objects} if
  whenever $\Vect E,\Vect F\in\COH^n$ satisfy $E_i\Subcoh F_i$ for
  $i=1,\dots,n$, one has $\cF(\Vect E) \Subcoh\cF(\Vect F)$ and
  \( \cF(\Cohemb{E_1}{F_1},\dots,\Cohemb{E_n}{F_n}) =\Cohemb{\cF(\Vect
    E)}{\cF(\Vect F)}\) and
  \(\cF(\Cohret{E_1}{F_1},\dots,\Cohret{E_n}{F_n}) =\Cohret{\cF(\Vect
    E)}{\cF(\Vect F)}\ \).  Moreover, $\cF$ commutes with the lubs of
  countable directed families of coherence spaces. In other words, for
  any countable directed families of coherence spaces $\List\cE 1n$,
  one has
  \( \cF(\Cohlub{\cE_1},\dots,\Cohlub{\cE_n})= \Cohlub\{\cF (\List
  E1n)\St E_1\in\cE_1,\dots,E_n\in\cE_n\} \).

  One says that $\cF$ is \emph{continuous on morphisms} if when
  $\Vect f,\Vect g\in\LCAT^n(\Vect E,\Vect F)$ satisfy
  $\Vect f\subseteq\Vect g$ (that is $\forall i\ f_i\subseteq g_i$)
  one has $\cF(\Vect f)\subseteq\cF(\Vect g)$ and, if $D$ is a
  directed subset of $\LCAT^n(\Vect E,\Vect F)$, one has
  $\cF(\cup D)=\cup_{\Vect f\in D}\cF(\Vect f)$ (equivalently
  $\cF(\cup D)\supseteq\cup_{\Vect f\in D}\cF(\Vect f)$).

  Last one says that $\cF$ is \emph{continuous} if it is both
  continuous on objects and on morphisms.
\end{defi}
Notice that this property is preserved by composition and duality
(setting, consistently with Section~\ref{sec:gen-strong-functors},
$(\Orth\cF)(\Vect E)=\Orthp{\cF(\Orth{\Vect E})}$ and similarly for
morphisms).

\begin{defi}\label{def:VCS}
  A ($n$-ary) variable coherence space (VCS) is a strong functor
  $\Vcsnot E:\COH^n\to\COH$ such that $\Strfun{\Vcsnot E}$ is
  monotonic and continuous.
\end{defi}

\begin{prop}\label{prop:VCS-operations}
  The operations $\ITens$, $\IPar$, $\IPlus$, $\IWith$, $\oc$ and
  $\wn$ are VCSs and VCSs are closed under De Morgan duality and
  composition.
\end{prop}
\proof This results immediately from the properties of strong
functors stated in Section~\ref{sec:gen-strong-functors} and from
straightforward computations (for the continuity statement).
\qed

% general terms are not compulsory anymore, 
% you may leave them out
% \terms
% term1, term2
%
% \keywords
% lambda-calculus, call by push value, LL, denotational semantics,
% Scott semantics

\subsection{Least fixed point of a VCS, universal properties
  wrt.~algebras and coalgebras}\label{sec:VCS-fixpoints}
Let $\cF:\LCAT\to\LCAT$ be continuous on objects (in the sense of
Definition~\ref{def:coh-cont-obj}). Then we have
\(
  \Cohempty\Subcoh\cF(\Cohempty)\Subcoh\cdots
  \Subcoh\cF^n(\Cohempty)
  \Subcoh\cF^{n+1}(\Cohempty)\Subcoh\cdots
\)
as shown by an easy induction on $n$. We set
\(
  \Vcsfp{\cF}=\Cohlub_{n=0}^\infty
  \cF^n(\Cohempty)
\).
By Scott continuity of $\cF$, we have
$\cF(\Vcsfp{\cF})=\Vcsfp{\cF}$.

\begin{lem}\label{lemma:vcs-fixpoint-alg-coalg}
  One has $\Vcsfp{(\Orth{\cF})}=\Orthp{\Vcsfp{\cF}}$. If moreover
  $\cF$ is continuous on morphisms then $\Vcsfp{\cF}$ is at the same
  time the initial object of $\ALGFUN{\LCAT}{\cF}$ and the final
  object of $\COALGFUN{\LCAT}{\cF}$.
\end{lem}
\proof The first statement results from the observation that
$(\Orth{\cF})^n=\Orthp{\cF^n}$. For the second statement, since
$\Vcsfp{(\Orth{\cF})}=\Orthp{\Vcsfp{\cF}}$, it suffices to prove that
$(\Vcsfp{\cF},\Id)$ is initial in $\ALGFUN{\LCAT}{\cF}$. This results
easily from Lemma~\ref{lemma:coh-union-limits} and from continuity on
morphisms. \qed

Let $\Vcsnot E$ be an $n+1$-ary VCS. Applying
Lemma~\ref{lemma:vcs-fixpoint-alg-coalg} to the functors
$\Strfun{\Vcsnot E}_{\Vect E}$ for all $\Vect E\in\LCAT^n$,
Lemma~\ref{lemma:strfun-lfp-general} shows that there is an $n$-ary
strong functor $\Funlfp{\Vcsnot E}$ uniquely determined by the
following equations
\begin{itemize}
% \item
%   $\Strfun{\Vcsnot E}(\Vect E,\Strfun{\Fungfp{\Vcsnot E}}(\Vect
%   E))=\Strfun{\Fungfp{\Vcsnot E}}(\Vect E)$
\item
  $\Strfun{\Funlfp{\Vcsnot E}}(\Vect E)=\Vcsfp{\Strfun{\Vcsnot E}_{\Vect E}}$
\item
  $\Strfun{\Vcsnot E}(\Vect f,\Strfun{\Funlfp{\Vcsnot E}}(\Vect
  f))=\Strfun{\Funlfp{\Vcsnot E}}(\Vect f)$ for
  $\Vect f\in\LCAT(\Vect E,\Vect{E'})$
\item  and
    $\Strfun{\Vcsnot E}(\Tens{\Excl F}{\Vect
      E},\Strnat{\Funlfp{\Vcsnot E}}_{F,\Vect E})\Compl\Strnat{\Vcsnot
      E}_{F,(\Vect E,\Strfun{\Funlfp{\Vcsnot E}}(\Vect
      E))}=\Strnat{\Funlfp{\Vcsnot E}}_{F,\Vect E}$.
\end{itemize}

\begin{comment}
, we
define
$\Strnat{\Vcsfp{\Vcsnot E}}_{G,\Vect E} \in\LCAT(\Tens{\Excl
  G}{\Strfun{\Vcsfp{\Vcsnot E}}(\Vect E)}, \Strfun{\Vcsfp{\Vcsnot
    E}}(\Tens{\Excl G}{\Vect E}))$.

as the following morphism,
setting $F=\Strfun{\Vcsfp{\Vcsnot E}}(\Tens{\Excl G}{\Vect E})$ for
readability
% \begin{align*}
%   \Strnat{\Vcsnot E}_{G,(\Vect E,F)}\in
%   \LCAT(\Tens{\Excl G}{\Strfun{\Vcsnot E}(\Vect E,F)})
% \end{align*}
\begin{center}
    \begin{tikzpicture}[->, >=stealth]
      \node (1) {$\Tens{\Excl G}{\Strfun{\Vcsnot E}(\Vect E,F)}$};
      \node (2) [right of=1, node distance=3.9cm]
      {$\Strfun{\Vcsnot E}(\Tens{\Excl G}{\Vect E},\Tens{\Excl G}F)$};
      \node (3) [right of=2, node distance=4.9cm]
      {$\Strfun{\Vcsnot E}(\Tens{\Excl G}{\Vect E},F)=F$};
      \tikzstyle{every node}=[midway,auto,font=\scriptsize]
      \draw (1) -- node {$\Strnat{\Vcsnot E}_{G,(\Vect E,F)}$} (2);
      \draw (2) -- node {$\Strfun{\Vcsnot E}(\Vect{\Id},
        \Tens{\Weak{\Excl G}}{F})$} (3);
    \end{tikzpicture}   
\end{center}
\end{comment}

\begin{prop}\label{prop:coh-fp-init-final}
  The functor $\Funlfp{\Vcsnot E}$ is a variable coherence
  space. Defining the dual operation as
  $\Fungfp{\Vcsnot E}=\Orthp{\Funlfp{(\Orth{\Vcsnot E})}}$, one has
  $\Fungfp{\Vcsnot E}=\Funlfp{\Vcsnot E}$. We use $\Vcsfp{\Vcsnot E}$ for 
  this unique (final and initial) fixed point VCS.
\end{prop}
\proof
% Assume first that $\Vect E\Subcoh\Vect F$ in $\COHPO^n$, we prove that, for all $k\in\Nat$, we have
% $\Vcsnot E_{\Vect E}^k(\Cohempty)\Subcoh\Vcsnot E_{\Vect F}^k(\Cohempty)$. For $k=0$, this is obvious, so assume that the property holds for $k$. We have 
% \begin{align*}
%   \Vcsnot E_{\Vect E}^{k+1}(\Cohempty)
%   &=\Vcsnot E(\Vect E,\Vcsnot E_{\Vect E}^{k}(\Cohempty))\\
%   &\Subcoh \Vcsnot E(\Vect E,\Vcsnot E_{\Vect F}^{k}(\Cohempty))\quad\text{by
%     inductive hypothesis}\\
%   &\Subcoh \Vcsnot E(\Vect F,\Vcsnot E_{\Vect F}^{k}(\Cohempty))
%     \quad\text{because $\Vcsnot E$ is a VCS.}
% \end{align*}
% Therefore
% $\Vcsfp{\Vcsnot E}(\Vect E)\Subcoh\Vcsfp{\Vcsnot E}(\Vect F)$. Next
% one proves that, given a directed family $\Vect\cE$ in $\COHPO^n$,
% it holds that
The proof that $\Strfun{\Funlfp{\Vcsnot E}}$ is monotonic and Scott
continuous on $\COHPO^n$ is a standard domain-theoretic
verification. We are left with proving that given $\Vect E,\Vect F$ in
$\LCAT^n$ such that $\Vect E\Subcoh\Vect F$, one has
\begin{align*}
  \Strfun{\Funlfp{\Vcsnot E}}(\Cohemb{\Vect E}{\Vect F})
  =\Cohemb{\Strfun{\Funlfp{\Vcsnot E}}(\Vect E)}
  {\Strfun{\Funlfp{\Vcsnot E}}(\Vect F)}\quad\quad
    \Strfun{\Funlfp{\Vcsnot E}}(\Cohret{\Vect E}{\Vect F})
  =\Cohret{\Strfun{\Funlfp{\Vcsnot E}}(\Vect E)}
  {\Strfun{\Funlfp{\Vcsnot E}}(\Vect F)}\,.
\end{align*}
Let use prove the first equation, the proof of the second one being
completely similar.  By Lemma~\ref{lemma:equations-final-coalgebra},
it suffices to prove
$ \Vcsnot E({\Cohemb{\Vect E}{\Vect
    F}},{\Cohemb{\Strfun{\Funlfp{\Vcsnot E}}(\Vect E)}
  {\Strfun{\Funlfp{\Vcsnot E}}(\Vect F)}}) ={\Cohemb{\Strfun{\Funlfp{\Vcsnot
      E}}(\Vect E)}{\Strfun{\Funlfp{\Vcsnot E}}(\Vect F)}} $ which in turn
results from the assumption that $\Vcsnot E$ is a VCS.

The identity $\Funlfp{\Vcsnot E}=\Fungfp{\Vcsnot E}$ results from the
uniqueness statements of Lemmas~\ref{lemma:strfun-gfp-general}
and~\ref{lemma:strfun-lfp-general} and from the fact that
$\Strfun{\Fungfp{\Vcsnot E}}(\Vect E)=\Vcsfp{\Strfun{\Vcsnot E}_{\Vect
    E}}=\Strfun{\Funlfp{\Vcsnot E}}(\Vect E)$. \qed

\begin{rem}
  The methods developed in this section are by no way specific to
  coherence spaces and could be used in many other models of $\LL$
  (relational semantics, Scott semantics, hypercoherence spaces,
  probabilistic coherence spaces, game models, up to some adaptation
  since these are not models of classical $\LL$, etc).
\end{rem}

\section{Coherence spaces with totality}\label{sec:coh-tot}
Let $E$ be a coherence space and let $\cA\subseteq\Cl E$. We set
\[
  \Orth\cA=\{x'\in\Cl{\Orth E}\St\forall x\in\cA\ x\cap x'\not=\emptyset\}\,.
\]
Observe that if $x\cap x'\not=\emptyset$ then this intersection has
exactly one element, due to the fact that $x$ and $x'$ are cliques in
$E$ and $\Orth E$ respectively.

If $\cA,\cB\subseteq\Cl E$ and $\cA\subseteq\cB$, we have
$\Orth\cB\subseteq\Orth\cA$, and also
$\cA\subseteq\Biorth\cA$. Therefore $\Orth\cA=\Triorth\cA$.

A \emph{totality candidate} on $E$ is a set $\cT\subseteq\Cl E$ such
that $\Biorth\cT=\cT$, or equivalently $\Biorth\cT\subseteq\cT$. This
property is equivalent to the existence of a ``predual'' of $\cT$,
that is, of a set $\cA\subseteq\Cl{\Orth E}$ such that
$\cT=\Orth\cA$. We use $\Tot E$ for the set of all totality candidates
of the coherence space $E$, and we consider this set as a poset,
equipped with inclusion.

\begin{lem}
  The poset $\Tot E$ is a complete lattice.
\end{lem}
\proof
Let $\Theta\subseteq\Tot E$ and let
$\Theta'=\{\Orth\cT\St\cT\in\Theta\}$, which is a subset of
$\Tot{\Orth E}$. Given $x\in\Cl{E}$, we have $x\in\cap\Theta$ iff for
all $\cT'\in\Theta'$ and all $x'\in\cT'$, $x\cap x'\not=\emptyset$.
in other words
\(
  \cap\Theta=\Orthp{\bigcup_{\cT\in\Theta}\Orth\cT}\in\Tot E
\).
\qed

The greatest element of $\Tot E$ is $\Cl E$ and its least element is
$\Orthp{\Cl{\Orth E}}=\emptyset$ as easily checked. Any subset
$\Theta$ of $\Tot E$ has a least upper bound $\vee\Theta$ which is
given by
\(
  \vee\Theta=\Biorth{(\cup\Theta)}
\)
and this biorthogonal closure cannot be disposed of in general
($\cup\Theta$ is not necessarily a totality candidate). It is useful
to observe that the map $\cT\mapsto\Orth\cT$ is an isomorphism between
the complete lattices $\Tot{\Orth E}$ and $\Op{\Tot E}$.

A \emph{coherence space with totality} is a pair
$X=(\Tcohca X,\Tcoht X)$ where $\Tcohca X$ is a coherence space (the
carrier) and $\Tcoht X\in\Tot{\Tcohca X}$.

\subsection{Coherence spaces with totality as a model of $\LL$}
Let $X$ and $Y$ be coherence spaces with totality, we define a
coherence space with totality $\Limpl XY$ by
$\Tcohca{\Limpl XY}=\Limpl{\Tcohca X}{\Tcohca Y}$ and
$\Tcoht{(\Limpl XY)}=\{t\in\Cl{\Limpl{\Tcohca X}{\Tcohca Y}}\St
\forall x\in\Tcoht X\ \Matapp tx\in\Tcoht Y\}=\Orth{\{\Tens x{y'}\St
  x\in\Tcoht X\text{ and }y'\in\Tcoht{\Orth Y}\}}$, this latter
equation resulting from the equivalence
$(\Matapp tx)\cap y'\not=\emptyset\Equiv t\cap(x\times
y')\not=\emptyset$. It is clear that if $s\in\Tcoht{(\Limpl XY)}$ and
$t\in\Tcoht{(\Limpl YZ)}$ then $\Matapp ts\in\Tcoht{(\Limpl XZ)}$, and
also that $\Id\in\Tcoht{(\Limpl XX)}$, hence we have defined a
category that we denote as $\COHT$.

The following is a useful tool for proving that a linear morphism (a
morphism in $\COH$ between the carriers of two coherence spaces with
totality) is total.
\begin{lem}\label{lemma:total-morph-charact}
  Let $t\in\COH(\Tcohca X,\Tcohca Y)$ and let
  $\cA\subseteq\Cl{\Tcohca X}$ be such that $\Tcoht X=\Biorth\cA$. If
  $\forall x\in\cA\ \Matapp tx\in\Tcoht Y$ then $t\in\COHT(X,Y)$.
\end{lem}
\proof Let $x\in\Tcoht X$, we have to prove that
$\Matapp tx\in\Tcoht Y$ so let $y'\in\Tcoht{\Orth Y}$, we must prove
that $(\Matapp tx)\cap y'\not=\emptyset$. This statement is equivalent
to
$t\cap(\Tens x{y'})\not=\emptyset\Equiv (\Matapp{\Orth t}{y'})\cap
x\not=\emptyset$. So we must prove
$\forall x\in\Tcoht X\,\forall y'\in\Tcoht{\Orth Y}\ (\Matapp{\Orth
  t}{y'})\cap x\not=\emptyset$, that is
$\Orth t\in\COHT(\Orth Y,\Orth X)$. So let
$y'\in\Tcoht{\Orth Y}$, we must prove that
$\Matapp{\Orth t}{y'}\in\Tcoht{\Orth X}=\Orth\cA$ which results
from our assumption by the same reasoning. \qed

We equip $\Sone$ and $\Sbot$ with the same totality, namely
$\{\{\Onelem\}\}$. We define $\Tens{X_1}{X_2}$ by
$\Tcohca{\Tens{X_1}{X_2}}=\Tens{\Tcohca{X_1}}{\Tcohca{X_2}}$ and
$\Tcoht{(\Tens{X_1}{X_2})}=\Biorth{\{\Tens{x_1}{x_2}\St
  x_i\in\Tcoht{X_i}\text{ for }i=1,2\}}$, so that
$\Tens XY=\Orthp{\Limpl{X}{\Orth Y}}$. Then it is easy to check that
$\COHT$ is *-autonomous, \emph{with the same operations on morphisms
  as in $\COH$} (for instance one checks that if
$t_i\in\COHT(X_i,Y_i)$ then
$\Tens{t_1}{t_2}\in\COHT(\Tens{X_1}{X_2},\Tens{Y_1}{Y_2})$ which is
easy using Lemma~\ref{lemma:total-morph-charact}; in the same way one
proves easily that $\Evlin\in\COHT(\Tens{(\Limpl XY)}{X},Y)$
etc). Similarly one shows that the cartesian structure on $\COH$ gives
rise to a cartesian structure on $\COHT$:
$\Tcohca{\With{X_1}{X_2}}=\With{\Tcohca{X_1}}{\Tcohca{X_2}}$ and
$\{1\}\times x_1\cup\{2\}\times x_2\in\Tcoht{(\With{X_1}{X_2})}$ if
$x_i\in\Tcoht{X_i}$ for $i=1,2$. The total cliques of
$\Plus{X_1}{X_2}$ are the $\{i\}\times z$ for $i=1,2$ and
$z\in\Tcoht{X_i}$. Notice that $\Top$ and $\Zero$ are different
coherence spaces with totality: $\Tcoht\Top=\{\emptyset\}$ and
$\Tcoht\Zero=\emptyset$.

Last $\Excl X$ is given by $\Coalgca{\Excl X}=\Excl{\Coalgca X}$ and
$\Tcoht{(\Excl X)}=\Biorth{\{\Prom x\St x\in\Tcoht X\}}$ (where
$\Prom x=\Partfin x$). Then one proves easily that
$t\in\COHT(X,Y)\Implies\Excl t\in\COHT(\Excl X,\Excl Y)$ again using
Lemma~\ref{lemma:total-morph-charact}. It is also easy to check that
$\Der{\Coalgca X}\in\COHT(\Excl X,X)$ and that
$\Digg{\Coalgca X}\in\COHT(\Excl X,\Excl{\Excl X})$ so we denote these
morphisms as $\Der X$ and $\Digg X$ turning ``$\oc$'' into a comonad
on $\COHT$. The same holds for the monoidal structure (Seely
isomorphisms): $\Seelyz\in\COHT(\Sone,\Excl\Top)$ and
$\Seelyt_{X_1,X_2}\in\COHT(\Tens{\Excl{X_1}}{\Excl{X_2}},
\Excl{(\With{X_1}{X_2})})$.

\subsection{Variable coherence spaces with totality (VCST)}
\label{sec:VCST}
We first recall the well-known Knaster-Tarski's Theorem.
\begin{thm}\label{th:KT}
  Let $S$ be a complete lattice and let
  $f:S\to S$ be a monotonic function. Then $f$  has a least
  fixed point in $S$.  Let $(s_\alpha)$ be the ordinal-indexed
  family of elements of $S$ defined by $s_{\alpha+1}=f(s_\alpha)$ and
  $s_\lambda=\bigvee_{\alpha<\lambda}s_\alpha$ for $\lambda$ limit
  ordinal. Then this sequence is monotonic and there is an ordinal
  $\theta$ such that $s_\theta=s_{\theta+1}$. Moreover $s_\theta$ is
  the least fixed point of $f$ in $S$.
\end{thm}

We denote $s_\alpha$ as $f^\alpha(0)$ where $0$ is the least element
of $S$; this notation coincides with a finite iterations of $f$ when
$\alpha$ is a finite ordinal (we have $s_0=0$ considering $0\in\Nat$
as a limit ordinal).

\subsubsection{General definition of a VCST}
% We introduce the main concept of this paper.
To make the notations more readable, when $\Vcsnot E$ is a VCS (see
Definition~\ref{def:VCS}), we use $\Vcsnot E$ (instead of
$\Strfun{\Vcsnot E}$) to denote its functorial part. We keep denoting
as $\Strnat{\Vcsnot E}$ the associated strength natural
transformation.

\begin{defi}\label{def:VCST}
  An $n$-ary \emph{variable coherence space with totality} (VCST) is a
  pair $\Vcstnot X=(\Vcstca{\Vcstnot X},\Vcstot{\Vcstnot X})$ where
  \begin{itemize}
  \item $\Vcstca{\Vcstnot X}:\COHLIN^n\to\COHLIN$ is an $n$-ary VCS
    called the \emph{carrier of $\Vcstnot X$}
  \item and $\Vcstot{\Vcstnot X}$ is an operation, called the
    \emph{totality of $\Vcstnot X$}, which, with each $n$-tuple
    $\Vect X$ of coherence spaces with totality, associates
    $\Vcstot{\Vcstnot X}(\Vect X)\in\Tot{\Vcstca{\Vcstnot
        X}(\Tcohca{\Vect X})}$ ---~and we use the notation
    $\Vcstnot X(\Vect X)$ for the coherence space with totality
    $(\Vcstca{\Vcstnot X}(\Tcohca{\Vect X}),\Vcstot{\Vcstnot X}(\Vect
    X))$.
  \end{itemize}
  Moreover the two following properties must hold.
  \begin{itemize}
  \item If $\Vect X$ and $\Vect Y$ are objects of $\COHLINT^n$ and
    $\Vect f\in\COHLINT^n(\Vect X,\Vect Y)$, then the $\COH$ morphism
    $\Vcstca{\Vcstnot X}(\Vect f)$ belongs actually to
    $\COHLINT({\Vcstnot X}(\Vect X),{\Vcstnot X}(\Vect Y))$, so that
    $\Vcstnot X$ defines a functor $\COHLINT^n\to\COHLINT$ (denoted
    simply as $\Vcstnot X$).
  \item If $\Vect X$ is an object of $\COHT^n$ and $Y$ is an object of
    $\COHT$ then the $\COH$ morphism
    $\Strnat{\Vcstca{\Vcstnot X}}_{\Tcohca Y,\Tcohca{\Vect X}}$
    belongs actually to
    $\COHT(\Tens{\Excl Y}{{\Vcsnot X}}(\Vect X),{\Vcsnot
      X}(\Tens{\Excl Y}{\Vect X}))$. We denote this total morphism as
    $\Strnat{\Vcstnot X}_{Y,\Vect X}$.
  \end{itemize}
\end{defi}
So we can consider $\Vcstnot X$ as a strong functor $\COHT^n\to\COHT$
(the monoidality diagram commutations of
Figure~\ref{fig:strength-monoidality} hold because the $\LL$
operations on morphisms are interpreted in the same way in $\COHT$ and
in $\COH$).

\begin{rem}
  To fully understand this definition, it is essential to keep in mind
  that, if $\Vect f\in\COHLINT(\Vect X,\Vect Y)$ then actually
  $\Vect f\in\COHLIN(\Tcohca{\Vect X},\Tcohca{\Vect Y})$ so that the
  morphism
  $\Vcstca{\Vcstnot X}(\Vect f)\in\COHLIN(\Vcstca{\Vcstnot
    X}(\Tcohca{\Vect X}),\Vcstca{\Vcstnot X}(\Tcohca{\Vect Y}))$ is
  defined, \emph{independently of the notions of totality on $\Vect X$
    and $\Vect Y$} and similarly for
  $\Strnat{\Vcstnot X}_{Y,\Vect X}$. This decoupling of the
  totality-free part of the notions involved from the totality
  dependent ones makes life much simpler. This situation can certainly
  be axiomatized categorically, around the obvious forgetful functor
  $\COHT\to\COH$ which commutes with all $\LL$ constructs; this
  abstract categorical analysis is postponed to further work.
\end{rem}

\begin{rem}
  Strictly speaking, an $n$-ary VCST $\Vcstnot X$ is not a strong
  functor $\COHT^n\to\COHT$ but a structure which induces ---~as
  explained above~--- such a strong functor $\cF$, that we have
  denoted simply as $\Vcstnot X$. This choice of notation is motivated
  by the fact that $\Vcstnot X$ can very simply be recovered from
  $\cF$. We have indeed a forgetful functor $\Tcohforg:\COHT\to\COH$
  which maps $X$ to $\Tcohca X$ and acts as the identity on
  morphisms. This functor has a left adjoint $\Tcohempty:\COH\to\COHT$
  which maps a coherence space $E$ to $(E,\emptyset)$ (no cliques of
  $E$ are total) and acts as the identity on morphisms. Then we have
  $\Vcstca{\Vcstnot
    X}=\Tcohforg\Comp\cF\Comp\Tcohempty=\Tcohforg\Comp\cF\Comp\Orth\Tcohempty$
  (for the functorial part of $\Vcstca{\Vcstnot X}$) and for the
  strength
  $\Strnat{\Vcstca{\Vcstnot X}}_{F,\Vect
    E}=\Strnat\cF_{\Tcohempty(F),\Tcohempty(\Vect E)}$, and
  $\cT=\Vcstot{\Vcstnot X}(\Vect X)$ is defined by the fact that the
  coherence space with totality $\cF(\Vect X)$ is of shape
  $(F,\cT)$. In these definitions, the choice of $\Tcohempty$ as
  ``inverse'' of $\Tcohforg$ is arbitrary. By the definition of VCSTs
  we could have used the right adjoint $\Orth\Tcohempty$ (it maps $E$
  to $(E,\Cl E)$ where all cliques are total) or any other functor
  $\cY:\COH\to\COHT$ such that $\Tcohforg\Comp\cY=\Id$ instead: the
  resulting $\Vcstnot X$ would have been the same. For these reasons,
  it is meaningful to consider VCSTs as strong functors
  $\COHT^n\to\COHT$, what we do now.

  This observation also motivates our general notion of model
  presented in Definition~\ref{def:categorical-muLL-models}.
\end{rem}

\begin{prop}\label{prop:VCST-operations}
  The operations $\ITens$, $\IPar$, $\IPlus$, $\IWith$, $\oc$ and
  $\wn$ are VCSTs and VCSTs are closed under De Morgan duality and
  composition.
\end{prop}
This is a consequence of Proposition~\ref{prop:VCS-operations}.

% Just as VCSs, VCSTs are closed under composition, De~Morgan
% dualization, and all the connectives of $\LL$ can be seen as VCSTs. We
% address now their main feature: they have least and greatest
% fixed points.

\subsubsection{Fixed Points of VCST's}\label{sec:fix-VCST}
% Let $\Vcstnot X$ be an $n+1$-ary VCST. Then $\Vcstca{\Vcstnot X}$ is
% an $n+1$-ary VCS so that we have an $n$-ary VCS
% $\Vcsfp{\Vcstca{\Vcstnot X}}$ as defined in
% Section~\ref{sec:VCS-fixed points}.

We deal first with least fixed points of unary VCST's, so let $\Vcsnot X$
be a unary VCST (whose strength is not used in this first step). We
define a coherence space with totality $\Vcstmuu{\Vcstnot X}$.  First,
we set $\Tcohca{\Vcstmuu{\Vcstnot X}}=\Vcsfp{\Vcstca{\Vcstnot X}}$.

We define a map
$\Totop{\Vcstnot X}:\Tot{\Tcohca{\Vcstmuu{\Vcstnot
      X}}}\to\Tot{\Tcohca{\Vcstmuu{\Vcstnot X}}}$ as follows: if
$\cT\in\Tot{\Tcohca{\Vcstmuu{\Vcstnot X}}}$, then
$\Vcstot{\Vcstnot X}(\Tcohca{\Vcstmuu{\Vcstnot
    X}},\cT)\in\Tot{\Vcstca{\Vcstnot X}(\Tcohca{\Vcstmuu{\Vcstnot
      X}})}=\Tot{\Tcohca{\Vcstmuu{\Vcstnot X}}}$ and we set
$\Totop{\Vcstnot X}(\cT)=\Vcstot{\Vcstnot X}(\Tcohca{\Vcstmuu{\Vcstnot
    X}},\cT)$.  We contend that this mapping is monotonic on the
lattice $\Tot{\Tcohca{\Vcstmuu{\Vcstnot X}}}$. Assume that
$\cT,\cT'\in\Tot{\Tcohca{\Vcstmuu{\Vcstnot X}}}$ with
$\cT\subseteq\cT'$. Then
$\Id\in\COHT((\Tcohca{\Vcstmuu{\Vcstnot
    X}},\cT),(\Tcohca{\Vcstmuu{\Vcstnot X}},\cT'))$ (see
Section~\ref{sec:coh-tot}) and hence
$\Id=\Vcstca{\Vcstnot X}(\Id)\in\COHT((\Tcohca{\Vcstmuu{\Vcstnot
    X}},\Totop{\Vcstnot X}(\cT)),(\Tcohca{\Vcstmuu{\Vcstnot
    X}},\Totop{\Vcstnot X}(\cT')))$ by Definition~\ref{def:VCST}, from
which it follows that
$\Totop{\Vcstnot X}(\cT)\subseteq \Totop{\Vcstnot X}(\cT')$.

Let $\cU$ be the least fixed point of $\Totop{\Vcstnot X}$ (applying
Theorem~\ref{th:KT}), we set $\Tcohtp{\Vcstmuu{\Vcstnot X}}=\cU$ and
this ends the definition of the coherence space with totality
$\Vcstmuu{\Vcstnot X}$, which satisfies
$\Vcstnot X(\Vcstmuu{\Vcstnot X})=\Vcstmuu{\Vcstnot X}$. Now we prove
that it is initial in $\ALGFUN{\COHT}{\Coalgca{\Vcstnot X}}$.

For this we shall use the following sequence of candidates of totality
for $\Tcohca{\Vcstmuu{\Vcstnot X}}$, indexed by ordinals:
$\cU_{\alpha+1}=\Totop{\Vcstnot X}(\cU_\alpha)$ and
$\cU_\lambda=\Biorth{(\bigcup_{\alpha<\lambda}\cU_\alpha)}$ when
$\lambda$ is a limit ordinal. Remember that there is an ordinal
$\theta$ such that $\cU_{\theta+1}=\cU_\theta$, and that we have
$\cU=\cU_\theta$ (see Theorem~\ref{th:KT}).

\begin{prop}\label{prop:VCST-mu-init}
  $\Vcstmuu{\Vcstnot X}$ is initial in the category
  $\ALGFUN{\COHLINT}{\Vcstnot X}$.
\end{prop}
\proof
Let $(X,g)$ be an object in $\ALGFUN{\COHLINT}{\Vcstnot X}$, that is
$g\in\COHLINT(\Vcstnot X(X),X)$. This means in particular that
$g\in\COHLIN(\Vcstca{\Vcstnot X}(\Tcohca X),\Tcohca X)$ so that, by
Proposition~\ref{prop:coh-fp-init-final}, we know that there is
exactly one morphism
$\hat g\in\COHLIN(\Tcohca{\Vcstmuu{\Vcstnot X}},\Tcohca X)$ such that
\begin{align*}
  \Matapp g{\Vcstca{\Vcstnot X}(\hat g)}=\hat g\,.
\end{align*}
We have to prove that $\hat g\in\COHLINT(\Vcstmuu{\Vcstnot X},X)$. By
induction on the ordinal $\alpha$, we prove that
\begin{align*}
  \hat g\in\COHLINT((\Tcohca{\Vcstmuu{\Vcstnot X}},\cU_\alpha),X)
\end{align*}
for all ordinal $\alpha$. Assume first that the property holds for
$\alpha$ and let us prove it for $\alpha+1$. By
Definition~\ref{def:VCST} we get
$\Vcstca{\Vcstnot X}(\hat g)\in\COHLINT((\Tcohca{\Vcstmuu{\Vcstnot
    X}},\cU_{\alpha+1}),\Vcstnot X(X))$ and hence
$\hat g=\Matapp g{\Vcstnot X(\hat
  g)}\in\COHLINT((\Tcohca{\Vcstmuu{\Vcstnot X}},\cU_{\alpha+1}),X)$.
Let now $\lambda$ be a limit ordinal and assume that
$\hat g\in\COHLINT((\Tcohca{\Vcstmuu{\Vcstnot X}},\cU_\alpha),X)$ for
all $\alpha<\lambda$. It will be sufficient to prove that
$\Orth{\hat g}\in\COHLINT(\Orth
X,\Orthp{\bigcup_{\alpha<\lambda}\cU_\alpha})$ so let
$x'\in\Orth{\Tcoht X}$, we must prove that
$\Matapp{\Orth{\hat
    g}}{x'}\in\Orthp{\bigcup_{\alpha<\lambda}\cU_\alpha}$ so let
$y\in\cU_\alpha$ for some $\alpha<\lambda$, we must prove that
$(\Matapp{\Orth{\hat g}}{x'})\cap y\not=\emptyset$, that is
$x'\cap\Matapp gy\not=\emptyset$ which results from our inductive
hypothesis applied to ordinal $\alpha$.

% Given $x\in\cU_\lambda$, we must prove that
% $\Matapp{\hat g}x\in\Tcoht X$. Since
% $\cU_\lambda=\bigcup_{\alpha<\lambda}\cU_\alpha$, there is an
% $\alpha<\lambda$ such that $x\in\cU_\alpha$ and we have
% $\Matapp{\hat g}x\in\Tcoht X$ by inductive hypothesis.

So we have proven the existence of
$\hat g\in\COHLINT(\Vcstmuu{\Vcstnot X},X)$ such that
$\Matapp g{\Vcstca{\Vcstnot X}(\hat g)}=\hat g$. Uniqueness follows
from the uniqueness property for $\Tcohca{\Vcstmuu{\Vcstnot X}}$.
\qed

We consider now the case of several variables, so let $\Vcstnot X$ be
an $n+1$-ary VCST. Given $\Vect X\in\COHT^n$ consider the unary VCST
$\Vcstnot X_{\Vect X}$ defined as follows:
$\Vcstca{\Vcstnot X_{\Vect X}}=\Vcstca{\Vcstnot X}_{\Tcohca{\Vect X}}$
% (see Section~\ref{sec:VCS-parameters})
and
$\Vcstot{(\Vcstnot X_{\Vect X})}(X)=\Vcstot{\Vcstnot X}(\Vect X,X)$
(the strength can be defined in a similar way though this is not
needed actually because the proof of
Proposition~\ref{prop:VCST-mu-init} does not involve the
strength). Then by Proposition~\ref{prop:VCST-mu-init} applied to
$\Vcstnot X_{\Vect X}$ and Lemma~\ref{lemma:strfun-lfp-general} we
have an $n$-ary strong functor $\Phi=(\Strfun\Phi,\Strnat\Phi)$ on $\COHT$
such that $\Strfun\Phi(\Vect X)=\Funlfp{(\Vcstnot X_{\Vect X})}$ and
whose action on morphisms and strength are uniquely characterized by
  \begin{itemize}
  \item
    $\Strfun{\Vcsnot X}(\Vect f,\Strfun\Phi(\Vect
    f))=\Strfun\Phi(\Vect f)$ for all
    $\Vect f\in\COHT^n(\Vect X,\Vect{X'})$
  \item and
    $\Strfun{\Vcsnot X}(\Tens{\Excl Y}{\Vect
      X},\Strnat\Phi_{Y,\Vect X})\Compl\Strnat{\Vcsnot
      X}_{Y,(\Vect X,\Strfun{\Funlfp{\Vcsnot F}}(\Vect
      X))}=\Strnat\Phi_{Y,\Vect X}$.
  \end{itemize}
  By Proposition~\ref{prop:coh-fp-init-final}, the first equation
  implies that
  $\Strfun\Phi(\Vect f)={\Strfun{\Vcsfp{\Vcsnot X}}}(\Vect f)$
  (remember that actually $\Vect f\in\COH^n(\Vect X,\Vect{X'})$ and
  that $\Vcsfp{\Vcsnot X}$ is an $n$-ary VCS characterized by that
  proposition) and the second equation shows that
  $\Strnat\Phi_{Y,\Vect X}=\Strnat{\Vcsfp{\Vcsnot X}}_{\Tcohca
    Y,\Tcohca{\Vect X}}$. This proves that
  ${\Strfun{\Vcsfp{\Vcsnot X}}}(\Vect f)\in\COHT(\Funlfp{(\Vcstnot
    X_{\Vect X})},\Funlfp{(\Vcstnot X_{\Vect Y})})$ and that
  $\Strnat{\Vcsfp{\Vcsnot X}}_{\Tcohca Y,\Tcohca{\Vect
      X}}\in\COHT(\Tens{Y}{\Funlfp{(\Vcstnot X_{\Vect
        X})}},\Funlfp{(\Vcstnot X_{\Tens Y{\Vect X}})})$. Therefore we
  have defined a VCST $\Funlfp{\Vcstnot X}$ whose carrier
  $\Vcstca{\Funlfp{\Vcstnot X}}$ is the VCS
  $\Vcsfp{\Vcstca{\Vcstnot X}}$ and whose totality
  $\Vcstot{(\Funlfp{\Vcstnot X})}$ is such that
  $(\Vcsfp{\Vcstca{\Vcstnot X}}(\Tcohca{\Vect
    X}),\Vcstot{(\Funlfp{\Vcstnot X})}(\Vect X))=\Funlfp{(\Vcstnot
    X_{\Vect X})}$ for all $\Vect X\in\COHT^n$. We can summarize our
  constructions as follows.
\begin{thm}\label{th:lfp-vcst-functorial}
  Let $\Vcstnot X$ be an $n+1$-ary VCST. There is a unique VCST
  $\Vcstmu{\Vcstnot X}$ whose carrier is $\Vcsfp{\Vcstca{\Vcstnot X}}$
  and whose totality is such that
  $\Vcstnot X(\Vect X,\Vcstmu{\Vcstnot X}(\Vect X)) =\Vcstmuu{\Vcstnot
    X}(\Vect X)$ and $(\Vcstmu{\Vcstnot X}(\Vect X),\Id)$ is initial
  in the category $\ALGFUN{\COHLINT}{\Vcstnot X_{\Vect X}}$.
\end{thm}
Moreover, we have provided a ``concrete'' way for defining this
operation (which involves an ordinal iteration).

Now we can define ``greatest fixed points'' by De Morgan duality. So let
$\Vcstnot X$ be an $n+1$-ary VCST. Given an $n$-tuple of coherence
spaces with totality $\Vect X$, we set
\(
  \Vcstnu{\Vcstnot X}(\Vect X)=\Orthp{\Vcstmu{(\Ddvcst{\Vcstnot X})}
  (\Orth{\Vect X})}
\).
More precisely, this means that the carrier of $\Vcstnu{\Vcstnot X}$
is the VCS $\Vcsfp{\Vcstca{{\Vcstnot X}}}$ (the \emph{very same} as
for $\Vcstmu{\Vcstnot X}$), and that
$\Vcstot{(\Vcstnu{\Vcsnot X})}(\Vect X)\in\Tot{\Vcsfp{\Vcstca{\Vcstnot
      X}}(\Tcohca{\Vect X})}$ is given by
\(
  \Vcstot{(\Vcstnu{\Vcsnot X})}(\Vect X)
  =\Orthp{\Vcstotp{\Vcstmu{(\Ddvcstnp{\Vcstnot X})}}(\Orth{\Vect X})}
\)
which indeed makes sense because
$\Vcstot{\Vcstmu{(\Ddvcstnp{\Vcstnot X})}}(\Orth{\Vect X})
\in\Tot{\Vcsfp{\Vcstca{\Ddvcstnp{\Vcstnot X}}}(\Tcohca{\Orth{\Vect
      X}})}$
and
$\Vcsfp{\Vcstca{\Ddvcstnp{\Vcstnot X}}}(\Tcohca{\Orth{\Vect
    X}})=\Orthp{\Vcsfp{\Vcstca{\Vcstnot X}}(\Tcohca{\Vect X})}$
by definition of the De Morgan dual of a VCS.

More concretely, this means that
$\Vcstotp{\Vcstnu{\Vcsnot X}}(\Vect X)=(\Vcsfp{\Vcstca{\Vcstnot
    X}}(\Tcohca{\Vect X}),\cV)$
where $\cV$ is the greatest totality candidate of
$\Vcsfp{\Vcstca{\Vcstnot X}}(\Tcohca{\Vect X})$ such that $F(\cV)=\cV$ where
\(
  F(\cT)=\Vcstot{\Vcstnot X}(\Vect X,(\Vcsfp{\Vcstca{\Vcstnot
  X}}(\Tcohca{\Vect X}),\cT))
\).
In other words,
$\cV=\bigcap_{\alpha}F^\alpha(\Cl{\Vcsfp{\Vcstca{\Vcstnot
      X}}(\Tcohca{\Vect X})})$.

\section{$\MULL$ and its interpretation}\label{sec:MULL}

We assume to be given an infinite set of propositional variables
$\LLvars$ (ranged over by Greek letters $\zeta,\xi\dots$). We
introduce a language of propositional $\LL$ formulas with least and
greatest fixed points.
\begin{align*}
  A,B,\dots &\Bnfeq \One \Bnfor \Fbot \Bnfor \Tens AB \Bnfor \Par AB\\
  &\quad \Bnfor \Zero \Bnfor \Top \Bnfor \Plus AB \Bnfor \With AB\\
  &\quad \Bnfor \Excl A \Bnfor \Int A\\
  &\quad\Bnfor\zeta \Bnfor \Lfpll\zeta A \Bnfor \Gfpll\zeta A\,.
\end{align*}
The notion of closed types is defined as usual, the two last
constructions being the only binders.

\begin{rem}
  In contrast with second-order linear logic or dependent type systems
  where open formulas or types play a crucial role ---~and it is
  necessary there to provide an interpretation of proofs or programs
  having non-closed types~---, in the case of fixed points, all formulas
  appearing in sequents and other syntactical objects allowing to give
  types to programs will be closed. In our setting, open types appear
  only locally, for allowing the expression of (least and greatest)
  fixed points. This is made possible by the fact that the deduction
  rules we consider for these operations preserve closeness of
  formulas in both directions (upwards and downwards) unlike the rule
  for quantifiers (think of $\forall$-intro).
\end{rem}

We can define two basic operations on formulas.
\begin{itemize}
\item \emph{Substitution}: $\Subst AB\zeta$, taking care of not
  binding free variables (uses $\alpha$-conversion).
\item \emph{Negation} or \emph{dualization}: defined by induction on formulas
  $\Orth\One=\Fbot$, $\Orth\Fbot=\One$,
  $\Orthp{\Par AB}=\Tens{\Orth A}{\Orth B}$,
  $\Orthp{\Tens AB}=\Par{\Orth A}{\Orth B}$, $\Orth\Zero=\Top$,
  $\Orth\Top=\Zero$, $\Orthp{\With AB}=\Plus{\Orth A}{\Orth B}$,
  $\Orthp{\Plus AB}=\With{\Orth A}{\Orth B}$,
  $\Orthp{\Excl A}=\Int{\Orth A}$, $\Orthp{\Int A}=\Excl{\Orth A}$,
  $\Orth\zeta=\zeta$, $\Orthp{\Lfpll\zeta A}=\Gfpll\zeta{\Orth A}$ and
  $\Orthp{\Gfpll\zeta A}=\Lfpll\zeta{\Orth A}$.
\end{itemize}

\begin{rem}
  The only subtle point of this definition is negation of
  propositional variables: $\Orth\zeta=\zeta$. The purpose of this is
  that we have $\Orthp{\Subst BA\zeta}=\Subst{\Orth B}{\Orth A}\zeta$
  as easily proven by induction on $B$. If we consider $B$ as a
  compound connective with placeholders labeled by variables then
  $\Orth B$ is its De~Morgan dual. This is also a very natural way of
  preventing the introduction of fixed points wrt.~variables with
  negative occurrences. For instance
  $D=\Lfpll\zeta{(\With\One{(\Limpl{\Excl\zeta}\zeta}))}$ \emph{is
    not} a formula of $\MULL$. Indeed $D$ should be written
  $\Lfpll\zeta{(\With\One{(\Par{\Int{(\Orth\zeta)}}{\zeta})})}$ which
  is not a formula of $\MULL$ (if we do not remove the linear
  negation) or has not the intended meaning if we apply the equation
  $\Orth\zeta=\zeta$. The rejection of such general recursive types is
  coherent with the fact that $D$ allows, for instance, to type all
  pure lambda-terms and hence also non-normalizing ones.
\end{rem}

% We do not record the standard rules of the $\LL$ sequent calculus, here
% are the new rules, essentially borrowed from~\cite{Baelde12} (apart
% for the $\Ngfp$ rule which is slightly generalized by introducing the
% context $\Int\Gamma$) to which we refer for all of its syntactic
% properties.
We give now the deduction rules, in a standard unilateral Linear Logic
sequent calculus as in~\cite{Baelde12}.

The identity fragment:
{\footnotesize
\begin{align*}
  \infer[\Naxiom]{\Seq{\Orth A,A}}{}
\qquad
  \infer[\Ncut]{\Seq{\Gamma,\Delta}}{\Seq{\Gamma,A} & \Seq{\Orth A,\Delta}}
\end{align*}}

The multiplicative fragment:
{\footnotesize
\begin{align*}
  \infer[\None]{\Seq\One}{}
\qquad
  \infer[\Ntens]{\Seq{\Gamma,\Delta,\Tens AB}}{\Seq{\Gamma,A} & \Seq{\Delta,B}}
\qquad
  \infer[\Nbot]{\Seq{\Gamma,\Fbot}}{\Seq\Gamma}
\qquad
  \infer[\Npar]{\Seq{\Gamma,\Par AB}}{\Seq{\Gamma,A,B}}
\end{align*}}

The additive fragment:
{\footnotesize
\begin{align*}
  \infer[\Ntop]{\Seq{\Gamma,\Top}}{}
\qquad
  \infer[\Nplusl]{\Seq{\Gamma,\Plus AB}}{\Seq{\Gamma,A}}
\qquad
  \infer[\Nplusr]{\Seq{\Gamma,\Plus AB}}{\Seq{\Gamma,B}}
\qquad
  \infer[\Nwith]{\Seq{\Gamma,\With AB}}{\Seq{\Gamma,A} & \Seq{\Gamma,B}}
\end{align*}}

The exponential fragment:
{\footnotesize
\begin{align*}
  \infer[\Nweak]{\Seq{\Gamma,\Int A}}{\Seq\Gamma}
\qquad
  \infer[\Ncontr]{\Seq{\Gamma,\Int A}}{\Seq{\Gamma,\Int A,\Int A}}
\qquad
  \infer[\Nder]{\Seq{\Gamma,\Int A}}{\Seq{\Gamma,A}}
\qquad
  \infer[\Nprom]{\Seq{\Int\Gamma,\Excl A}}{\Seq{\Int\Gamma,A}}
\end{align*}}

The fixed point fragment:
{\footnotesize
\begin{align*}
  &\infer[\Nlfp]{\Seq{\Gamma,\Lfpll\zeta F}}
  {\Seq{\Gamma,\Subst{F}{\Lfpll\zeta F}{\zeta}}}
\qquad
  \infer[\Ngfpfold]{\Seq{\Gamma,\Gfpll\zeta F}}
  {\Seq{\Gamma,\Subst{F}{\Gfpll\zeta F}{\zeta}}}\\
  &\hspace{1.2cm}\infer[\Ngfp]{\Seq{\Delta,\Int\Gamma,\Gfpll\zeta F}}
  {\Seq{\Delta,A} & \Seq{\Int\Gamma,\Orth A,\Subst{F}{A}{\zeta}}}
                    \end{align*}
}

By taking, in the last rule, $\Delta=\Orth A$ and proving the left premise by an axiom, we obtain the following derived rule
{\footnotesize
\begin{center}
  \AxiomC{$\Seq{\Int\Gamma,\Orth A,\Subst{F}{A}{\zeta}}$}
  \RightLabel{$\Ngfpbis$}
  \UnaryInfC{$\Seq{\Int\Gamma,\Orth A,\Gfpll\zeta F}$}
  \DisplayProof
\end{center}}

The only cut-elimination rule that we give is $\Nlfp/\Ngfp$, in
Section~\ref{sec:synt-functoriality}; for the other ones, see for
instance~\cite{Girard87} or any other presentation of the classical
$\LL$ Sequent Calculus. We refer to~\cite{Baelde12} for a proof that
this system admits cut-elimination\footnote{The system considered by
  Baelde is slightly different: no exponentials, no context in
  the~$\Ngfp$ rule. Though it seems quite clear that his proof can be
  adapted to the system presented here which has the same
  as Baelde's, in terms of provability. Observe however that,
  denotationally, our extension $\MULL$ is quite meaningful as
  explained in Section~\ref{sec:exclmu}}. Observe that a cut-free
proof has not the sub-formula property in general because of rule
$\Ngfp$. But Baelde's theorem makes sure that a proof of a sequent
\emph{which does not contain any $\nu$-formula} has a cut-free proof
with the sub-formula property. This is the main motivation for this
apparently weird formulation of the $\nu$-rule.

\subsection{Functoriality of formulas.}\label{sec:synt-functoriality}
Let $\zeta$ be a variable, $F$ a formula and $\tau$ be a proof of
$\Seq{\Int\Gamma,\Orth A,B}$, let $\Vect\xi=(\List\xi 1n)$ and
$\Vect C=(\List C1n)$ be closed formulas. Then one can define a proof
$\Substbis F{\tau/\zeta,\Vect C/\Vect\xi}$ of
$\Seq{\Int\Gamma,\Orth{\Substbis F{A/\zeta,\Vect
      C/\Vect\xi}},\Substbis F{B/\zeta,\Vect C/\Vect\xi}}$ by
induction on $F$, see~\cite{Baelde12}. As an example, assume that
$F=\Lfpll\xi G$. The proof $\Substbis F{\tau/\zeta,\Vect C/\Vect\xi}$ is
defined by (setting $G'=\Subst G{\Vect C}{\Vect\zeta}$)
\begin{center}{\footnotesize
    \AxiomC{$\Substbis {G}{\tau/\zeta,
        \Subst{(\Lfpll\xi {G'})}{B}{\zeta}/\xi,\Vect C/\Vect\xi}$}
  \noLine
  \UnaryInfC{$\Seq{\Int\Gamma,\Orth{\Substbis {G'}{A/\zeta,\Subst{(\Lfpll\xi {G'})}{B}{\zeta}/\xi}},\Substbis {G'}{B/\zeta,\Subst{(\Lfpll\xi {G'})}{B}{\zeta}/\xi}}$}
  \RightLabel{$\Nlfp$}
  \UnaryInfC{$\Seq{\Int\Gamma,\Orth{\Substbis {G'}{A/\zeta,\Subst{(\Lfpll\xi {G'})}{B}{\zeta}/\xi}},\Subst{(\Lfpll\xi {G'})}{B}{\zeta}}$}
  \RightLabel{$\Ngfpbis$}
  \UnaryInfC{$\Seq{\Int\Gamma,\Orth{\Subst{(\Lfpll\xi {G'})}{A}{\zeta}},\Subst{(\Lfpll\xi {G'})}{B}{\zeta}}$}
  \DisplayProof}
\end{center}
Let us also deal with the case $F=\Excl G$. Then, with the same
conventions as above, $\Substbis F{\tau/\zeta,\Vect C/\Vect\xi}$ is
defined as
\begin{center}{\footnotesize
  \AxiomC{$\Substbis{G}{\tau/\zeta,\Vect G/\Vect\xi}$}
  \noLine
  \UnaryInfC{$\Seq{\Int\Gamma,\Orth{{\Subst{G'}{A}{\zeta}}},{\Subst{G'}{B}{\zeta}}}$}
  \RightLabel{$\Nder$}
  \UnaryInfC{$\Seq{\Int\Gamma,\Orthp{\Excl{\Subst{G'}{A}{\zeta}}},{\Subst{G'}{B}{\zeta}}}$}
  \RightLabel{$\Nprom$}
  \UnaryInfC{$\Seq{\Int\Gamma,\Orthp{\Excl{\Subst{G'}{A}{\zeta}}},\Excl{\Subst{G'}{B}{\zeta}}}$}
  \DisplayProof}
\end{center}
Where we crucially use the fact that the context is made of
$\wn$-formulas. This feature is also essential in the case where
$F=\Tens{G_1}{G_2}$, for instance.

\subsection{Cut elimination}
The only two reductions that we will mention here are the
$\Nlfp/\Ngfp$ and $\Nlfp/\Ngfpfold$.  Consider first a proof $\theta$ of
shape
\begin{center}{\footnotesize
  \AxiomC{$\pi$}
  \noLine
  \UnaryInfC{$\Seq{\Lambda,\Subst F{\Lfpll\zeta F}\zeta}$}
  \RightLabel{$\Nlfp$}
  \UnaryInfC{$\Seq{\Lambda,\Lfpll\zeta F}$}
  \AxiomC{$\lambda$}
  \noLine
  \UnaryInfC{$\Seq{\Delta,\Orth A}$}
  \AxiomC{$\rho$}
  \noLine
  \UnaryInfC{$\Seq{\Int\Gamma,A,\Orthp{\Subst FA\zeta}}$}
  \RightLabel{$\Ngfp$}
  \BinaryInfC{$\Seq{\Delta,\Int\Gamma,\Orthp{\Lfpll\zeta F}}$}
  \BinaryInfC{$\Seq{\Lambda,\Delta,\Int\Gamma}$}
  \DisplayProof}
\end{center}
and let $\rho'$ be the proof
%\begin{center}
\raisebox{5pt}{{\footnotesize\AxiomC{}
  \noLine
  \UnaryInfC{$\Seq{A,\Orth A}$}
  \AxiomC{$\rho$}
  \noLine
  \UnaryInfC{$\Seq{\Int\Gamma,A,\Orthp{\Subst FA\zeta}}$}
  \BinaryInfC{$\Seq{\Int\Gamma,A,\Orthp{\Lfpll\zeta F}}$}
  \DisplayProof}}.
%\end{center}
Then $\theta$ reduces to
\begin{center}
  {\scriptsize
  \AxiomC{$\Subst F{\rho'}\zeta$}
  \noLine
  \UnaryInfC{$\Seq{\Int\Gamma,\Subst FA\zeta},\Orth{\Subst F{\Lfpll\zeta F}\zeta}$}
  \AxiomC{$\pi$}
  \noLine
  \UnaryInfC{$\Seq{\Lambda,\Subst F{\Lfpll\zeta F}\zeta}$}
  \BinaryInfC{$\Seq{\Lambda,\Int\Gamma},\Subst FA\zeta$}
  \AxiomC{$\rho$}
  \noLine
  \UnaryInfC{$\Seq{\Int\Gamma,A,\Orthp{\Subst FA\zeta}}$}
  \BinaryInfC{$\Seq{\Lambda,\Int\Gamma,\Int\Gamma,A}$}
  \RightLabel{$\Ncontr$}
  \doubleLine
  \UnaryInfC{$\Seq{\Lambda,\Int\Gamma,A}$}
  \AxiomC{$\lambda$}
  \noLine
  \UnaryInfC{$\Seq{\Delta,\Orth A}$}
  \BinaryInfC{$\Seq{\Delta,\Lambda,\Int\Gamma}$}
  \DisplayProof}
\end{center}
This reduction rule uses the functoriality of formulas as well as the
$\wn$-contexts in the $\Ngfp$ rule.

Next, a proof of shape
\begin{center}
  {\footnotesize
    \AxiomC{$\lambda$}
    \noLine
    \UnaryInfC{$\Seq{\Gamma,\Subst{A}{\Lfpll\zeta A}{\zeta}}$}
    \RightLabel{$\Nlfp$}
    \UnaryInfC{$\Seq{\Gamma,\Lfpll\zeta A}$}
    \AxiomC{$\rho$}
    \noLine
    \UnaryInfC{$\Seq{\Delta,\Orthp{\Subst{A}{\Lfpll\zeta A}{\zeta}}}$}
    \RightLabel{$\Ngfpfold$}
    \UnaryInfC{$\Seq{\Delta,\Orthp{\Lfpll\zeta A}}$}
    \BinaryInfC{$\Seq{\Gamma,\Delta}$}
    \DisplayProof
  }
\end{center}
reduces to
\begin{center}
  {\footnotesize
    \AxiomC{$\lambda$}
    \noLine
    \UnaryInfC{$\Seq{\Gamma,\Subst{A}{\Lfpll\zeta A}{\zeta}}$}
    \AxiomC{$\rho$}
    \noLine
    \UnaryInfC{$\Seq{\Delta,\Orthp{\Subst{A}{\Lfpll\zeta A}{\zeta}}}$}
    \BinaryInfC{$\Seq{\Gamma,\Delta}$}
    \DisplayProof
  }
\end{center}

\begin{rem}
  In terms of \emph{provability}, the rule $\Ngfpfold$ is redundant
  since it can be derived as follows
  \begin{center}
    {\footnotesize
      \AxiomC{$\Seq{\Gamma,\Subst A{\Gfpll\zeta A}\zeta}$}
      \AxiomC{$\Subst A\pi\zeta$}
      \noLine
      \UnaryInfC{$\Seq{\Orthp{\Subst A{\Gfpll\zeta A}\zeta},
          \Subst A{\Subst A{\Gfpll\zeta A}\zeta}\zeta}$}
      \RightLabel{$\Ngfp$}
      \BinaryInfC{$\Seq{\Gamma,\Gfpll\zeta A}$}
      \DisplayProof
    }
  \end{center}
  where $\pi$ is the following proof
  \begin{center}
    {\footnotesize
      \AxiomC{}
      \UnaryInfC{$\Seq{\Orthp{\Subst A{\Gfpll\zeta A}\zeta},
          \Subst A{\Gfpll\zeta A}\zeta}$}
      \RightLabel{$\Nlfp$}
      \UnaryInfC{$\Seq{\Orthp{\Gfpll\zeta A},\Subst A{\Gfpll\zeta A}\zeta}$}
      \DisplayProof
    }
  \end{center}
  Though, in terms of \emph{algorithmic expressiveness}, the
  $\Ngfpfold$ rule is essential since it corresponds to pattern
  matching. For instance, with the type
  $\Tnat=\Lfpll\zeta{(\Plus\One\zeta)}$ of integers
  (see~\ref{sec:strict-int}), the $\Ngfpfold$ allows to define a
  function of type $\Limpl\Tnat\Tnat$ which computes the predecessor
  of an integer $n$ in a fixed number of reduction steps whereas the
  predecessor function defined using the above $\Ngfp$-based
  definition of $\Ngfpfold$ requires a number of steps proportional to
  $n$.
\end{rem}

\subsection{Interpreting formulas and proofs}
\label{sec:interp-form-proofs}
% SPELL CHK à reprendre ici

With any formula $F$ and any repetition-free sequence
$\Vect\zeta=(\List\zeta 1n)$ of variables containing all variables free
in $F$, one can associate an $n$-ary VCST $\Tsem F_{\Vect\zeta}$ by
induction on $F$, using straightforwardly the constructions of
Section~\ref{sec:VCST}. If $F$ is closed (again, this holds for any
formula occurring in a sequent) then $\Tsem F$ is simply an object of
$\COHT$. If $\Gamma=(\List C1n)$ is a sequence of closed formulas then
$\Tsem\Gamma=\Tsem{C_1}\IPar\cdots\IPar\Tsem{C_n}$.

\begin{lem}\label{lemma:interp-subst-form}
  Let $F$ be a formula and $\Vect\zeta=(\List\zeta 1n)$ be a
  repetition-free list of variables containing all free variables of
  $F$. Let $\List G1n$ be a list of formulas and let
  $\Vect\xi=(\List\xi 1k)$ be a repetition-free list of variables
  containing all free variables of $\List G1n$. Then
  \begin{align*}
    \Tsem{\Substbis F{G_1/\zeta_1,\dots,G_n/\zeta_n}}_{\Vect\xi}
    =\Tsem F_{\Vect\zeta}\Comp(\Psem{G_1}_{\Vect\xi},
    \dots,\Psem{G_n}_{\Vect\xi})\,.
  \end{align*}
\end{lem}
The proof is a straightforward induction on $F$.

Next, with any proof $\pi$ of a sequent $\Seq\Gamma$, we can associate
$\Psem\pi\in\Tcoht{\Tsem\Gamma}$. More precisely,
$\Psem\pi\in\Cl{\Tcohca{\Tsem\Gamma}}$ and it turns out that this
clique is total in $\Tsem\Gamma$. We describe now this
interpretation. For simplifying the notations, we drop the
``$\Tcohca\ $'' notation and write simply
``$\Psem\pi\in\Cl{{\Tsem\Gamma}}$'' instead of
``$\Psem\pi\in\Cl{\Tcohca{\Tsem\Gamma}}$''.

\subsubsection{Reminder: interpreting the rules of $\LL$}
If $\pi$ is
\begin{center}
  \footnotesize{
    \AxiomC{}
    \UnaryInfC{$\Seq{\Orth A,A}$}
    \DisplayProof
  }
\end{center}
then $\Psem\pi=\Id_{\Psem A}=\{(a,a)\St a\in\Web{\Psem A}\}$. If $\pi$ is 
\begin{center}
  \footnotesize{
    \AxiomC{$\lambda$}
    \noLine
    \UnaryInfC{$\Seq{\Gamma,A}$}
    \AxiomC{$\rho$}
    \noLine
    \UnaryInfC{$\Seq{\Orth A,\Delta}$}
    \BinaryInfC{$\Seq{\Gamma,\Delta}$}
    \DisplayProof
  }
\end{center}
then, considering that
$\Psem\lambda\in\COH(\Orth{\Tsem\Gamma},\Tsem A)$ and
$\Psem\rho\in\COH(\Psem A,\Psem\Delta)$ then
$\Psem\pi=\Psem\rho\Compl\Psem\lambda$ that is
$\Psem\pi=\{(\gamma,\delta\St\exists a\in\Web{\Tsem A\
  (\gamma,a)\in\Psem\lambda\text{ and }(a,\delta)\in\Psem\rho})\}$. If $\pi$ is
\begin{center}
  \footnotesize{
  \AxiomC{}
  \UnaryInfC{$\Seq\One$}
  \DisplayProof}
\end{center}
then $\Psem\pi=\{\Oneelem\}$. If $\pi$ is 
\begin{center}
  \footnotesize{
    \AxiomC{$\lambda$}
    \noLine
    \UnaryInfC{$\Seq{\Gamma,A}$}
    \AxiomC{$\rho$}
    \noLine
    \UnaryInfC{$\Seq{\Delta,B}$}
    \BinaryInfC{$\Seq{\Gamma,\Delta,\Tens AB}$}
    \DisplayProof
  }
\end{center}
then $\Psem\pi=\Tens{\Psem\lambda}{\Psem\rho}$ considering that
$\Psem\lambda\in\COH(\Orth{\Psem\Gamma},\Psem A)$ and
$\Psem\rho\in\COH(\Orth{\Psem\Delta},\Psem B)$, that is
$\Psem\pi=\{(\gamma,\delta,(a,b))\St(\gamma,a)\in\Psem\lambda\text{
  and }(\delta,b)\in\Psem\rho\}$. If $\pi$ is 
\begin{center}
  \footnotesize{
    \AxiomC{$\lambda$}
    \noLine
    \UnaryInfC{$\Seq{\Gamma}$}
    \UnaryInfC{$\Seq{\Gamma,\Fbot}$}
    \DisplayProof
  }
\end{center}
then $\Psem\pi=\{(\gamma,\Oneelem)\St\gamma\in\Psem\lambda\}$.  If
$\pi$ is
\begin{center}
  \footnotesize{
    \AxiomC{$\lambda$}
    \noLine
    \UnaryInfC{$\Seq{\Gamma,A,B}$}
    \UnaryInfC{$\Seq{\Gamma,\Par AB}$}
    \DisplayProof
  }
\end{center}
then $\Psem\pi=\Psem\lambda$ if we consider that
$\Psem\lambda\in\COH(\Orth{\Psem\Gamma},\Psem{\Par AB})$  that is
$\Psem\pi=\{(\gamma,(a,b))\St(\gamma,a,b)\in\Psem\lambda\}$. If $\pi$ is
\begin{center}
  \footnotesize{
    \AxiomC{}
    \UnaryInfC{$\Seq{\Gamma,\Top}$}
    \DisplayProof
  }
\end{center}
then $\Psem\pi=\emptyset$. If $\pi$ is 
\begin{center}
  \footnotesize{
    \AxiomC{$\lambda$}
    \noLine
    \UnaryInfC{$\Seq{\Gamma,A}$}
    \AxiomC{$\rho$}
    \noLine
    \UnaryInfC{$\Seq{\Gamma,B}$}
    \BinaryInfC{$\Seq{\Gamma,\With AB}$}
    \DisplayProof
  }
\end{center}
then $\Psem\pi=\Pair{\Psem\lambda}{\Psem\rho}$ considering that
$\Psem\lambda\in\COH(\Orth{\Psem\Gamma},A)$ and
$\Psem\rho\in\COH(\Orth{\Psem\Gamma},B)$, that is
$\Psem\pi=\{(\gamma,(1,a))\St(\gamma,a)\in\Psem\lambda\}
\cup\{(\gamma,(2,b))\St(\gamma,b)\in\Psem\rho\}$. If $\pi$ is 
\begin{center}
  \footnotesize{
    \AxiomC{$\lambda$}
    \noLine
    \UnaryInfC{$\Seq{\Gamma,A}$}
    \UnaryInfC{$\Seq{\Gamma,\Plus AB}$}
    \DisplayProof
  }
\end{center}
then $\Psem\pi=\Inj 1\Compl\Psem\lambda$ where
$\Inj 1\in\COH(\Tsem A,\Plus{\Tsem A}{\Tsem B})$, that is
$\Psem\pi=\{(\gamma,(1,a))\St(\gamma,a)\in\Psem\lambda\}$, considering
that $\Psem\lambda\in\COH(\Orth{\Psem\Gamma},\Psem A)$. If $\pi$ ends with a
right $\oplus$-rule, the interpretation is similar. If $\pi$ ends with 
\begin{center}
  \footnotesize{
    \AxiomC{$\lambda$}
    \noLine
    \UnaryInfC{$\Seq{\Gamma,A}$}
    \UnaryInfC{$\Seq{\Gamma,\Int A}$}
    \DisplayProof
  }
\end{center}
then $\Psem\pi=\Orth{\Der{\Orth{\Psem A}}}\Compl\Psem\lambda$, that is
$\Psem\pi=\{(\gamma,\{a\})\St(\gamma,a)\in\Psem\lambda\}$. If $\pi$ is 
\begin{center}
  \footnotesize{
    \AxiomC{$\lambda$}
    \noLine
    \UnaryInfC{$\Seq{\Gamma}$}
    \UnaryInfC{$\Seq{\Gamma,\Int A}$}
    \DisplayProof
  }
\end{center}
then $\Psem\pi=\Orth{\Weak{\Orth{\Psem A}}}\Compl\Psem\lambda$, that is
$\Psem\pi=\{(\gamma,\emptyset)\St(\gamma,a)\in\Psem\lambda\}$. If $\pi$ is 
\begin{center}
  \footnotesize{
    \AxiomC{$\lambda$}
    \noLine
    \UnaryInfC{$\Seq{\Gamma,\Int A,\Int A}$}
    \UnaryInfC{$\Seq{\Gamma,\Int A}$}
    \DisplayProof
  }
\end{center}
then $\Psem\pi=\Orth{\Contr{\Orth{\Psem A}}}\Compl\Psem\lambda$, that
is
$\Psem\pi=\{(\gamma,x_1\cup x_2)\St(\gamma,x_i)\in\Psem\lambda\text{
  for }i=1,2\text{ and }x_1\cup x_2\in\Web{\Int{\Psem{A}}}\}$,
considering that
$\Psem\lambda\in\COH(\Orth{\Psem\Gamma},\Par{\Int{\Psem A}}{\Int{\Psem
    A}})$. If $\pi$ is
\begin{center}
  \footnotesize{
    \AxiomC{$\lambda$}
    \noLine
    \UnaryInfC{$\Seq{\Int\Gamma,A}$}
    \UnaryInfC{$\Seq{\Int\Gamma,\Excl A}$}
    \DisplayProof
  }
\end{center}
then, considering that
$\Psem\lambda\in\COH(\Excl{E_1}\ITens\cdots\ITens\Excl{E_k},\Psem A)$
where $\Gamma=(\List C1k)$ and $E_j=\Orth{\Psem{C_j}}$ for
$j=1,\dots,k$, we set $\Psem\pi=\Prom{\Psem\lambda}$ that is
$\Psem\pi=\{(\gamma^1\cup\cdots\cup\gamma^p,\{\List
a1p\})\St(\gamma^l,a_l)\in\Psem\lambda\text{ for }l=1,\dots,p\text{
  and }\gamma^1_j\cup\cdots\cup\gamma^p_j\in\Cl{E_j}\text{ for
}j=1,\dots,k\}$.

\subsubsection{Interpreting the fixed point rules}
\label{sec:inter-fixed-points}
Assume that $\pi$ is
\begin{center}
  {\footnotesize
    \AxiomC{$\lambda$}
    \noLine
    \UnaryInfC{$\Seq{\Gamma,\Subst A{\Lfpll\zeta A}\zeta}$}
    \RightLabel{$\Nlfp$}
    \UnaryInfC{$\Seq{\Gamma,\Lfpll\zeta A}$}
    \DisplayProof
  }
\end{center}
We have
$\Psem\lambda\in\COHT(\Orth{\Tsem\Gamma},\Tsem A_\zeta(\Vcstmuu{\Tsem
  A_\zeta}))$ by inductive hypothesis and
Lemma~\ref{lemma:interp-subst-form}, that is
$\Psem\lambda\in\COHT(\Orth{\Tsem\Gamma},\Vcstmuu{\Tsem A_\zeta})$, so
that we simply set $\Psem\pi=\Psem\lambda$. The interpretation is
similar if $\pi$ ends with $\Ngfpfold$. Last, assume that $\pi$ is
% \begin{center}
%   {\footnotesize
%     \AxiomC{$\lambda$}
%     \noLine
%     \UnaryInfC{$\Seq{\Int\Gamma,\Orthp{\Subst AB\zeta},B}$}
%     \RightLabel{$\Ngfpbis$}
%     \UnaryInfC{$\Seq{\Int\Gamma,\Orthp{\Lfpll\zeta A},B}$}
%     \DisplayProof
%   }
% \end{center}
\begin{center}
  {\footnotesize
    \AxiomC{$\lambda$}
    \noLine
    \UnaryInfC{$\Seq{\Int\Gamma,\Orth B,\Subst AB\zeta}$}
    \RightLabel{$\Ngfpbis$}
    \UnaryInfC{$\Seq{\Int\Gamma,\Orth B,\Gfpll\zeta A}$}
    \DisplayProof
  }
\end{center}
Let $(\List C1k)=\Gamma$, let $X_i=\Orth{\Tsem{C_i}}$ for
$i=1,\dots,k$, and let $Y=X_1\IWith\cdots\IWith X_k$ so that setting $P=\Excl{X_1}\ITens\cdots\ITens\Excl{X_k}$ we have a
(generalized) Seely iso $\Seely_{\Vect X}\in\COHT(P,\Excl Y)$. Then we have
$\Psem\lambda\in\COHT(\Tens{P}{\Tsem B}, \Tsem A_\zeta(\Tsem B))$ by
inductive hypothesis, using also
Lemma~\ref{lemma:interp-subst-form}. Then we consider the object
$(\Tens{P}{\Tsem B},f)$ of
$\ALGFUN\COHT{\Tsem A_\zeta}$ where $f$ is defined as the following
composition of morphisms
\begin{equation*}
    \begin{tikzpicture}[->, >=stealth]
      \node (1) {$\Tens{P}{\Tsem B}$};
      \node (2) [below of=1, node distance=1cm]
      {$P\ITens\Tens{P}{\Tsem B}$};
      \node (3) [below of=2, node distance=1cm]
      {$\Excl{Y}\ITens\Tsem A_\zeta(\Tsem B)$};
      \node (4) [below of=3, node distance=1.2cm]
      {$\Tsem A_\zeta(\Excl{Y}\ITens\Tsem B)$};
      \node (5) [below of=4, node distance=1.2cm]
      {$\Tsem A_\zeta(P\ITens\Tsem B)$};
      \tikzstyle{every node}=[midway,auto,font=\scriptsize]
      \draw (1) -- node {$\Contr{}\ITens\Tsem B$} (2);
      \draw (2) -- node {$\Seely_{\Vect X}\ITens\Psem\lambda$} (3);
      \draw (3) -- node {$\Strnat{\Tsem A_\zeta}_{Y,\Tsem B}$} (4);
      \draw (4) -- node
      {$\Tsem A_\zeta(\Funinv{\Seely_{\Vect X}}\ITens\Tsem B)$} (5);
     \end{tikzpicture}   
\end{equation*}
Then $\Psem \pi$ is the unique morphism of
$\ALGFUN\COHT{\Tsem A_\zeta}$ from the coalgebra
$(\Tens{P}{\Tsem B},f)$ to the final coalgebra
$(\Tsem{\Gfpll\zeta A},\Id)$. So $\Psem\pi=\bigcup_{n=0}^\infty g_n$
where the morphisms
$g_n\in\COH(\Tcohca P\ITens\Tcohca{\Tsem B},\Vcsfp{\Vcstca{\Tsem
    A_\zeta}})$ are defined by: $g_0=\emptyset$ and $g_{n+1}$ is
\begin{equation*}
  \begin{tikzpicture}[->, >=stealth]
    \node (b) {$\Tens{\Tcohca P}{\Tcohca{\Tsem B}}$};
    \node (a) [below of=b, node distance=1cm]
    {$\Tcohca P\ITens\Tens{\Tcohca P}{\Tcohca {\Tsem B}}$};
    \node (1) [below of=a, node distance=1cm]
      {$\Tcohca{\Excl Y}\ITens\Vcstca{\Tsem A_\zeta}(\Tcohca{\Tsem B})$};
    \node (2) [below of=1, node distance=1.2cm]
      {$\Vcstca{\Tsem A_\zeta}(\Tens{\Tcohca{\Excl Y}}{\Tcohca{\Tsem B}})$};
    \node (3) [below of=2, node distance=1.2cm]
      {$\Vcstca{\Tsem A_\zeta}(\Tens{\Tcohca P}{\Tcohca{\Tsem B}})$};
    \node (4) [below of=3, node distance=1.2cm]
    {$\Tsem A_\zeta(\Vcsfp{\Vcstca{\Tsem A_\zeta}})
      =\Vcsfp{\Vcstca{\Tsem A_\zeta}}$};
    \tikzstyle{every node}=[midway,auto,font=\scriptsize]
    \draw (b) -- node {$\Contr{}\ITens\Tcohca{\Tsem B}$} (a);
    \draw (a) -- node {$\Seely_{\Vect X}\ITens\Psem\lambda$} (1);
    \draw (1) -- node {$\Strnat{\Tsem A_\zeta}_{Y,\Tsem B}$} (2);
    \draw (2) -- node {$\Vcstca{\Tsem A_\zeta}(\Funinv{\Seely_{\Vect X}}\ITens\Tcohca{\Tsem B})$} (3);
    \draw (3) -- node {$\Vcstca{\Tsem A_\zeta}(g_n)$} (4);
  \end{tikzpicture}   
\end{equation*}
Indeed $(g_n)_{n\in\Nat}$ is a monotonic sequence of cliques and so
$g=\cup_{n\in\Nat}g_n\in\COH(\Tens{P}{\Tsem B},\Vcsfp{\Vcstca{\Tsem
    A_\zeta}})$. To prove our contention that $\Psem\pi=g$, it
suffices to prove that
$g\in\ALGFUN\COH{\Vcstca{\Tsem A_\zeta}}((\Tens{\Tcohca
  P}{\Tcohca{\Tsem B}},f), (\Vcsfp{\Vcstca{\Tsem A_\zeta}},\Id))$,
that is $g={\Vcstca{\Tsem A_\zeta}}(g)\Compl f$ which follows readily
from the observation that
${\Vcstca{\Tsem A_\zeta}}(g_n)\Compl f=g_{n+1}$ by continuity of
$\Vcstca{\Tsem A_\zeta}$ (remember that the functor
$\Vcstca{\Tsem A_\zeta}$ is continuous in the sense of
Definition~\ref{def:coh-cont-obj}).

\begin{lem}\label{lemma:funct-formula-sem}
  Let $\pi$ be a proof of $\Seq{\Orthp{\Excl\Gamma},\Orth A,B}$ and
  consider $\Psem\pi$ as an element of
  $\Kcomod\COHT{\Excl{\Tsem\Gamma}}(\Tsem A,\Tsem B)$ (see the
  definition of this category of free comodules in
  Section~\ref{sec:EM-Kl-category}; we use the canonical structure of
  $\oc$-coalgebra of $\Excl{\Tsem\Gamma}$). Let $F$ be a formula and
  $\Vect C=(\List C1k)$ be closed formulas and $\zeta,\List\zeta 1k$
  be pairwise distinct variables. Then
  $\Psem{\Substbis F{\pi/\zeta,\Vect C/\Vect\zeta}}=\Tsem
  F_{\zeta,\Vect\zeta}(\Psem\pi,\Tsem{\Vect C}/\Vect\zeta)$ (see
  Section~\ref{sec:gen-strong-functors} for the action of a strong
  functor on a category of free comodules).
\end{lem}
This means that our definition of the functorial action of formulas on
proofs in Section~\ref{sec:synt-functoriality} is compatible with the
definition of the category of free comodules and of the extension of a
strong functor to this category explained in
Section~\ref{sec:gen-strong-functors}. The proof is a simple induction
on $F$.

As usual a main feature of this interpretation is the following.
\begin{thm}\label{th:muLL-sem-invariant}
  If $\pi$ reduces to $\pi'$, then $\Psem\pi=\Psem{\pi'}$.
\end{thm}
The proof is a lengthy and boring verification which uses crucially
Lemma~\ref{lemma:funct-formula-sem} in the most interesting case,
which is the $\Nlfp/\Ngfp$ cut reduction.
%\todot{Should we add something here or consider this theorm as ``trivial''?}

\section{Examples}\label{sec:examples}
\subsection{Some data and non-data types}

\paragraph{Strict integers.}\label{sec:strict-int}
The type of strict integers is $\Tnat=\Lfpll\zeta{(\Plus\One\zeta)}$.
The following deduction rules are derivable in $\MULL$:
\begin{center}{\footnotesize
    \AxiomC{}
    \RightLabel{$\Nzero$}
    \UnaryInfC{$\Seq{\Tnat}$}
    \DisplayProof
    \quad
    \AxiomC{$\Seq{\Delta,\Tnat}$}
    \RightLabel{$\Nsucc$}
    \UnaryInfC{$\Seq{\Delta,\Tnat}$}
    \DisplayProof
    \quad
    \AxiomC{$\Seq{\Int\Gamma,A}$}
    \AxiomC{$\Seq{\Int\Gamma,\Orth{C},{C}}$}
    \RightLabel{$\Nintit$}
    \BinaryInfC{$\Seq{\Int\Gamma,\Orth\Tnat,C}$}
    \DisplayProof
  }
\end{center}
The first rule corresponds to the constant $0$, the second one to the
successor function and the third one should be understood as an iteration
principle: the first premise is the base case and the second one is
the ``inductive step''.

More precisely these proofs are defined as follows.
\begin{center}
  {\footnotesize
    \AxiomC{}
    \UnaryInfC{$\Seq\One$}
    \UnaryInfC{$\Seq{\Plus\One\Tnat}$}
    \RightLabel{$\Nlfp$}
    \UnaryInfC{$\Seq{\Tnat}$}
    \DisplayProof
    \quad
    \AxiomC{$\Seq{\Delta,\Tnat}$}
    \UnaryInfC{$\Seq{\Delta,\Plus\One\Tnat}$}
    \RightLabel{$\Nlfp$}
    \UnaryInfC{$\Seq{\Delta,\Tnat}$}
    \DisplayProof
    \quad
    \AxiomC{$\Seq{\Int\Gamma,C}$}
    \UnaryInfC{$\Seq{\Int\Gamma,\Fbot,C}$}
    \AxiomC{$\Seq{\Int\Gamma,\Orth C,C}$}
    \BinaryInfC{$\Seq{\Int\Gamma,\With\Fbot{\Orth C},C}$}
    \RightLabel{$\Ngfpbis$}
    \UnaryInfC{$\Seq{\Int\Gamma,\Orth\Tnat,C}$}
    \DisplayProof
  }
\end{center}

The coherence space $\Tcohca{\Tsem\Tnat}$ is the least fixed point of the VCS
$E\mapsto\Plus\One E$ so that, up to trivial iso,
$\Tcohca{\Tsem\Tnat}$ is the coherence space $\Tcohca\Snat=(\Nat,=)$. For
computing $\Tcoht{\Tsem\Tnat}$, the method explained in
Section~\ref{sec:fix-VCST} boils down to computing the least
fixed point of the map
$\Theta:\Tot{\Tcohca\Snat}\to\Tot{\Tcohca\Snat}$ such that
$\Theta(T)=\{\{0\}\}\cup\{x+1\St x\in T\}$ (where
$x+1=\{n+1\St n\in x\}$). We have
$\Theta^n(\emptyset)=\{\{0\},\{1\},\dots,\{n-1\}\}$ so
$T=\bigcup_{n\in\Nat}\Theta^n(\emptyset)=\{\{n\}\St n\in\Nat\}$. Hence
$\Orth T=\{\Nat\}$ and it follows that $T=\Biorth T$, hence $T$ is the
least fixed point of $\Theta$ in $\Tot{\Tcohca\Snat}$. Finally
$\Tcoht{\Tsem\Tnat}=\{\{n\}\St n\in\Nat\}$.

We specialize the definition of Section~\ref{sec:inter-fixed-points}
taking (with the notations of that section) $A=\With\Fbot\zeta$ and
$B=\Orth C$. If $\pi$ is
\begin{center}
  {\footnotesize
    \AxiomC{}
    \RightLabel{$\Nzero$}
    \UnaryInfC{$\Seq\Tnat$}
    \DisplayProof
  }
\end{center}
then $\Psem\pi=\{0\}$. If $\pi$ is
\begin{center}
  {\footnotesize
    \AxiomC{$\lambda$}
    \noLine
    \UnaryInfC{$\Seq{\Delta,\Tnat}$}
    \RightLabel{$\Nsucc$}
    \UnaryInfC{$\Seq{\Delta,\Tnat}$}
    \DisplayProof
  }
\end{center}
then $\pi=\{(\delta,n+1)\St(\delta,n)\in\Psem\lambda\}$. For the
$\Nintit$ rule, assume that $\pi$ is 
\begin{center}
  {\footnotesize
    \AxiomC{$\lambda$}
    \noLine
    \UnaryInfC{$\Seq{\Int\Gamma,C}$}
    \AxiomC{$\rho$}
    \noLine
    \UnaryInfC{$\Seq{\Int\Gamma,\Orth C,C}$}
    \RightLabel{$\Nintit$}
    \BinaryInfC{$\Seq{\Int\Gamma,\Orth\Tnat,C}$}
    \DisplayProof
  }
\end{center}
One has first to compute $\Strnat{\Tsem{\With\Fbot\zeta}_\zeta}$ in
this special case applying the general definition of formula
interpretation as explained in
Section~\ref{sec:interp-form-proofs}. Then given coherence spaces $E$
and $F$, the morphism
$(\Strnat{\Tsem{\With\Fbot\zeta}_\zeta})_{F,E}\in\COH(\Tens{\Excl
  F}{(\With{\Fbot}{E})},\With\Fbot{(\Tens{\Excl F}{E})})$ is
$\{((\emptyset,(1,\Onelem)),(1,\Onelem))\}\cup\{((y_0,(2,a)),(2,(y_0,a)))\St
y_0\in\Web{\Excl F}\text{ and }a\in\Web E\}$. Next applying the recipe
at the end of Section~\ref{sec:inter-fixed-points} we obtain that
$\Psem{\pi}$ is the least
$g\in\COH(\Excl{\Orth{\Tsem\Gamma}}\ITens\Snat,\Psem C)$ such that
\begin{align*}
  g=&\{(\gamma,0,c)\St (\gamma,c)\in\Psem\lambda\}\\
  &\cup
  \{(\gamma^1\cup\gamma^2,n+1,c)\St\exists c'\in\Web{\Tsem C}
    \ (\gamma^1,n,c')\in g\text{, }\\
  &\hspace{4cm}(\gamma^2,c',c)\in\Psem\rho
    \text{ and }\gamma^1\cup\gamma^2\in\Web{\Excl{\Orth{\Tsem\Gamma}}}\}\,.
\end{align*}

Setting $(\List A1k)=\Orth\Gamma$ we can consider $\Psem\lambda$ as a
stable function
$h:\Cl{\Tsem{A_1}}\times\cdots\times\Cl{\Tsem{A_1}}\to\Cl{\Tsem C}$
and $\Psem\rho$ as a stable function
$k:\Cl{\Tsem{A_1}}\times\cdots\times\Cl{\Tsem{A_1}}\times\Cl{\Tsem
  C}\to\Cl{\Tsem C}$ which is \emph{linear in its last argument}. Then
$g$ can be seen as the stable function
$g:\Cl{\Tsem{A_1}}\times\cdots\times\Cl{\Tsem{A_1}}\times\Cl{\Tsem
  \Snat}\to\Cl{\Tsem C}$, linear in its last argument, such that
$g(\Vect x,\{0\})=h(\Vect x)$ and
$g(\Vect x,\{n+1\})=k(\Vect x,g(\Vect x,\{n\}))$.

\begin{rem}
  If $\pi$ is a proof of $\Seq{\Orth\Tnat,\Tnat}$ then $\Psem\pi$ is a
  clique of $\Limpl\Snat\Snat$ such that
  $\forall n\,\exists p\ (n,p)\in\Psem\pi$, that is, a total function
  $\Nat\to\Nat$. Using the derived rules above, we shall see in
  Section~\ref{sec:rep-syst-T} that a version of Gödel's System T can
  be represented in $\MULL$, which shows that all functions
  $\Nat\to\Nat$ representable in System T appear as interpretations of
  such proofs $\pi$.
\end{rem}

% We conjecture the converse.
% This
% representation uses crucially the existence of a proof of
% $\Limpl\Tnat{\Excl\Tnat}$ which means essentially that integers are
% freely duplicable and erasable ($\Tsem\Tnat$ is a $\oc$-coalgebra).

\paragraph{Lazy integers.} One can define a type of lazy integers as
$\Lnat=\Lfpll\zeta{(\Plus\One{\Excl\zeta})}$. The following deduction
rules are derivable in $\MULL$:
\begin{center}{\footnotesize
  \AxiomC{}
  \UnaryInfC{$\Seq{\Lnat}$}
  \DisplayProof
  \quad
  \AxiomC{$\Seq{\Delta,\Excl\Lnat}$}
  \UnaryInfC{$\Seq{\Delta,\Lnat}$}
  \DisplayProof
  \quad
  \AxiomC{$\Seq{\Int\Gamma,A}$}
  \AxiomC{$\Seq{\Int\Gamma,\Limpl{\Excl A}{A}}$}
  \BinaryInfC{$\Seq{\Int\Gamma,\Limpl\Lnat A}$}
  \DisplayProof
  }
\end{center}
Then $\Tsem{\Lnat}$ is an object of $\COHT$, that is, a coherence
space $L$ equipped with a totality $T$ which can be described as
follows. One defines a sequence of coherence spaces $L_n$ by: $L_0$ is
the coherence space which has $\{(1,*)\}$ as web (where $*$ is the
unique element of $\Web\One$) and, if $L_n$ is known then an element
of $\Web{L_{n+1}}$ is either $(1,*)$ or a pair $(2,x_0)$ where $x_0$
is a finite clique of $L_n$. The coherence relation given by
$(i,a)\scoh(j,b)$ iff $i=j=2$ and $a\scoh_{\Excl{L_n}} b$. Last
$\Web L$ is the union of all the $\Web{L_n}$ (which form a monotone
sequence of sets) and $a\coh_L b$ if $a\coh_{L_n}b$ for some $n$. With
each integer $k$ is associated a total clique $x(k)$ of $L$:
$x(0)=\{(1,*)\}$ and
$x(k+1)=\{(2,x_0)\St x_0\subseteq x(k)\text{ and }x_0\text{ finite}\}$
so that $x(1)=\{(2,\emptyset),(2,\{(1,*)\}\}$,
$x(2)=\{(2,\emptyset),(2,\{(2,\emptyset)\}),(2,\{(2,\{(1,*)\})\}),(2,\{(2,\emptyset),(2,\{(1,*)\})\})\}$
etc. The clique $\{(2,\emptyset)\}$, which is not total, represents a
``partial integer'' that one could denote as
$\mathsf{succ}(\Omega)$ and $\{((1,*),(1,*)),((2,\emptyset),(2,*)\}$
is the interpretation of a proof of $\Limpl\Lnat{\Plus\One\One}$ which
tests whether an integer is zero or not without evaluating it completely:
\begin{center}{\footnotesize
  \AxiomC{}
  \UnaryInfC{$\Seq\One$}
  \RightLabel{$\Nplusl$}
  \UnaryInfC{$\Seq{\Plus\One\One}$}
  \AxiomC{}
  \UnaryInfC{$\Seq\One$}
  \RightLabel{$\Nplusr$}
  \UnaryInfC{$\Seq{\Plus\One\One}$}
  \RightLabel{$\Nweak$}
  \UnaryInfC{$\Seq{\Int{(\With\bot\bot)},\Plus\One\One}$}
  \UnaryInfC{$\Seq{\Limpl{\Excl{(\Plus\One\One)}}{\Plus\One\One}}$}
  \BinaryInfC{$\Seq{\Limpl{\Lnat}{\Plus\One\One}}$}
  \DisplayProof}
\end{center}

\paragraph{Streams of booleans.} Let
$\gamma=\Gfpll\zeta{(\With\One{(\Plus\zeta\zeta)})}$, let $C$ be the
carrier of $\Tsem\gamma$ and $T$ be the totality. Then $\Web C$ can be
described as the set $\{0,1\}^{<\omega}$ of finite sequences of
booleans, and $s\coh_Ct$ if $s$ is a prefix of $t$ or conversely. So a
clique of $C$ is set $x$ of finite sequences which is totally ordered
by the prefix order, and one checks easily that $T$ is the set of all
maximal such sets: up to iso, $T=\{0,1\}^\omega$. Dually a total
clique of $\Tsem{\Orth\gamma}$ is a set $x'$ of pairwise incomparable
sequences such that each element of $\{0,1\}^\omega$ has a prefix in
$x'$. Such an $x'$ is finite by compactness of the Cantor Space.

Another definition is considered in the literature for the type of
streams of booleans, namely
$\gamma'=\Gfpll\zeta{(\Plus\zeta\zeta)}$. A simple computation shows
that $\Tsem{\gamma'}=\Top$ which may seem weird since the formula
$\gamma'$ has many different proofs in $\MULL$. This is due to the
fact that there is no way to write a proof $\pi$ of
$\Limpl{\gamma'}{\Plus\One\One}$ which would allow to extract a finite
information from a stream of type $\gamma'$ (for instance, its first
element). Actually, this model provides a proof of this fact: if such
a proof would exist, it would map $\emptyset$ (total in
$\Tsem{\gamma'}=\Top$) to $\emptyset$ (not total in $\Plus\One\One$).

\subsection{Polarization in $\MULL$}
Among all formulas of $\MULL$, we define two subsets: \emph{positive}
and \emph{negative} formulas. We use letters $P,Q,\dots$ to denote
positive formulas and letters $M,N,\dots$ for negative formulas and as
before $A,B,\dots$ for general formulas.
\begin{align*}
  P,Q,\dots &\Bnfeq \Zero \Bnfor \Plus PQ \Bnfor \One \Bnfor \Tens PQ
              \Bnfor \Excl A \Bnfor \Lfpll\zeta P\Bnfor \zeta\\
  M,N,\dots & \Bnfeq \Top \Bnfor \With MN \Bnfor \Fbot \Bnfor \Par MN
              \Bnfor \Int A \Bnfor \Gfpll\zeta M \Bnfor \zeta
\end{align*}
The only formulas which are at the same time positive and negative are
the variables $\zeta$'s. So there is no closed formula which is both
negative and positive, and $M$ is negative iff $\Orth M$ is
positive. There are of course $\MULL$ formulas, like
$\Tens\Zero\Fbot$, which are neither positive nor negative.

From our
Curry-Howard point of view, the main property of positive formulas is
the following, which expresses that positive formulas have
``structural rules'', meaning that they are types of storable values.
\begin{thm}\label{th:mull-pos-coalg}
  For each positive formula $P$, each sequence of pairwise distinct
  variables $\Vect\zeta=(\List\zeta 1n)$ and each sequence of closed
  formulas $\Vect A=(\List A1n)$ there is a proof
  $\Pprom P{\Vect A}{\Vect\zeta}$ in $\MULL$ of
  $\Seq{\Orth{P'},\Excl{P'}}$ where
  $P'=\Substbis P{\Excl{\Vect A}/\Vect\zeta}$.
\end{thm}
\proof By induction on $P$. If $A$ is a formula, we use $A'$ for
$\Substbis A{\Vect A/\Vect\zeta}$. The proof
$\Pprom\Zero{\Vect A}{\Vect\zeta}$ is an instance of the $\Ntop$
axiom. The proof $\Pprom{\zeta_i}{\Vect A}{\Vect\zeta}$ is
\begin{center}
  \AxiomC{}
  \UnaryInfC{$\Seq{\Orthp{\Excl{A_i}},{\Excl{A_i}}}$}
  \UnaryInfC{$\Seq{\Orthp{\Excl{A_i}},\Excl{\Excl{A_i}}}$}
  \DisplayProof
\end{center}

The proof $\Pprom{\Plus{P_1}{P_2}}{\Vect A}{\Vect\zeta}$ is
\begin{center}
  {\footnotesize
    \AxiomC{$\Pprom{P'_1}{\Vect A}{\Vect\zeta}$}
    \noLine
    \UnaryInfC{$\Seq{\Orth{{P'_1}},\Excl{P'_1}}$}
    \AxiomC{}
    \UnaryInfC{$\Seq{{\Orth{{P'_1}},{{P'_1}}}}$}
    \UnaryInfC{$\Seq{{\Orth{{P'_1}},{\Plus{P'_1}{P'_2}}}}$}
    \UnaryInfC{$\Seq{\Int{\Orth{{P'_1}},{\Plus{P'_1}{P'_2}}}}$}
    \UnaryInfC{$\Seq{\Int{\Orth{{P'_1}},\Exclp{\Plus{P'_1}{P'_2}}}}$}
    \RightLabel{$\Ncut$}
    \BinaryInfC{$\Seq{\Orth{{P'_1}},\Exclp{\Plus{P'_1}{P'_2}}}$}
    \AxiomC{$\Pprom{P'_1}{\Vect A}{\Vect\zeta}$}
    \noLine
    \UnaryInfC{$\Seq{\Orth{{P'_2}},\Excl{P'_2}}$}
    \AxiomC{}
    \UnaryInfC{$\Seq{{\Orth{{P'_2}},{{P'_2}}}}$}
    \UnaryInfC{$\Seq{{\Orth{{P'_2}},{\Plus{P'_1}{P'_2}}}}$}
    \UnaryInfC{$\Seq{\Int{\Orth{{P'_2}},{\Plus{P'_1}{P'_2}}}}$}
    \UnaryInfC{$\Seq{\Int{\Orth{{P'_2}},\Exclp{\Plus{P'_1}{P'_2}}}}$}
    \RightLabel{$\Ncut$}
    \BinaryInfC{$\Seq{\Orth{{P'_2}},\Exclp{\Plus{P'_1}{P'_2}}}$}
    \RightLabel{$\Nwith$}
    \BinaryInfC{$\Seq{\With{\Orth{{P'_1}}}{\Orth{{P'_2}}},
     \Exclp{\Plus{P'_1}{P'_2}}}$}
    \DisplayProof
  }
\end{center}

The proof $\Pprom\One{\Vect A}{\Vect\zeta}$ is
\begin{center}
  {\footnotesize
    \AxiomC{}
    \UnaryInfC{$\Seq{\One}$}
    \UnaryInfC{$\Seq{\Excl\One}$}
    \UnaryInfC{$\Seq{\Fbot,\Excl\One}$}
    \DisplayProof
  }
\end{center}

The proof $\Pprom{\Tens{P'_1}{P'_2}}{\Vect A}{\Vect\zeta}$ is
\begin{center}
  {\footnotesize
    \AxiomC{$\Pprom{P'_1}{\Vect A}{\Vect\zeta}$}
    \noLine
    \UnaryInfC{$\Seq{\Orth{{P'_1}},\Excl{P'_1}}$}
    \AxiomC{$\Pprom{P'_2}{\Vect A}{\Vect\zeta}$}
    \noLine
    \UnaryInfC{$\Seq{\Orth{{P'_2}},\Excl{P'_2}}$}
    \AxiomC{}
    \UnaryInfC{$\Seq{{\Orth{{P'_1}}},P'_1}$}
    \UnaryInfC{$\Seq{\Int{\Orth{{P'_1}}},P'_1}$}
    \AxiomC{}
    \UnaryInfC{$\Seq{{\Orth{{P'_2}}},P'_2}$}
    \UnaryInfC{$\Seq{\Int{\Orth{{P'_2}}},P'_2}$}
    \BinaryInfC{$\Seq{\Int{\Orth{{P'_1}}},\Int{\Orth{{P'_2}}},{\Tens{P'_1}{P'_2}}}$}
    \UnaryInfC{$\Seq{\Int{\Orth{{P'_1}}},\Int{\Orth{{P'_2}}},\Exclp{\Tens{P'_1}{P'_2}}}$}
    \BinaryInfC{$\Seq{\Int{\Orth{{P'_1}}},\Orth{{P'_2}},\Exclp{\Tens{P'_1}{P'_2}}}$}
    \BinaryInfC{$\Seq{{\Orth{{P'_1}}},\Orth{{P'_2}},\Exclp{\Tens{P'_1}{P'_2}}}$}
    \UnaryInfC{$\Seq{\Par{\Orth{{P'_1}}}{\Orth{{P'_2}}},\Exclp{\Tens{P'_1}{P'_2}}}$}
    \DisplayProof
  }
\end{center}

The proof $\Pprom{\Excl A}{\Vect A}{\Vect\zeta}$ is
\begin{center}
  {\footnotesize
    \AxiomC{}
    \UnaryInfC{$\Seq{\Int{\Orth{{A'}}},{\Excl{A'}}}$}
    \UnaryInfC{$\Seq{\Int{\Orth{{A'}}},\Excl{\Excl{A'}}}$}
    \DisplayProof
  }
\end{center}

The proof $\Pprom{\Lfpll\zeta P}{\Vect A}{\Vect\zeta}$ is
\begin{center}
  {\footnotesize
    \AxiomC{$\Pprom{P}{(\Vect A,\Lfpll\zeta{P})}{\Vect\zeta,\zeta}$}
    \noLine
    \UnaryInfC{$\Seq{\Orth{\Subst P{\Excl{\Lfpll\zeta{P}}}\zeta},\Excl{\Subst P{\Excl{\Lfpll\zeta{P}}}\zeta}}$}
    \AxiomC{$\Subst P\delta\zeta$}
    \noLine
    \UnaryInfC{$\Seq{\Orth{{\Subst P{\Excl{\Lfpll\zeta{P}}}\zeta}},{\Subst P{\Lfpll\zeta P}\zeta}}$}
    \UnaryInfC{$\Seq{\Orth{{\Subst P{\Excl{\Lfpll\zeta{P}}}\zeta}},{\Lfpll\zeta P}}$}
    \UnaryInfC{$\Seq{\Orthp{\Excl{\Subst P{\Excl{\Lfpll\zeta{P}}}\zeta}},{\Lfpll\zeta P}}$}
    \UnaryInfC{$\Seq{\Orthp{\Excl{\Subst P{\Excl{\Lfpll\zeta{P}}}\zeta}},\Exclp{\Lfpll\zeta P}}$}
    \BinaryInfC{$\Seq{\Orth{\Subst P{\Excl{\Lfpll\zeta{P}}}\zeta},\Exclp{\Lfpll\zeta P}}$}
    %\UnaryInfC{$\Seq{\Orth{\Subst P{\Excl{\Lfpll\zeta{P}}}\zeta},\Exclp{\Lfpll\zeta P}}$}
    \RightLabel{$\Ngfpbis$}
    \UnaryInfC{$\Seq{\Gfpll\zeta{\Orth P},\Exclp{\Lfpll\zeta P}}$}
    \DisplayProof
  }
\end{center}
where $\delta$ is the proof
\begin{center}
  {\footnotesize
    \AxiomC{}
    \UnaryInfC{$\Seq{\Orth{{\Lfpll\zeta{P}}},\Lfpll\zeta{P}}$}
    \UnaryInfC{$\Seq{\Orthp{\Excl{\Lfpll\zeta{P}}},\Lfpll\zeta{P}}$}
  \DisplayProof}
\end{center}
\qed

As an example (up to some cut-eliminations) the proof $\Ppromcl\Tnat$ is
\begin{center}
  {\footnotesize
    \AxiomC{}
    \RightLabel{$\Nzero$}
    \UnaryInfC{$\Seq{\Tnat}$}
    \UnaryInfC{$\Seq{\Excl\Tnat}$}
    \UnaryInfC{$\Seq{\Fbot,\Excl\Tnat}$}
    \AxiomC{}
    \UnaryInfC{$\Seq{\Orth{\Tnat},\Tnat}$}
    \RightLabel{$\Nsucc$}
    \UnaryInfC{$\Seq{\Orth{\Tnat},\Tnat}$}
    \UnaryInfC{$\Seq{\Orthp{\Excl\Tnat},\Tnat}$}
    \UnaryInfC{$\Seq{\Orthp{\Excl\Tnat},\Excl\Tnat}$}
    \RightLabel{$\Nwith$}
    \BinaryInfC{$\Seq{\With{\Fbot}{\Orthp{\Excl\Tnat}},\Excl\Tnat}$}
    \RightLabel{$\Ngfpbis$}
    \UnaryInfC{$\Seq{\Orth\Tnat,\Excl\Tnat}$}
    \DisplayProof
  }
\end{center}

\begin{thm}
  Let $\List N1k$ be closed negative formulas and let $A$ be a closed
  formula. Then the generalized promotion rule and the generalized
  structural rules
  \begin{center}
    \AxiomC{$\Seq{\List N1k,A}$}
    \RightLabel{$\Ngprom$}
    \UnaryInfC{$\Seq{\List N1k,\Excl A}$}
    \DisplayProof
    \quad
    \AxiomC{$\Seq\Gamma$}
    \RightLabel{$\Ngweak$}
    \UnaryInfC{$\Seq{N,\Gamma}$}
    \DisplayProof
    \quad
    \AxiomC{$\Seq{N,N,\Gamma}$}
    \RightLabel{$\Ngcontr$}
    \UnaryInfC{$\Seq{N,\Gamma}$}
    \DisplayProof
  \end{center}
  are derivable in $\MULL$.
\end{thm}
\proof
Cut the proof
\begin{center}
  \AxiomC{$\Seq{\List {\Int N}1k,A}$}
  \UnaryInfC{$\Seq{\List {\Int N}1k,\Excl A}$}
  \DisplayProof
\end{center}
against the proofs $\Ppromcl{\Orth{N_i}}$ (for $i=1,\dots,k$) and
similarly for the two other proofs.  \qed

\subsection{Representing a System T in $\MULL$}\label{sec:rep-syst-T}
We define a version of Gödel's System T where integers can be handled
in a strict way thanks to an additional $\mathsf{let}$
construct\footnote{One motivation for focusing on such a language is
  explained in~\cite{EhrhardPaganiTasson18}, in a context of
  probabilistic computing. Indeed the setting presented here can
  easily and meaningfully be applied to such computations whereas a
  purely CBN version of system T wouldn't really make sense in a
  probabilistic setting from an algorithmic expressiveness viewpoint.}
(in contrast with the system presented in~\cite{Girard89}) that we
call $\GODELT$. We first define the syntax of our language:
\begin{align*}
  \sigma,\tau\dots &\Bnfeq \Gtnat \Bnfor \Timpl\sigma\tau \quad\text{(types)}\\
  s,t,u\dots &\Bnfeq \Num n \Bnfor \Var x \Bnfor \App st \Bnfor \Abst x\sigma s
               \Bnfor \Succ s\Bnfor \Rec stu
               \Bnfor \Let xst\quad \text{(terms)}
\end{align*}
The typing rules are
\begin{center}{\footnotesize
  \AxiomC{$n\in\Nat$}
  \UnaryInfC{$\Tseq{\Phi}{\Num n}{\Gtnat}$}
  \DisplayProof
  \quad
  \AxiomC{}
  \UnaryInfC{$\Tseq{\Phi,\Var x:\sigma}{\Var x}{\sigma}$}
  \DisplayProof
  \quad
  \AxiomC{$\Tseq{\Phi}{s}{\Timpl\sigma\tau}$}
  \AxiomC{$\Tseq{\Phi}{t}{\sigma}$}
  \BinaryInfC{$\Tseq{\Phi}{\App st}{\tau}$}
  \DisplayProof
  }
\end{center}

\begin{center}
  {\footnotesize
    \AxiomC{$\Tseq{\Phi,\Var x:\sigma}{s}{\tau}$}
    \UnaryInfC{$\Tseq{\Phi}{\Abst x\sigma s}{\Timpl\sigma\tau}$}
    \DisplayProof
    \quad
    \AxiomC{$\Tseq\Phi s\Gtnat$}
    \UnaryInfC{$\Tseq\Phi{\Succ s}\Gtnat$}
    \DisplayProof
    \quad
    \AxiomC{$\Tseq\Phi s\Gtnat$}
    \AxiomC{$\Tseq{\Phi,x:\Gtnat}t\sigma$}
    \BinaryInfC{$\Tseq\Phi{\Let xst}\sigma$}
    \DisplayProof
  }
\end{center}

\begin{center}
  {\footnotesize
    \AxiomC{$\Tseq{\Phi}{s}{\Gtnat}$}
    \AxiomC{$\Tseq{\Phi}{t}{\sigma}$}
    \AxiomC{$\Tseq{\Phi}{u}{\Timpl{\Gtnat}{\Timpl\sigma\sigma}}$}
    \TrinaryInfC{$\Tseq{\Phi}{\Rec stu}{\sigma}$}
    \DisplayProof
  }
\end{center}
In these rules $\Phi$ denotes a typing context
$\Phi=(\Var x_1:\sigma_1,\dots,\Var x_k:\sigma_k)$. We define an operational
semantics by means of a weak-head reduction relation specified by the
following deduction rules.
\begin{center}
  {\footnotesize
    % \AxiomC{$\sigma\not=\Gtnat$}
    % This is no more necessary because we have added a let construct
    % to this calulus
    \AxiomC{}
    \UnaryInfC{$\App{\Abst x\sigma s}{t}\Rel\Gredwh\Subst st{\Var x}$}
    \DisplayProof
    \quad
    \AxiomC{}
    \UnaryInfC{$\Let x{\Num n}s\Rel\Gredwh\Subst s{\Num n}x$}
    \DisplayProof
    \quad
    \AxiomC{}
    \UnaryInfC{$\Succ{\Num n}\Rel\Gredwh\Num{n+1}$}
    \DisplayProof
  }
\end{center}

\begin{center}
  {\footnotesize
    \AxiomC{}
    \UnaryInfC{$\Rec{\Num 0}{t}{u}\Rel\Gredwh t$}
    \DisplayProof
    \quad
    \AxiomC{}
    \UnaryInfC{$\Rec{\Num{n+1}}{t}{u}\Rel
      \Gredwh\App u{\Num n\Appsep\Rec{\Num n}{t}{u}}$}
    \DisplayProof
    \quad
    \AxiomC{$s\Rel\Gredwh s'$}
    \UnaryInfC{$\App st\Rel\Gredwh\App{s'}t$}
    \DisplayProof
    }
\end{center}

\begin{center}
  {\footnotesize
    \AxiomC{$s\Rel\Gredwh s'$}
    \UnaryInfC{$\Let xst\Rel\Gredwh\Let x{s'}t$}
    \DisplayProof
    \quad
    \AxiomC{$s\Rel\Gredwh s'$}
    \UnaryInfC{$\Succ s\Rel\Gredwh\Succ{s'}$}
    \DisplayProof
    \quad
    \AxiomC{$s\Rel\Gredwh s'$}
    \UnaryInfC{$\Rec stu\Rel\Gredwh\Rec{s'}tu$}
    \DisplayProof
  }
\end{center}
Of course $\Gredwh$ enjoys Subject Reduction as easily checked.  The
first five rules describe the three forms of redexes in our System
$\GODELT$, the $\mathsf{let}$ construct offering the ability to handle
integers in a CBV manner. One can also define a general reduction
relation allowing to reduce these redexes anywhere in a term and it
can be proved that this general reduction is Church-Rosser, but here
we focus on $\Gredwh$ which is a deterministic reduction strategy
turning $\GODELT$ into a programming language.

\paragraph{Normalization of $\GODELT$.} It is well known that
System T is strongly normalizing, but, since our presentation of this
system is slightly different from the usual one (by our CBV handling
of integers) it is meaningful to give a direct proof of this fact. We
actually don't prove strong normalization, but normalization of the
$\Gredwh$ strategy, and only for closed terms of type $\Gtnat$, which
will be enough for our purpose and can be done in a few lines.

\begin{lem}\label{lemma:Greduc-antibeta-closed}
  If $\Tseq{}{s}{\Gtnat}$ and $s$ is $\Gredwh$-normal then there is an
  integer $n$ such that $s=\Num n$.
\end{lem}
\proof Simple inductive analysis of the structure of $s$.
\qed

\begin{thm}
  The reduction $\Gredwh$ is normalizing for closed terms of type $\Tnat$.
\end{thm}
\proof We adapt the standard reducibility proof. For each type
$\sigma$ we define a set $\Greduc\sigma$ of \emph{reducible} closed
terms $s$ such that $\Tseq{}{s}{\sigma}$. The definition, by
induction on $\sigma$, is: $\Greduc\Gtnat$ is the set of all normalizing
closed terms of type $\Gtnat$, and $\Greduc{\Timpl\sigma\tau}$ is the
set of all $s$ such that $\Tseq{}{s}{\Timpl\sigma\tau}$ and, for all
$t\in\Greduc\sigma$, one has $\App st\in\Greduc\tau$.

One proves first that if $\Tseq{}{s}{\sigma}$ and
$s\Rel\Gredwh s'\in\Greduc\sigma$, then $s\in\Greduc\sigma$. This is
done by induction on $\sigma$. The case $\sigma=\Gtnat$ results
immediately from the definition of $\Greduc\Gtnat$ so assume that
$\sigma=(\Timpl\tau\phi)$. Let $t\in\Greduc\tau$, we have
$\App st\Rel\Gredwh\App{s'}t\in\Greduc\tau$ by definition of $\Gredwh$
and hypothesis on $s'$, and hence $\App st\in\Greduc\tau$ by inductive
hypothesis.

Last we prove by induction on $s$ that if
$\Tseq{\Var x_1:\sigma_1,\dots,\Var x_k:\sigma_k}{s}{\tau}$, then for
all $s_1\in\Greduc{\sigma_1}$,\dots,$s_k\in\Greduc{\sigma_k}$, one has
$\Substbis s{s_1/\Var x_1,\dots,s_k/\Var x_k}\in\Greduc\tau$.  We set
$\Phi=(\Var x_1:\sigma_1,\dots,\Var x_k:\sigma_k)$ and
$\bar w=\Substbis w{s_1/\Var x_1,\dots,s_k/\Var x_k}$ for all term
$w$.
\begin{itemize}
\item If $s=\Var x_i$ or $s=\Num n$ the reasoning is straightforward.
\item Assume $s=\Succ t$ with $\Tseq\Phi t\Gtnat$. By inductive
  hypothesis $\bar t\in\Greduc\Gtnat$ and hence
  $\bar t\Reltr\Gredwh\Num n$ for some $n\in\Nat$, and hence
  $\bar s\Reltr\Gredwh\Num{n+1}$ by definition of $\Gredwh$.
\item Assume $s=\App tu$ with $\Tseq\Phi t{\Timpl\sigma\tau}$ and
  $\Tseq\Phi u\sigma$ so that, by inductive hypothesis
  $\bar t\in\Greduc{\Timpl\sigma\tau}$ and $\bar u\in\Greduc\sigma$
  and hence $\bar s=\App{\bar t}{\bar u}\in\Greduc\tau$ by definition of
  $\Greduc{\Timpl\sigma\tau}$.
\item Assume $s=\Abst x\sigma t$ with
  $\Tseq{\Phi,\Var x:\sigma}{t}{\tau}$. We must prove that
  $\bar s=\Abst x\sigma{\bar t}\in\Greduc{\Timpl\sigma\tau}$, so let
  $u\in\Greduc\sigma$ and let us prove that
  $\App{\bar s}{u}=\App{\Abst x\sigma{\bar t}}u\in\Greduc\tau$. We have
  $\App{\Abst x\sigma{\bar t}}u\Rel\Gredwh\Subst{\bar t}u{\Var x}$ and we know
  that $\Subst{\bar t}u{\Var x}\in\Greduc\tau$ by inductive hypothesis
  applied to $t$. We conclude that
  $\bar s\in\Greduc{\Timpl\sigma\tau}$ by
  Lemma~\ref{lemma:Greduc-antibeta-closed}.
\item Assume $s=\Rec tuv$ with $\Tseq\Phi t\Gtnat$,
  $\Tseq\Phi u\sigma$ and
  $\Tseq\Phi v{\Timpl\Gtnat{\Timpl\sigma\sigma}}$ so that
  $\bar t\in\Greduc\Gtnat$, $\bar u\in\Greduc\sigma$ and
  $\bar v\in\Greduc{\Timpl\Gtnat{\Timpl\sigma\sigma}}$ by inductive
  hypothesis. By definition of $\Greduc\Gtnat$ there is
  % a\footnote{uniquely defined since $\Gredwh$ is a reduction strategy}
  $n\in\Nat$ such that $\bar t\Reltr\Gredwh\Num n$ and therefore
  $\bar s\Reltr\Gredwh\Rec{\Num n}{\bar u}{\bar v}$ by definition of
  $\Gredwh$. An easy induction on $n$ (using
  Lemma~\ref{lemma:Greduc-antibeta-closed}) shows that
  $\Rec{\Num n}{\bar u}{\bar v}\in\Greduc\sigma$ and hence
  $\bar s\in\Greduc\sigma$ by Lemma~\ref{lemma:Greduc-antibeta-closed}
  again.
\item If $s=\Let xtu$ with $\Tseq\Phi t\Gtnat$ and
  $\Tseq{\Phi,x:\Gtnat}u\sigma$ we have $\bar t\in\Greduc\Gtnat$ and hence
  $\bar t\Reltr\Gredwh\Num n$ for some $n\in\Nat$. Therefore
  $\Let x{\bar t}{\bar u}\Reltr\Gredwh\Let x{\Num n}{\bar
    u}\Rel\Gredwh\Subst{\bar u}{\Num n}{\Var x}$ by definition of
  $\Gredwh$ and we have $\Subst{\bar u}{\Num n}{\Var x}\in\Greduc\tau$
  by inductive hypothesis on $u$. It follows that
  $\bar s\in\Greduc\sigma$ by
  Lemma~\ref{lemma:Greduc-antibeta-closed} again. \qed
\end{itemize}

Therefore, given $s$ such that $\Tseq{}{s}{\Timpl\Gtnat\Gtnat}$, there
is a uniquely defined total function $f:\Nat\to\Nat$ such that
$\forall n\in\Nat\ \App s{\Num
  n}\Reltr\Gredwh\Num{f(n)}$. Following~\cite{Girard89}, one can show
that these functions which are definable in $\GODELT$ are exactly
those whose totality can be proved in Peano's Arithmetics.

\paragraph{Denotational semantics of $\GODELT$ in $\COHT$.} With each
type $\sigma$ one associates an object $\Tsem\sigma$ of $\COHT$ using
the structure of $\LL$ model of this category and with each $s$ such
that $\Tseq\Phi s\sigma$ (with
$\Phi=(\Var x_1:\sigma_1,\dots,\Var x_k:\sigma_k)$) one associates an
element of $\COHT(\Tsem\Phi,\Tsem\sigma)$ where
$\Tsem\Phi=\Excl{\Tsem{\sigma_1}}\ITens\cdots\ITens\Excl{\Tsem{\sigma_k}}$. For
$\Tsem\Gtnat$ we take the coherence space with totality $\Snat$ such
that $\Tcohca\Snat=(\Nat,\mathord=)$ and
$\Tcoht\Snat=\{\{n\}\St n\in\Nat\}$. Then we set
$\Tsem{\Timpl\sigma\tau}=(\Limpl{\Excl{\Tsem\sigma}}{\Tsem\tau})$.
There is a morphism $\Coalgstr\Snat\in\COHT(\Snat,\Excl\Snat)$ which
turns $\Snat$ into a $\oc$-coalgebra, namely
$\Coalgstr\Snat=\{(n,x)\in\Nat\times\Part\Nat\St x=\emptyset\text{ or
}x=\{n\}\}$. Given $n\in\Nat$, we set $\Snum n=\{n\}\in\Cl{\Snat}$,
that we also consider as an element of $\COHT(\One,\Snat)$.

If $\Tseq\Phi s\sigma$ then
$\Psem s_\Phi\in\COHT(\Tsem\Phi,\Tsem\sigma)$ is defined by induction
on $s$ using standard algebraic constructs. We give here a direct
concrete description of this interpretation.
\begin{itemize}
\item
  $\Psem{\Var x_i}_\Phi=\{(\List x1k,n)\St n\in\Nat\text{,
  }x_i=\{n\}\text{ and }x_j=\emptyset\text{ if }j\not=i\}$.
\item $\Psem{\Num n}_\Phi=\{(\emptyset,\dots,\emptyset,n)\}$.
\item $\Psem{\Succ s}_\Phi=\{(\Vect x,n+1)\St (\Vect x,n)\in\Psem s_\Phi)\}$.
% \item
%   $\Psem{\Abst x\Tnat s}_\Phi=\{(\Vect x,(n,a))\St (\Vect
%   x,y,a)\in\Psem s_{\Phi,\Var x:\Tnat}\text{ and }y=\emptyset\text{ or
%   }y=\{n\}\}$. Expressed categorically, this definition invloves a use
%   of $\Coalgstr\Snat$: by inductive hypothesis we have
%   $\Psem s_{\Phi,\Var
%     x:\Tnat}\in\COHT(\Tens{\Psem\Phi}{\Excl\Snat},\Psem\tau)$ (where
%   $\tau$ is such that $\Tseq{\Phi,\Var x:\Tnat}{s}{\tau}$), then
%   $\Psem s_{\Phi,\Var x:\Tnat}\Compl(\Tens{\Tsem\Phi}{h_\Snat})
%   \in\COHT(\Tens{\Psem\Phi}{\Snat},\Psem\tau)$ and then
%   $\Psem{\Abst x\Tnat s}_\Phi=\Curlin{(\Psem s_{\Phi,\Var
%       x:\Tnat}\Compl(\Tens{\Tsem\Phi}{h_\Snat}))}$.
\item %If $\sigma\not=\Tnat$ then
  $\Psem{\Abst x\sigma s}_\Phi=\{(\Vect x,(x_0,a))\St (\Vect
  x,x_0,a)\in\Psem s_{\Phi,\Var x:\sigma}\}$.
\item Assume $\Tseq\Phi s{\Timpl\sigma\tau}$ and $\Tseq\Phi t\sigma$
  so that $\Tseq\Phi{\App st}\tau$. % If $\sigma=\Tnat$ then
  % $\Psem{\App st}_\Phi=\{(\Vect x\cup\Vect
  % y,a)\in\Web{\Limpl{\Tsem\Phi}{\Tsem\tau}}\St \exists n\,(\Vect
  % x,(n,a))\in\Psem s_\Phi\text{ and }(\Vect y,n)\in\Psem t_\Phi\}$. If
  % $\sigma\not=\Tnat$, then
  Then $\Psem{\App st}_\Phi=\{(\Vect x\cup \Vect{y^1}\cup\cdots\cup
  \Vect{y^p},a)\in\Web{\Limpl{\Tsem\Phi}{\Tsem\tau}}\St \exists \List
  a1p\in\Web{\Tsem\sigma}\,(\Vect x,(\{\List a1p\},a))\in\Psem
  s_\Phi\text{ and }(\Vect{y^j},a_j)\in\Psem t_\Phi\text{ for
  }j=1,\dots,p\}$.
\item Let $X$ be a coherence space with totality. We define by
  induction on $n\in\Nat$ a family of morphisms
  $r_n\in\COHT(\Excl X\ITens\Exclp{\Limpl{\Excl\Snat}{\Limpl{\Excl
        X}{X}}},X)$ as follows (keeping monoidal isos implicit). The
  morphism $r_0$ is
\begin{center}
\begin{tikzpicture}[->, >=stealth]
  \node (1) {$\Excl X\ITens\Exclp{\Limpl{\Excl\Snat}{\Limpl{\Excl
            X}{X}}}$};
  \node (2) [ right of=1, node distance=3.4cm ]
  {$\Excl X$};
  \node (3) [ right of=2, node distance=1.4cm ]
  {$X$};
  \tikzstyle{every node}=[midway,auto,font=\scriptsize]
  \draw (1) -- node {$\Tens{\Excl X}{\Weak X}$} (2);
  \draw (2) -- node {$\Der X$} (3);
\end{tikzpicture}    
\end{center}
and the morphism $r_{n+1}$ is 
\begin{center}
\begin{tikzpicture}[->, >=stealth]
  \node (1) {$\Excl X\ITens\Exclp{\Limpl{\Excl\Snat}{\Limpl{\Excl
            X}{X}}}$};
  \node (2) [ below of=1, node distance=1cm ]
  {$\Excl X\ITens\Exclp{\Limpl{\Excl\Snat}{\Limpl{\Excl
            X}{X}}}\ITens\Exclp{\Limpl{\Excl\Snat}{\Limpl{\Excl
            X}{X}}}\ITens\Excl\Snat$};
    \node (3) [ below of=2, node distance=1cm ]
    {$\Excl X\ITens(\Limpl{\Excl\Snat}{\Limpl{\Excl X}{X}})\ITens\Excl\Snat$};
    \node (4) [below of=3, node distance=0.9cm] {$X$};
  \tikzstyle{every node}=[midway,auto,font=\scriptsize]
  \draw (1) -- node {${\Excl X}\ITens{\Contr{}}\ITens\Prom{\Snum{n}}$} (2);
  \draw (2) -- node {$\Prom{r_n}\ITens\Der{\Limpl{\Snat}{\Limpl{\Excl X}{X}}}
    \ITens\Excl\Snat$} (3);
  \draw (3) -- node {$e$} (4);
\end{tikzpicture}    
\end{center}
where $e$
uses twice the evaluation morphism $\Evlin$. Stated otherwise, if we
consider $r_n$ as a binary stable function
$\Cl X\times\Cl{\Limpl{\Excl\Snat}{\Limpl{\Excl X}{X}}}\to\Cl X$, we have
$r_0(x,f)=x$ and $r_{n+1}(x,f)=f(\{n\},r_n(x,f))$ (considering
$f\in\Cl{\Limpl{\Excl\Snat}{\Limpl{\Excl X}{X}}}$ as a binary stable
function $\Cl\Snat\times\Cl X\to\Cl X$). Then one defines
$r\in\COHT(\Snat\ITens\Excl X\ITens\Exclp{\Limpl{\Snat}{\Limpl{\Excl
      X}{X}}},X)$ by
\[
  r=\{((n,x_0,f_0),a)\St n\in\Nat\text{ and }((x_0,f_0),a)\in
  r_n\}\,.
\]
Indeed, the elements of $\Tcoht\Snat$ are the $\{n\}$ for
$n\in\Nat$ and we have
$r'(\{n\})=r_n\in\Tcohtp{\Limpl{\Excl
    X\ITens\Exclp{\Limpl{\Snat}{\Limpl{\Excl X}{X}}}}{X}}$ where $r'$
is the Curry transpose of $r$; it follows that
$r\in\COHT(\Snat\ITens\Excl X\ITens\Exclp{\Limpl{\Snat}{\Limpl{\Excl
      X}{X}}},X)$.  Assume that $\Tseq\Phi s\Gtnat$,
$\Tseq\Phi t\sigma$ and $\Tseq\Phi u{\Timpl\Gtnat{\Timpl\sigma\sigma}}$
and take $X=\Tsem\sigma$, then we have
$\Psem s_\Phi\in\COHT(\Tsem\Phi,\Snat)$,
$\Psem t_\Phi\in\COHT(\Tsem\Phi,X)$ and
$\Psem u_\Phi\in\COHT(\Tsem\Phi,\Limpl{\Excl\Snat}{\Limpl{\Excl X}{X}})$
and we define $\Psem{\Rec stu}_\Phi$ as the following composition of
morphisms is $\COHT$
\begin{center}
\begin{tikzpicture}[->, >=stealth]
  \node (1) {$\Tsem\Phi$};
  \node (2) [ below of=1, node distance=1cm ]
  {$\Tsem\Phi\ITens\Tsem\Phi\ITens\Tsem\Phi$};
    \node (3) [ below of=2, node distance=1cm ]
    {$\Snat\ITens\Excl X\ITens\Exclp{\Limpl{\Excl\Snat}{\Limpl{\Excl
      X}{X}}}$};
    \node (4) [below of=3, node distance=0.9cm] {$X$};
  \tikzstyle{every node}=[midway,auto,font=\scriptsize]
  \draw (1) -- node {$c$} (2);
  \draw (2) -- node {$\Psem s_\Phi\ITens\Prom{\Psem t_\Phi}
    \ITens\Prom{\Psem u_\Phi}$} (3);
  \draw (3) -- node {$r$} (4);
\end{tikzpicture}    
\end{center}
where $c$ is obtained by combining two copies of $\Contr{\Psem\Phi}$.
\item Assume last that $s=\Let xtu$ where $\Tseq\Phi t\Gtnat$ and
  $\Tseq{\Phi,x:\Gtnat}u\sigma$. By inductive hypothesis we have
  $\Psem t_\Phi\in\COHT(\Tsem\Phi,\Snat)$ and
  $\Psem
  s_{\Phi,x:\Gtnat}\in\COHT(\Tsem\Phi\ITens\Excl\Snat,\Tsem\sigma)$. Then
  $\Psem{\Let xtu}_\Phi$ is
\begin{center}
\begin{tikzpicture}[->, >=stealth]
  \node (1) {$\Tsem\Phi$};
  \node (2) [ right of=1, node distance=1.7cm ]
  {$\Tsem\Phi\ITens\Tsem\Phi$};
    \node (3) [ right of=2, node distance=3cm ]
    {$\Tsem\Phi\ITens\Snat$};
    \node (4) [right of=3, node distance=2.8cm] {$\Tsem\Phi\ITens\Excl\Snat$};
    \node (5) [right of=4, node distance=2.6cm] {$\Tsem\sigma$};
  \tikzstyle{every node}=[midway,auto,font=\scriptsize]
  \draw (1) -- node {$c$} (2);
  \draw (2) -- node {$\Tsem\Phi\ITens\Psem t_\Phi$} (3);
  \draw (3) -- node {$\Tsem\Phi\ITens\Coalgstr\Snat$} (4);
  \draw (4) -- node {$\Psem u_{\Phi,x:\Gtnat}$} (5);
\end{tikzpicture}    
\end{center}
\end{itemize}

\begin{prop}\label{prop:}
  Assume that $\Tseq\Phi s\sigma$ and that $s\Rel\Gredwh s'$ (so that
  $\Tseq\Phi{s'}{\sigma}$). Then $\Psem{s'}_\Phi=\Psem s_\Phi$.
\end{prop}
\proof We only sketch the proof which is quite standard and relies
on a preliminary Substitution Lemma: if
$\Tseq{\Phi,\Var x:\sigma}{s}{\tau}$ and $\Tseq{\Phi}{t}{\sigma}$, the
morphism $\Psem{\Subst stx}_\Phi$ is equal to
\begin{center}
\begin{tikzpicture}[->, >=stealth]
  \node (1) {$\Tsem\Phi$};
  \node (2) [ right of=1, node distance=2.4cm ]
  {$\Tsem\Phi\ITens\Tsem\Phi$};
    \node (3) [ right of=2, node distance=3.2cm ]
    {$\Tens{\Tsem\Phi}{\Excl{\Tsem\sigma}}$};
    \node (4) [ right of=3, node distance=2.4cm] {$\Tsem\tau$};
  \tikzstyle{every node}=[midway,auto,font=\scriptsize]
  \draw (1) -- node {$\Contr{\Tsem\Phi}$} (2);
  \draw (2) -- node {$\Tens{\Tsem\Phi}{\Prom{\Psem t_\Phi}}$} (3);
  \draw (3) -- node {$\Psem s_{\Phi,\Var x:\sigma}$} (4);
\end{tikzpicture}    
\end{center}
which is proven by a simple induction on $s$.  In the proof of the
proposition itself the only delicate case is when $s=\Let x{\Num n}t$
where $\Tseq{\Phi,\Var x:\Tnat}t\sigma$. To deal with it it suffices
to observe that $h_\Snat\Compl\Psem{\Num n}=\Prom{\Psem{\Num n}}$, in
other words, $\Psem{\Num n}$ is a $\oc$-coalgebra morphism from $\One$
to $\Snat$ (see~\cite{Ehrhard16a} to read more on this point of view on
values).  \qed

Notice that the elements of $\Tcoht{(\Limpl\Snat\Snat)}$ are exactly
the total functions $\Nat\to\Nat$. Those which are of shape
$\Tsem s\Compl\Coalgstr\Snat$ for some $s$ such that
$\Tseq{}{s}{\Timpl\Gtnat\Gtnat}$ are exactly the functions which are
provably total in Peano Arithmetics. We prove now that these functions
are definable in $\MULL$.

\paragraph{Translating $\GODELT$ into $\MULL$.}
First we translate types: given a type $\sigma$ we define a $\MULL$
formula $\Tgtrad\sigma$ by: $\Tgtrad\Gtnat=\Tnat$ and
$\Tgtrad{(\Timpl\sigma\tau)}=\Limpl{\Excl{\Tgtrad\sigma}}{\Tgtrad\tau}$.
% The following is no more necessary, because of the let construct
% 
% This latter dichotomy reflects the fact that $\Gtnat$ is the only
% type which has values in our System $\GODELT$ (these values are of
% course the $\Num n$'s).
Observe that $\Psem\sigma$ (in the sense of the
semantics of $\GODELT$) coincides with $\Psem{\Tgtrad\sigma}$ (in the
sense of the semantics of $\MULL$).

Now we define a translation of terms: given $s$,
$\Phi=(\Var x_1:\sigma_1,\dots,\Var x_k:\sigma_k)$ and $\tau$ such
that $\Tseq{\Phi}{s}{\tau}$ (our typing system is such that when
$\tau$ exists, it is unique), a proof $\Tgtrad s_\Phi$ of the sequent
$\Seq{\Orthp{\Excl{\Tgtrad{\sigma_1}}},\dots,
  \Orthp{\Excl{\Tgtrad{\sigma_k}}},\Tgtrad\tau}$. We use the notation
$\Tgtradc\Phi$ for the sequence
$\Orthp{\Excl{\Tgtrad{\sigma_1}}},\dots,
\Orthp{\Excl{\Tgtrad{\sigma_k}}}$. This translation is defined by
induction on $s$ as follows.  In the course of this inductive
definition, we argue that $\Psem{\Tgtrad s_\Phi}=\Psem
s_\Phi$. Observe indeed that
$\Psem{\Tgtrad
  s_\Phi}\in\Cl{\Orth{\Psem{\Excl{\Tgtrad{\sigma_1}}}}\IPar\cdots
  \IPar\Orth{\Psem{\Excl{\Tgtrad{\sigma_k}}}}\IPar\Psem{\Tgtrad\tau}}$,
$\Psem
s_\Phi\in\COH(\Excl{\Psem{\sigma_1}}\ITens\cdots\ITens\Excl{\Psem{\sigma_k}},
\Psem\tau)$ and that
$\Cl{\Orth{\Psem{\Excl{\Tgtrad{\sigma_1}}}}\IPar\cdots
  \IPar\Orth{\Psem{\Excl{\Tgtrad{\sigma_k}}}}\IPar\Psem{\Tgtrad\tau}}
=\COH(\Excl{{\Psem{\sigma_1}}}\ITens\cdots\ITens\Excl{{\Psem{\sigma_k}}},
\Psem\tau)$ up to trivial iso (using the fact that
$\Psem{\Tgtrad\sigma}=\Psem\sigma$ for all type $\sigma$ of $\GODELT$).

If $s=\Var x_i$ then $\Tgtrad s_\Phi$ is
\begin{center}
  {\footnotesize
  \AxiomC{$\Seq{\Orthp{\Tgtrad{\sigma_i}},\Tgtrad{\sigma_i}}$}
  \UnaryInfC{$\Seq{\Orthp{\Excl{\Tgtrad{\sigma_i}}},\Tgtrad{\sigma_i}}$}
  \RightLabel{$\Nweak$}
  \doubleLine
  \UnaryInfC{$\Seq{\Tgtrad\Phi,\Tgtrad{\sigma_i}}$}
  \DisplayProof}
\end{center}
Then $(\gamma,a)\in\Psem{\Tgtrad s_\Phi}$ iff $\gamma_i=\{a\}$ and
$\gamma_j=\emptyset$ if $j\not=i$,
$\Psem{\Tgtrad s_\Phi}=\Psem s_\Phi$ follows.

If $s=\Num 0$ then $\Tgtrad s_\Phi$ is
\begin{center}
  {\footnotesize
    \AxiomC{}
    \RightLabel{$\Nzero$}
    \UnaryInfC{$\Seq\Tnat$}
    \RightLabel{$\Nweak$}
    \doubleLine
    \UnaryInfC{$\Seq{\Tgtradc\Phi,\Tnat}$}
    \DisplayProof
  }
\end{center}
Then $(\gamma,k)\in\Psem{\Tgtrad s_\Phi}$ iff $k=0$ and
$\gamma_i=\emptyset$ for all $i$, $\Psem{\Tgtrad s_\Phi}=\Psem s_\Phi$
follows.

If $s=\Num{n+1}$ then $\Tgtrad s_\Phi$ is
\begin{center}
  {\footnotesize
    \AxiomC{$\Tgtrad{\Num n}_\Phi$}
    \noLine
    \UnaryInfC{$\Seq{\Tgtradc\Phi,\Tnat}$}
    \RightLabel{$\Nsucc$}
    \UnaryInfC{$\Seq{\Tgtradc\Phi,\Tnat}$}
    \DisplayProof
  }
\end{center}
Then $(\gamma,k)\in\Psem{\Tgtrad s_\Phi}$ iff $k=n+1$ and
$\gamma_i=\emptyset$ for all $i$, $\Psem{\Tgtrad s_\Phi}=\Psem s_\Phi$
follows.

If $s=\Succ t$ then $\Tgtrad s_\Phi$ is
\begin{center}
  {\footnotesize
    \AxiomC{$\Tgtrad{t}_\Phi$}
    \noLine
    \UnaryInfC{$\Seq{\Tgtradc\Phi,\Tnat}$}
    \RightLabel{$\Nsucc$}
    \UnaryInfC{$\Seq{\Tgtradc\Phi,\Tnat}$}
    \DisplayProof
  }
\end{center}
Then $(\gamma,k)\in\Psem{\Tgtrad s_\Phi}$ iff $k>0$ and
$(\gamma,k-1)\in\Psem{\Tgtrad t_\Phi}$, and
$\Psem{\Tgtrad s_\Phi}=\Psem s_\Phi$ follows since
$\Psem{\Tgtrad t_\Phi}=\Psem t_\Phi$ by inductive hypothesis.

% If $s=\App tu$ with $\Tseq\Phi t{\Timpl\Gtnat\tau}$ and
% $\Tseq\Phi u\Gtnat$ then $\Tgtrad s_\Phi$ is
% \begin{center}
%   {\footnotesize
%     \AxiomC{$\Tgtrad t_\Phi$}
%     \noLine
%     \UnaryInfC{$\Seq{\Tgtradc\Phi,\Par{\Orth\Tnat}{\Tgtrad\tau}}$}
%     \AxiomC{}
%     \UnaryInfC{$\Seq{\Orth\Tnat,\Tnat}$}
%     \AxiomC{}
%     \UnaryInfC{$\Seq{\Tgtrad\tau,\Orthp{\Tgtrad\tau}}$}
%     \BinaryInfC{$\Seq{\Orth\Tnat,\Tgtrad\tau,\Tens\Tnat{\Orthp{\Tgtrad\tau}}}$}
%     \BinaryInfC{$\Seq{\Tgtradc\Phi,\Orth\Tnat,\Tgtrad\tau}$}
%     \AxiomC{$\Tgtrad u_\Phi$}
%     \noLine
%     \UnaryInfC{$\Seq{\Tgtradc\Phi,\Tnat}$}
%     \BinaryInfC{$\Seq{\Tgtradc\Phi,\Tgtrad\tau}$}
%     \DisplayProof
%   }
% \end{center}
% Then $(\gamma,b)\in\Psem{\Tgtrad s_\Phi}$ iff there is $n\in\Nat$ and
% $\gamma^0,\gamma^1$ such that $\gamma=\gamma^0\cup\gamma^1$ with
% $(\gamma^0,(n,b))\in\Psem{\Tgtrad t_\Phi}$ and
% $(\gamma^1,n))\in\Psem{\Tgtrad u_\Phi}$ and the required equation
% follows by inductive hypothesis.

If $s=\App tu$ with $\Tseq\Phi t{\Timpl\sigma\tau}$ and
$\Tseq\Phi u\sigma$ then $\Tgtrad s_\Phi$ is
\begin{center}
  {\footnotesize
    \AxiomC{$\Tgtrad t_\Phi$}
    \noLine
    \UnaryInfC{$\Seq{\Tgtradc\Phi,\Par{\Orthp{\Excl{\Tgtrad\sigma}}}{\Tgtrad\tau}}$}
    \AxiomC{}
    \UnaryInfC{$\Seq{\Orthp{\Excl{\Tgtrad\sigma}},{\Excl{\Tgtrad\sigma}}}$}
    \AxiomC{}
    \UnaryInfC{$\Seq{\Tgtrad\tau,\Orthp{\Tgtrad\tau}}$}
    \BinaryInfC{$\Seq{\Orthp{\Excl{\Tgtrad\sigma}},\Tgtrad\tau,\Tens{\Excl{\Tgtrad\sigma}}{\Orthp{\Tgtrad\tau}}}$}
    \BinaryInfC{$\Seq{\Tgtradc\Phi,\Orth{\Excl{\Tgtrad\sigma}},\Tgtrad\tau}$}
    \AxiomC{$\Tgtrad u_\Phi$}
    \noLine
    \UnaryInfC{$\Seq{\Tgtradc\Phi,{{\Tgtrad\sigma}}}$}
    \RightLabel{$\Nprom$}
    \UnaryInfC{$\Seq{\Tgtradc\Phi,{\Excl{\Tgtrad\sigma}}}$}
    \BinaryInfC{$\Seq{\Tgtradc\Phi,\Tgtradc\Phi,\Tgtrad\tau}$}
    \doubleLine
    \RightLabel{$\Ncontr$}
    \UnaryInfC{$\Seq{\Tgtradc\Phi,\Tgtrad\tau}$}
    \DisplayProof
  }
\end{center}
Then $(\gamma,b)\in\Psem{\Tgtrad s_\Phi}$ iff there are
$\List a1k\in\Web{\Psem{\Tgtrad\sigma}}$ and $\gamma^0,\List\gamma 1k$
such that
$\gamma=\gamma^0\cup\gamma^1\cup\cdots\cup\gamma^k\in\Web{\Psem{\Tgtrad\Phi}}$
with $(\gamma^0,\{\List a1k\},b))\in\Psem{\Tgtrad t_\Phi}$ and
$(\gamma^j,a_j))\in\Psem{\Tgtrad u_\Phi}$ for $j=1,\dots,k$ and the
required equation follows by inductive hypothesis.

% If $s=\Abst x\Tnat t$ with $\Tseq{\Phi,\Var x:\Gtnat}{s}{\tau}$ then
% $\Tgtrad s$ is
% \begin{center}
%   {\footnotesize
%     \AxiomC{$\Tgtrad t_{\Phi,\Var x:\Gtnat}$}
%     \noLine
%     \UnaryInfC{$\Seq{\Tgtradc\Phi,\Orthp{\Excl\Tnat},\Tgtrad\tau}$}
%     \AxiomC{$\Ppromcl{\Tnat}$}
%     \noLine
%     \UnaryInfC{$\Seq{\Orth\Tnat,\Excl\Tnat}$}
%     \BinaryInfC{$\Seq{\Tgtradc\Phi,\Orth{\Tnat},\Tgtrad\tau}$}
%     \UnaryInfC{$\Seq{\Tgtradc\Phi,\Par{\Orth{\Tnat}}{\Tgtrad\tau}}$}
%     \DisplayProof
%   }
% \end{center}
% where the proof $\Ppromcl\Tnat$ is defined in
% Theorem~\ref{th:mull-pos-coalg}. Observing that
% $\Psem{\Ppromcl\Tnat}=\Coalgstr\Snat$, the inductive hypothesis
% implies that $\Psem{\Tgtrad s_\Phi}=\Psem s_\Phi$.

If $s=\Abst x\sigma t$ with $\Tseq{\Phi,\Var x:\sigma}{s}{\tau}$
then $\Tgtrad s$ is
\begin{center}
  {\footnotesize
    \AxiomC{$\Tgtrad t_{\Phi,\Var x:\sigma}$}
    \noLine
    \UnaryInfC{$\Seq{\Tgtradc\Phi,\Orthp{\Excl{\Tgtrad\sigma}},\Tgtrad\tau}$}
    \UnaryInfC{$\Seq{\Tgtradc\Phi,\Par{\Orthp{\Excl{\Tgtrad\sigma}}}{\Tgtrad\tau}}$}
    \DisplayProof
  }
\end{center}
and the required equation results straightforwardly from the inductive
hypothesis.

If $s=\Let xtu$ with $\Tseq\Phi t\Gtnat$ and
$\Tseq{\Phi,x:\Gtnat}{u}{\sigma}$. Then $\Tgtrad s_\Phi$ is
\begin{center}
  {\footnotesize
    \AxiomC{$\Tgtrad t_\Phi$}
    \noLine
    \UnaryInfC{$\Seq{\Tgtradc\Phi,\Tnat}$}
    \AxiomC{$\Ppromcl\Tnat$}
    \noLine
    \UnaryInfC{$\Seq{\Orth\Tnat,\Excl\Tnat}$}
    \BinaryInfC{$\Seq{\Tgtradc\Phi,\Excl\Tnat}$}
    \AxiomC{$\Tgtrad u_{\Phi,x:\Gtnat}$}
    \noLine
    \UnaryInfC{$\Seq{\Tgtradc\Phi,\Orthp{\Excl\Tnat},\Tgtrad\sigma}$}
    \BinaryInfC{$\Seq{\Tgtradc\Phi,\Tgtradc\Phi,\Tgtrad\sigma}$}
    \doubleLine
    \UnaryInfC{$\Seq{\Tgtradc\Phi,\Tgtrad\sigma}$}
    \DisplayProof
  }
\end{center}
and the required equation follows directly from the fact that
$\Psem{\Ppromcl\Tnat}=\Coalgstr\Snat$.

If $s=\Rec tuv$ with $\Tseq{\Phi}{t}{\Gtnat}$,
$\Tseq{\Phi}{u}{\sigma}$,
$\Tseq{\Phi}{v}{\Timpl{\Gtnat}{\Timpl\sigma\sigma}}$
then $\Tgtrad s_\Phi$ is
\begin{center}
  {\footnotesize
    \AxiomC{$\Tgtrad u_\Phi$}
    \noLine
    \UnaryInfC{$\Seq{\Tgtradc\Phi,\Tgtrad\sigma}$}
    \UnaryInfC{$\Seq{\Tgtradc\Phi,\Excl{\Tgtrad\sigma}}$}
    \AxiomC{}
    \RightLabel{$\Nzero$}
    \UnaryInfC{$\Seq\Tnat$}
    \UnaryInfC{$\Seq{\Excl\Tnat}$}
    \BinaryInfC{$\phi:\ \Seq{\Tgtradc\Phi,
        \Tens{\Excl\Tnat}{\Excl{\Tgtrad\sigma}}}$}
    \AxiomC{$\Tgtrad{v}_\Phi$}
    \noLine
    \UnaryInfC{$\Seq{\Tgtradc\Phi,\Par{\Orthp{\Excl\Tnat}}{(\Par{\Orthp{\Excl{\Tgtrad\sigma}}}{\Tgtrad\sigma})}}$}
    \LeftLabel{$(\rho)$}
    \doubleLine
%    \UnaryInfC{$\Seq{\Tgtradc\Phi,\Orth\Tnat,\Orthp{\Excl{\Tgtrad\sigma}},\Tgtrad\sigma}$}
    \UnaryInfC{$\Seq{\Tgtradc\Phi,\Orthp{\Excl\Tnat},\Orthp{\Excl{\Tgtrad\sigma}},\Tgtrad\sigma}$}
    \UnaryInfC{$\Seq{\Tgtradc\Phi,\Orthp{\Excl\Tnat},\Orthp{\Excl{\Tgtrad\sigma}},\Excl{\Tgtrad\sigma}}$}
    \AxiomC{}
    \UnaryInfC{$\Seq{\Orth{\Tnat},\Tnat}$}
    \RightLabel{$\Nsucc$}
    \UnaryInfC{$\Seq{\Orth{\Tnat},\Tnat}$}
    \UnaryInfC{$\Seq{\Orthp{\Excl\Tnat},\Tnat}$}
    \UnaryInfC{$\Seq{\Orthp{\Excl\Tnat},\Excl\Tnat}$}
    \insertBetweenHyps{\hskip 8pt}
    \BinaryInfC{$\Seq{\Tgtradc\Phi,\Orthp{\Excl\Tnat},\Orthp{\Excl\Tnat},\Orthp{\Excl{\Tgtrad\sigma}},\Tens{\Excl\Tnat}{\Excl{\Tgtrad\sigma}}}$}
    \UnaryInfC{$\psi:\ \Seq{\Tgtradc\Phi,\Orthp{\Excl\Tnat},\Orthp{\Excl{\Tgtrad\sigma}},\Tens{\Excl\Tnat}{\Excl{\Tgtrad\sigma}}}$}
    \UnaryInfC{$\Seq{\Tgtradc\Phi,\Par{\Orthp{\Excl\Tnat}}{\Orthp{\Excl{\Tgtrad\sigma}}},\Tens{\Excl\Tnat}{\Excl{\Tgtrad\sigma}}}$}
    \RightLabel{$\Nintit$}
    \insertBetweenHyps{\hskip -10pt}
    \BinaryInfC{$\theta:\ \Seq{\Tgtradc\Phi,\Orth\Tnat,\Tens{\Excl\Tnat}{\Excl{\Tgtrad\sigma}}}$}
    \AxiomC{}
    \UnaryInfC{$\Seq{{\Orthp{{\Tgtrad\sigma}}},\Tgtrad\sigma}$}
    \UnaryInfC{$\Seq{{\Orthp{\Excl{\Tgtrad\sigma}}},\Tgtrad\sigma}$}
    \UnaryInfC{$\Seq{{\Orthp{\Excl\Tnat}},{\Orthp{\Excl{\Tgtrad\sigma}}},\Tgtrad\sigma}$}
    \UnaryInfC{$\Seq{\Par{\Orthp{\Excl\Tnat}}{\Orthp{\Excl{\Tgtrad\sigma}}},\Tgtrad\sigma}$}
    \insertBetweenHyps{\hskip 2pt}
    \BinaryInfC{$\Seq{\Tgtradc\Phi,\Orth\Tnat,\Tgtrad\sigma}$}
    \AxiomC{$\Tgtrad t_\Phi$}
    \noLine
    \UnaryInfC{$\Seq{\Tgtrad\Phi,\Tnat}$}
    \insertBetweenHyps{\hskip 2pt}
    \BinaryInfC{$\Seq{\Tgtradc\Phi,\Tgtradc\Phi,\Tgtrad\sigma}$}
    \doubleLine
    \UnaryInfC{$\Seq{\Tgtradc\Phi,\Tgtrad\sigma}$}    
    \DisplayProof
  }
\end{center}
where $\rho$ consists of two cuts against tensor rules.  The fact that
$\Psem{\Tgtrad s_\Phi}=\Psem s_\Phi$ can be seen as follows (we
actually apply an old trick for encoding primitive recursion using
iteration).

First observe that $\Psem{\Tgtrad v_\Phi}$ can be seen as a stable
function
$\Cl{\Psem{\Tgtrad{\sigma_1}}}\times\cdots\times
\Cl{\Psem{\Tgtrad{\sigma_k}}}\times\Cl\Snat\times
\Cl{\Psem{\Tgtrad\sigma}}\to\Psem{\Tgtrad\sigma}$,
% linear in its $k+1$-th argument,
that $\Psem{\Tgtrad u_\Phi}$ can be
seen as a stable function
$\Cl{\Psem{\Tgtrad{\sigma_1}}}\times\cdots
\times\Cl{\Psem{\Tgtrad{\sigma_k}}}\to\Psem{\Tgtrad\sigma}$ and that
$\Psem{\Tgtrad t_\Phi}$ can be seen as a stable function
$\Cl{\Psem{\Tgtrad{\sigma_1}}}\times\cdots
\times\Cl{\Psem{\Tgtrad{\sigma_k}}}\to\Psem{\Snat}$.

Then the proof $\phi$ above the sequent
$\Seq{\Tgtradc\Phi, \Tens{\Excl\Tnat}{\Excl{\Tgtrad\sigma}}}$
satisfies
$\Psem\phi(\Vect x)=\Prom{\Psem{\Tgtrad u_\Phi}(\Vect
  x)}\ITens\Prom{\{0\}}$. The proof $\psi$ can be seen as a stable
function
$\Cl{\Psem{\Tgtrad{\sigma_1}}}\times\cdots\times
\Cl{\Psem{\Tgtrad{\sigma_k}}}\times\Cl\Snat\times
\Cl{\Psem{\Tgtrad\sigma}}\to{\Excl\Snat\ITens\Excl{\Psem{\Tgtrad\sigma}}}$
such that
$\Psem\psi(\Vect x,y,x)=\Prom{(y^+)}\ITens\Prom{\Psem{\Tgtrad
    v_\Phi}(\Vect x,y,x)}$ where $y^+=\{n+1\St n\in y\}$. Applying the
interpretation of the derived rule $\Nintit$ as described in
Section~\ref{sec:strict-int}, an easy induction shows that
\begin{align*}
\Psem\theta(\Vect x,\{0\})&=\Prom{\{0\}}\ITens\Prom{\Psem{\Tgtrad
    u_\Phi}(\Vect x)}\\
\Psem\theta(\Vect x,\{n+1\})&=\Tens{\Prom{\{n+1\}}}{\Prom{\Psem{\Tgtrad
    v_\Phi}(\Vect x,\{n\},\Psem\theta(\Vect x,\{n\}))}}
\end{align*}
from which the required equation follows. So we have proved the
following result.
\begin{thm}\label{th:sem-muLL-T-equal}
  If $s$ is a term of $\GODELT$ such that $\Tseq{\Phi}{s}{\tau}$ then
  $\Psem{\Tgtrad s_\Phi}=\Psem s_\Phi$.
\end{thm}

\begin{thm}
  For any function $f:\Nat\to\Nat$, if $f$ can be represented by a
  term of $\GODELT$ then there is a proof $\pi$ if $\MULL$ of
  $\Seq{\Orth\Tnat,\Tnat}$ such that for all $n\in\Nat$ the proof
  \begin{center}
    \AxiomC{$\pi$}
    \noLine
    \UnaryInfC{$\Seq{\Orth\Tnat,\Tnat}$}
    \AxiomC{$\Num n$}
    \noLine
    \UnaryInfC{$\Seq{\Tnat}$}
    \BinaryInfC{$\Seq{\Tnat}$}
    \DisplayProof
  \end{center}
  reduces to $\Num{f(n)}$.
\end{thm}
\proof Let $s$ be a closed term of $\GODELT$ such that
$\Tseq{}{s}{\Timpl\Gtnat\Gtnat}$ and $s$ represents $f$ (that is, for
all $n\in\Nat$, $\App s{\Num n}\Reltr\Gredwh\Num{f(n)}$). Let
$\pi=\Tgtrad s$ which is a proof of
$\Seq{\Par{\Orthp{\Excl\Tnat}}{\Tnat}}$, that is of
$\Seq{\Orthp{\Excl\Tnat},\Tnat}$ (up to a cut with a trivial
proof). Baelde's normalization theorem tells us that for all
$n\in\Nat$ the following proof\footnote{Observe that $\Tgtrad{\Num n}$ is the
  canonical representation of $n$ as a proof of $\MULL$ consisting of
  a stack of $n$ $\Nsucc$ rules topped by a $\Nzero$ axiom.} $\pi_n$
\begin{center}
  \AxiomC{$\pi$}
  \noLine
  \UnaryInfC{$\Seq{\Orthp{\Excl\Tnat},\Tnat}$}
  \AxiomC{$\Tgtrad{\Num n}$}
  \noLine
  \UnaryInfC{$\Seq{\Tnat}$}
  \RightLabel{$\Nprom$}
  \UnaryInfC{$\Seq{\Excl\Tnat}$}
  \BinaryInfC{$\Seq{\Tnat}$}
  \DisplayProof
\end{center}
reduces to a proof $\Tgtrad{\Num k}$ for some $k\in\Nat$
% (but \emph{a priori} there might be several such $k$'s)
so that $\Psem{\pi_n}=\{k\}$ by
Theorem~\ref{th:muLL-sem-invariant}. By
Theorem~\ref{th:sem-muLL-T-equal} we have
$\{f(n)\}=\Psem{\App s{\Num n}}=\Psem{\pi_n}=\{k\}$ and hence $k=f(n)$
which proves our contention. \qed
%\todot{Do we need more for System T?}

\subsection{Encoding of the exponentials}\label{sec:exclmu}
% In a Seely model $\cL$ of $\LL$, there is a comonad $\oc:\cL\to\cL$
% with counit $\Der X\in\cL(\Excl X,X)$ and comultiplication
% $\Digg X\in\cL(\Excl X,\Excl{\Excl X})$. This comonad is also
% equiped with a monoidal structure from the monoidal category
% $(\cL,\mathord{\IWith})$ to the monoidal category
% $(\cL,\mathord\ITens)$ and
In~\cite{Baelde12} it is noticed that the exponential connectives of
$\LL$ can be encoded using least and greatest fixed points. It is also
mentioned that this encoding is shallow in the sense that it allows,
on formulas involving exponentials, proofs which do not make sense in
$\LL$. We propose here a short denotational analysis of this encoding
showing that it gives rise to a resource comonad which however does
not validate the Seely isomorphism. This is our motivation for
including the exponentials in $\MULL$ from the beginning. So let us
set
\begin{align*}
  \Exclmu A=\Gfpll\zeta{(\One\IWith A\IWith\Tensp\zeta\zeta)}
  \quad\text{and}\quad
  \Intmu A=\Orthp{\Exclmu{\Orth A}}
  =\Lfpll\zeta{(\Fbot\IPlus A\IPlus\Parp\zeta\zeta)}\,.
\end{align*}

\subsubsection{Functoriality and comonadic structure} We consider the
binary VCST $\Vcstnot X$ given by
$\Vcstnot X(X,U)=\One\IWith X\IWith\Tensp UU$ and we set
$\Exclmu{X}=\Vcstnu{\Vcstnot X_X}$ (remember that this notation
$\Vcstnot X_X$ is introduced at the end of the Introduction for
arbitrary functors) so that $\Psem{\Exclmu A}=\Exclmu{\Psem A}$.  We
set $\Dermu X=\Proj 1\in\COHT(\Exclmu X,X)$ since
$\Vcstnot X(X,\Exclmu X)=\Exclmu X$ and similarly
$\Weakmu X=\pi_0\in\COHT(\Exclmu X,\One)$ and
$\Contrmu X=\pi_2\in\COHT(\Exclmu X,\Tens{\Exclmu X}{\Exclmu X})$,
these maps exhibit on $\Exclmu X$ a structure of ``free co-magma''
generated by $X$.

For
$g\in\COHT(Y,\Vcstnot X_X(Y))$, there is exactly one
$\Exclmuext g\in\COHT(Y,\Exclmu X)$ such that
$\Exclmuext g=\Vcstnot X_X(\Exclmuext g)\Compl g$, that is
\begin{align}\label{eq:exclmuext-charact}
  \Weakmu Y\Compl\Exclmuext g=\Proj 0\Compl g\quad
  \Dermu Y\Compl\Exclmuext g=\Proj 1\Compl g\quad
  \Contrmu Y\Compl\Exclmuext g=
  \Tensp{\Exclmuext g}{\Exclmuext g}\Compl\Proj 2\Compl g\,.
\end{align}
Let $f\in\COHT(\Exclmu X,Y)$, we have
$\Tuple{\Weakmu X,f,\Contrmu X}\in\COHT(\Exclmu X,\Vcstnot X_Y(\Exclmu
X))$ and so we can set
$\Prommu f=\Exclmuext{\Tuple{\Weakmu X,f,\Contrmu X}}\in\COHT(\Exclmu
X,\Exclmu Y)$ so that this morphism $\Prommu f$ is fully characterized
by the following equations
\begin{align}\label{eq:prommu-charact}
  \Weakmu Y\Compl\Prommu f=\Weakmu X\quad
  \Dermu Y\Compl\Prommu f=f\quad
  \Contrmu Y\Compl\Prommu f=\Tensp{\Prommu f}{\Prommu f}\Compl\Contrmu X\,.
\end{align}
Then the action of the exponential on morphisms is defined as
$\Exclmu f=\Prommu{(\Compl f\Dermu X)}$ for $f\in\COHT(X,Y)$ and is
easily seen to be functorial using
Equations~\Eqref{eq:prommu-charact}. Observe that $\Exclmu f$ is
characterized by the following equations.
\begin{align}\label{eq:exclmu-functor-charact}
  \Weakmu Y\Compl\Exclmu f=\Weakmu X\quad
  \Dermu Y\Compl\Exclmu f=f\Dermu Y\quad
  \Contrmu Y\Compl\Exclmu f=\Tensp{\Exclmu f}{\Exclmu f}\Compl\Contrmu X\,.
\end{align}

Last the comultiplication of the
exponential is defined as
\[
\Diggmu X=\Prommu{(\Id_{\Exclmu X})}\in\COHT(\Exclmu
X,\Exclmu{\Exclmu X})
\]
and is easily checked to satisfy the required commutations for the
comonad structure, using again Equations~\Eqref{eq:prommu-charact}.

Now we will argue that this exponential lacks the required Seely
isomorphisms that we mentioned in Section~\ref{sec:LL-models}. Indeed,
if such isomorphisms
$\Seemu_{X,Y}\in\COHT(\Tens{\Exclmu X}{\Exclmu Y},\Exclmu{(\With
  XY)})$ and
$\Seeinvmu_{X,Y}\in\COHT(\Exclmu{(\With XY)},\Tens{\Exclmu X}{\Exclmu
  Y})$ existed they would satisfy the following equations
\begin{align}\label{eq:Seelymu_charact}
  \begin{split}
  \Weakmu{\With XY}\Compl\Seemu_{X,Y}&=\Tens{\Weakmu X}{\Weakmu Y}\\
  \Dermu{\With XY}\Compl\Seemu_{X,Y}
  &=\Tuple{\Tens{\Dermu X}{\Weakmu Y},\Tens{\Weakmu X}{\Dermu Y}}\\
  \Contrmu{\With XY}\Compl\Seemu_{X,Y}
  &=\Tensp{\Seemu_{X,Y}}{\Seemu_{X,Y}}\Compl\Sym_{2,3}
  \Compl\Tensp{\Contrmu X}{\Contrmu Y}
  \end{split}
\end{align}
and
\begin{equation}
  \label{eq:seelyinvmu-charact}
  \Seeinvmu_{X,Y}=\Tensp{\Exclmu{\Proj 1}}{\Exclmu{\Proj 2}}
  \Compl\Contrmu{\With XY}\,.
\end{equation}
Equation~\Eqref{eq:seelyinvmu-charact} can be understood as a mere
definition of $\Seeinvmu_{X,Y}$ since we know how to compute its right
hand member. And by the
characterization~\Eqref{eq:exclmuext-charact}, the
equations~\Eqref{eq:Seelymu_charact} determine $\Seemu_{X,Y}$
uniquely. So it suffices to compute these two morphisms in our model
$\COHT$ to convince ourselves that they do not define
an isomorphism.

First of all, given a coherence space with totality $X$, observe that
the elements $\alpha,\alpha_1,\dots$ of the web of the coherence space
$\Tcohca{\Exclmu X}$ can be described by the following grammar
\begin{equation*}
  \alpha,\alpha_1,\dots \Bnfeq \Wweak \Bnfor \Wder a
  \Bnfor \Wcontr{\alpha_1}{\alpha_2}
\end{equation*}
where $a$ stands for the elements of $\Web{\Tcohca X}$. In other
words, an element of $\Web{\Tcohca{\Exclmu X}}$ is a finite binary
tree with two kinds of leaves: ``weakening leaves'' $\Wweak$ and
``dereliction leaves'' $\Wder a$. Coherence in $\Tcohca{\Exclmu X}$ is
given by: $\Coh{\Tcohca{\Exclmu X}}{\alpha}{\alpha'}$ if
$\alpha=\Wder a,\ \alpha'=\Wder{a'}\Implies\Coh{\Tcohca X}a{a'}$ and
$\alpha=\Wcontr{\alpha_1}{\alpha_2},\
\alpha'=\Wcontr{\alpha'_1}{\alpha'_2}\Implies\Coh{\Tcohca{\Exclmu
    X}}{\alpha_i}{\alpha'_i}\text{ for }i=1,2$.  Totality
$\Tcoht{\Exclmu X}$ is defined as a greatest fixed point as explained
in Section~\ref{sec:VCST}, and we don't know yet any more explicit
presentation of this totality. We call \emph{paths} the elements of
$\{1,2\}^{<\omega}$ and we define a partial function
$\Subtree:\Web{\Exclmu X}\times\{1,2\}^{<\omega}\to\Web{\Exclmu X}$ by
\begin{equation*}
  \begin{split}
    \Subtree(\alpha,\Emptypath)&=\alpha\\
    \Subtree(\Wcontr{\alpha_1}{\alpha_2},\Conspath
    i\theta)&=\Subtree(\alpha_i,\theta)
  \end{split}
\end{equation*}
and $\Pathes\alpha$ is the set of paths $\theta$ such that
$\Subtree(\alpha,\theta)$ is defined (it is a finite set).

Let $f\in\COHT(X,Y)$. By Equations~\Eqref{eq:exclmu-functor-charact}
and by the fact that $\Exclmuext g$ is defined as a least fixed point
(see Section~\ref{sec:VCST}) we get that $\Exclmu f$ is the least set
such that $(\alpha,\beta)\in\Exclmu f$ iff $\alpha=\beta=\Wweak$ or
$\alpha=\Wder a$, $\beta=\Wder b$ and $(a,b)\in f$ or
$\alpha=\Wcontr{\alpha_1}{\alpha_2}$,
$\beta=\Wcontr{\beta_1}{\beta_2}$ and $(\alpha_i,\beta_i)\in\Exclmu f$
for $i=1,2$. In other words $(\alpha,\beta)\in\Exclmu f$ iff
$\Pathes\alpha=\Pathes\beta=D$ and for all $\theta\in D$ which is
maximal for the prefix order:
\begin{itemize}
\item either $\Subtree(\alpha,\theta)=\Subtree(\beta,\theta)=\Wweak$
% \item or $\Subtree(\alpha,\theta)=\Wcontr{\alpha_1}{\alpha_2}$ and
%   $\Subtree(\beta,\theta)=\Wcontr{\beta_1}{\beta_2}$
\item or $\Subtree(\alpha,\theta)=\Wder a$ and
  $\Subtree(\beta,\theta)=\Wder b$ with $(a,b)\in f$.
\end{itemize}
That is, $\alpha$ and $\beta$ have exactly the same shape and
corresponding $\Wder\_$-leaves of $\alpha$ and $\beta$ are related by
$f$.

Now Equation~\Eqref{eq:seelyinvmu-charact} gives us an explicit
definition of $\Seeinvmu_{X,Y}$, namely:
$(\gamma,(\alpha,\beta))\in\Seeinvmu_{X,Y}$ iff
$\gamma=\Wcontr{\Winl\alpha}{\Winr\beta}$ where $\Winl\Wweak=\Wweak$,
$\Winl{\Wder a}=\Wder{1,a}$ and
$\Winl{\Wcontr{\alpha_1}{\alpha_2}}=\Wcontr{\Winl{\alpha_1}}{\Winl{\alpha_2}}$
and similarly for $\Winr\beta$. The map
$(\alpha,\beta)\mapsto\Wcontr{\Winl\alpha}{\Winr\beta}$ is injective
but not surjective and so $\Seeinvmu_{X,Y}$ is not an
iso\footnote{Remember that $f\in\COH(E,F)$ is an iso iff, as a
  relation, $f$ is the graph of a bijection $\phi:\Web E\to\Web F$
  such that $\Coh Ea{a'}\Equiv\Coh F{\phi(a)}{\phi(a')}$}.

This suffices to show that $\Exclmu\_$ does not endow $\COHT$ with a
structure of Seely category (and hence is not a categorical model of $\LL$ in
the usual sense), but for the sake of completeness we also give
$\Seemu_{X,Y}$. It is the least set such that
$((\alpha,\beta),\gamma)\in\Seemu_{X,Y}$ iff
\begin{itemize}
\item $\alpha=\beta=\gamma=\Wweak$
\item or $\gamma=\Wder{1,a}$, $\alpha=\Wder a$ and $\beta=\Wweak$
\item or $\gamma=\Wder{2,b}$, $\alpha=\Wweak$ and $\beta=\Wder b$
\item or $\gamma=\Wcontr{\gamma_1}{\gamma_2}$,
  $\alpha=\Wcontr{\alpha_1}{\alpha_2}$,
  $\beta=\Wcontr{\beta_1}{\beta_2}$ and
  $((\alpha_i,\beta_i),\gamma_i)\in\Seemu_{X,Y}$ for $i=1,2$.
\end{itemize}
In other words, $((\alpha,\beta),\gamma)\in\Seemu_{X,Y}$ iff
$\alpha=\Wprl\gamma$ and $\beta=\Wprr\gamma$ where
$\Pathes{\Wprl\gamma}=\Pathes\gamma=D$, and for all $\theta\in D$, if
$\theta$ is maximal (for the prefix order),
$\Subtree(\gamma,\theta)=\Wder{1,a}\Implies\Subtree(\Wprl\gamma,\theta)=\Wder
a$ and
$\Subtree(\gamma,\theta)=\Wder{2,b}\Implies\Subtree(\Wprl\gamma,\theta)
=\Wweak$. That is, $\Wprl\gamma$ is $\gamma$ where the $b$'s are
replaced with $\Wweak$. The definition of $\Wprr\gamma$ is
symmetrical. Again, the map $\gamma\mapsto(\Wprl\gamma,\Wprr\gamma)$
is injective but not surjective since $\Wprl\gamma$ and $\Wprr\gamma$
have the same ``internal'' tree structure (namely, that of $\gamma$).

\begin{rem}
  Such non-commutative exponentials in coherence spaces have been
  introduced by Myriam Quatrini in her PhD thesis~\cite{Quatrini95} as
  early as in~1995 and it would be interesting to compare them with
  this $\Exclmu\_$ construction. One major feature of this exponential
  is that, unlike the usual set-based and multiset-based exponentials
  in coherence spaces, the $\Exclmu\_$ is non-uniform, meaning that
  the elements of $\Web E$ occurring in an element $\alpha$ of
  $\Web{\Exclmu E}$ are not required to be pairwise coherent. There is
  a price to pay for this feature: the Seely isos are not
  available\footnote{They are replaced by injections, which are not so
    far from being bijections after all\dots} and so the Kleisli
  category $\COH_{\Exclmu{}}$ is not cartesian
  closed. In~\cite{BucciarelliEhrhard99} we exhibited another way of
  accommodating non-uniform exponentials with coherence spaces, in a
  true model of $\LL$ and there, the price to pay was the loss of
  coherence's reflexivity (on the web). It would certainly be
  interesting to explore the connection between these two kinds of
  non-uniform exponentials in coherence spaces.
\end{rem}

\section{Other models and generalizations}\label{sec:generalizations}

The main feature of our model construction is that it is based on a
two-level structure:
\begin{itemize}
\item The $\LL$ model $\COH$ of coherence spaces, where a notion of
  variable object can be defined for which least and greatest
  fixed-points coincide
\item and the $\LL$ model of $\COHT$ of coherence spaces with totality
  where we can define a notion of variable types by equipping the
  variable types of the first level with an additional structure
  (here, a totality structure). In this second level, least and
  greatest fixed points are interpreted in different ways.
\end{itemize}

The same two-level structure can certainly be developed on the model
of hypercoherence spaces~\cite{Ehrhard93}, on the Scott semantics of
$\LL$~\cite{Ehrhard11b} or on the model of probabilistic coherence
space~\cite{DanosEhrhard08}. In all these cases a notion of totality
similar to that of $\COHT$ can be defined.

A slightly different structure can be obtained by using the category
$\REL$ of sets and relations as base category and the category of
finiteness spaces~\cite{Ehrhard00b} (which are sets equipped with a
finiteness structure) as level-2 structure. In this case, an
interesting challenge will be to understand how the fixed point
constructs interact with the \emph{linearization} typical of
finiteness spaces: given a field, any finiteness space can canonically
be turned into a topological vector space over this field (considered
as a discrete space). In this case there might even be a third level
since these linearizations can be equipped with a totality which is
simply a closed affine subspace as explained in~\cite{Tasson09}.

\subsection{A categorical axiomatization models of $\MULL$}
\label{sec:cat-models}
In Section~\ref{sec:LL-models} we tried to give as much general
categorical definitions as possible, to serve as a background for the
concrete developments in further sections, and also for a general
categorical presentation of $\MULL$ models extending naturally the
concept of Seely category. Here is a tentative such definition,
compatible with the example we have presented in the paper and the
other examples that we mentioned.

\begin{defi}\label{def:categorical-muLL-models}
  A \emph{categorical model of $\MULL$} is a family
  $\Vect\cL=(\cL_n)_{n\in\Nat}$ where
  \begin{itemize}
  \item $\cL_0$ is a Seely category
  \item $\cL_n$ is a family of strong functors $\cL_0^n\to\cL_0$ and
    our choice of notations for $\cL_0$ means that all constant
    functors are in $\cL_0$ (see Section~\ref{sec:gen-strong-functors}
    for basic definitions on strong functors in our $\LL$ categorical
    setting)
  \item if $\Vcstnot X\in\cL_n$ and $\Vcstnot X_i\in\cL_k$ (for
    $i=1,\dots,n$) then $\Vcstnot X\Comp\Vect{\Vcstnot X}\in\cL_k$
  \item the strong functors $\ITens$ and $\IWith$ belong to $\cL_2$,
    the strong functor $\Excl\_$ belongs to $\cL_1$ and, if
    $\Vcstnot X\in\cL_n$, then $\Orth{\Vcstnot X}\in\cL_n$
  \item and last, for all $\Vcstnot X\in\cL_1$ the category
    $\COALGFUN{\cL_0}{\Vcstnot X}$ (see
    Section~\ref{sec:fixpoints-functors}) has a final object or,
    equivalently, the category $\ALGFUN{\cL_0}{\Vcstnot X}$ has an
    initial object.
  \end{itemize}
\end{defi}

Of course we should interpret, in such an abstract model, all sequents
and proofs of $\MULL$ and we are confident that this can be done
following the pattern we developed for the interpretation in $\COHT$
(where $\cL_n$ is of course the class of $n$-ary VCSTs, considered as
strong functors as explained in Section~\ref{sec:VCST}); such a formal
verification is postponed to further work.

\section{Conclusion}

One of the main goals of this work was to develop syntax-independent
tools to study new proof-systems for $\MULL$ and more specifically
infinite proof-systems as in~\cite{BaeldeDoumaneSaurin16}, and
denotational semantics is clearly a natural framework for such
tools. A first step in this study will be to prove that these infinite
proofs can be interpreted in the model of VCSTs. It would be also
quite interesting to understand how the notion of totality is related
with that of \emph{productiveness}, crucial in the study of
coinductive types. Our models should also suggest natural notions of
proof-nets for $\MULL$.

We would like to thank many people for exciting and quite helpful
discussions on these topics, and in particular Amina Doumane,
Paul-André Melliès, Rémy Nollet, Alexis Saurin and Christine Tasson.

\bibliographystyle{plain}

\bibliography{newbiblio}

% The bibliography should be embedded for final submission.

% \section{Appendix}

% \subsection{Proof of Lemma~\ref{lemma:functor-gfp-general}}

% \subsection{Proof of Lemma~\ref{lemma:strfun-gfp-general}}

% \subsection{Proof of Proposition~\ref{prop:VCST-mu-init}}

% \subsection{Standard deduction rules of $\LL$}

\end{document}